\documentclass{aa}

\usepackage{txfonts}
\usepackage{sistyle}
\usepackage{multirow}
\usepackage{graphicx,epstopdf}
\usepackage{natbib}
\bibpunct{(}{)}{;}{a}{}{,}

\graphicspath{{Images/}}

\newcommand{\ebv}{\ifmmode E_{\rm B-V} \else $E_{\rm B-V}$\fi}
\newcommand{\oh}{\ifmmode 12+\log({\rm O/H}) \else$12+\log({\rm O/H})$\fi}
\newcommand{\nh}{\ifmmode N_{\rm H~I} \else $N_{\ion{H}{i}}$\fi}
\newcommand{\nhthin}{\ifmmode N_{\rm H~I}^{\rm thin} \else $N_{\ion{H}{i}}^{\rm thin}$\fi}
\newcommand{\nhthick}{\ifmmode N_{\rm H~I}^{\rm thick} \else $N_{\ion{H}{i}}^{\rm thick}$\fi}
\newcommand{\nhchan}{\ifmmode N_{\rm H~I}^{\rm channel} \else $N_{\ion{H}{i}}^{\rm channel}$\fi}
\newcommand{\nhcloud}{\ifmmode N_{\rm H~I}^{\rm cloud} \else $N_{\ion{H}{i}}^{\rm cloud}$\fi}
\newcommand{\no}{\ifmmode N_{\rm O~I} \else $N_{\ion{O}{i}}$\fi}
\newcommand{\fesc}{\ifmmode f_{\rm esc} \else $f_{\rm esc}$\fi}
\newcommand{\fesclyc}{\ifmmode f_{\rm esc}({\rm LyC}) \else $f_{\rm esc}{\rm(LyC)}$\fi}
\newcommand{\fesclycobs}{\ifmmode f_{\rm esc}^{\rm obs}({\rm LyC}) \else $f_{\rm esc}^{\rm obs}{\rm(LyC)}$\fi}
\newcommand{\lya}{\ifmmode {\rm Ly}\alpha \else Ly$\alpha$\fi}
\newcommand{\fesclya}{\ifmmode f_{\rm esc}({\rm Ly}\alpha) \else $f_{\rm esc}{({\rm Ly}\alpha)}$\fi}
\newcommand{\vsep}{\ifmmode v^{\rm sep}_{\rm \lya} \else $v^{\rm sep}_{\rm \lya}$\fi}
\newcommand{\sone}{SGAS J1226} 
\newcommand{\stwo}{SGAS J1527} 
\newcommand{\megasaura}{M\textsc{eg}a\textsc{S}a\textsc{ura}}
\usepackage{arydshln}
\usepackage{amsmath}

\usepackage{xcolor}

 \makeatletter
\def\@fnsymbol#1{\ensuremath{\ifcase#1\or \dagger\or \ddagger\or
   \mathsection\or \mathparagraph\or \|\or **\or \dagger\dagger
   \or \ddagger\ddagger \else\@ctrerr\fi}}
    \makeatother

\begin{document} 

\title{The origin of the escape of Lyman $\alpha$ and ionizing photons in Lyman Continuum Emitters}
\author{S. Gazagnes\inst{1,2} \and J. Chisholm\inst{3,}\thanks{Hubble Fellow}  \and D. Schaerer\inst{2,4} \and A. Verhamme\inst{2} \and Y. Izotov \inst{5} }

\offprints{s.r.n.gazagnes@rug.nl}

\institute{
Kapteyn Astronomical Institute, University of Groningen, P.O Box 800, 9700 AV Groningen, The Netherlands\and
Observatoire de Gen\`eve, Universit\'e de Gen\`eve, 51 Ch. des Maillettes, 1290 Versoix, Switzerland
\and
University of California-Santa Cruz, 1156 High Street, Santa Cruz, CA, 95064, USA
\and
 CNRS, IRAP, 14 Avenue E. Belin, 31400 Toulouse, France \and
Bogolyubov Institute for Theoretical Physics, National Academy of Sciences of Ukraine, 14-b Metrolohichna str., Kyiv 03143, Ukraine
}

\date{Received <date> / Accepted <date>}

\titlerunning{The escape of \lya\ and LyC photons in porous ISM}
\abstract
{Identifying the physical mechanisms driving the escape of Lyman Continuum (LyC) photons is crucial to find Lyman Continuum Emitter (LCE) candidates.}
{To understand the physical properties involved in the leakage of LyC photons, we investigate the connection between the \ion{H}{i} covering fraction, \ion{H}{i} velocity width, the Lyman $\alpha$ (\lya) properties and the escape of LyC photons in a sample of 22 star-forming galaxies including 13 confirmed LCEs.}
{We fit the stellar continuum, dust attenuation, and absorption lines between 920 \AA\ and 1300 \AA\ to extract the \ion{H}{i} covering fractions and dust attenuation. Additionally, we measure the \ion{H}{i} velocity widths of the optically thick Lyman series and derive the \lya\ equivalent widths (EW), escape fractions (\fesc), peak velocities and fluxes at the minimum of the observed \lya\ profiles.}
{Overall, we highlight strong observational correlations between the presence of low \ion{H}{i} covering fractions and the observation of (1) low \lya\ peak velocities; (2) more flux at the profile minimum; and (3) larger EW(\lya), \fesclya, and \fesclycobs. Hence, low column density channels are crucial ISM ingredients for the leakage of \lya\ and LyC photons.  Additionally, galaxies with narrower \ion{H}{i} absorption velocity widths have higher \lya\ equivalent widths, larger \lya\ escape fractions, and lower \lya\ peak velocity separations. This may suggest that these galaxies have lower \ion{H}{i} column densities. Finally, we find that dust also regulates the amount of \lya\ and LyC radiation that actually escapes the ISM.}
{The ISM porosity is one origin of strong \lya\ emission and enables the escape of ionizing photons in low-z leakers. However, this is not enough to explain the largest \fesclycobs, which indicates that the most extreme LCEs are likely density-bounded along all lines of sight to the observer. Overall, the neutral gas porosity constrains a lower limit to the escape fraction of LyC and \lya\ photons, providing a key estimator of the leakage of ionizing photons. }

\maketitle

\section{Introduction}

The Epoch of Reionization (EoR) is a key transition phase in the history of the Universe, which is still
largely unconstrained from observations. In the upcoming era of large telescopes, numerous galaxies within the EoR will be observed, but determining the mechanisms responsible for the propagation of the ionizing radiation from the interstellar medium (ISM) to the intergalactic medium (IGM) is crucial. Although the ionizing contribution of AGN during reionization is actively discussed  \citep{fontanot2012, fontanot2014, robertson2015, madau2015, mitra2018}, many studies suggest that a population of low-mass star-forming galaxies with an average escape fraction of Lyman Continuum (LyC) photons of 10-20\% probably dominates the contribution to the ionizing budget of the EoR \citep{ouchi, robertson, dressler, finkelstein2019}. However, the quest to find LyC leaking galaxies at high redshift ($z > 6$) is very challenging, and direct observations of the ionizing flux at $\lambda < 912\ \AA$ are complicated (or statistically unfeasible) due the attenuation of the IGM and the presence of interlopers along the line of sight. 

These caveats have been overcome by searches for compact, low-mass star-forming galaxies in the local universe, which can serve as analogs, and significant progress has been made over the past few years, with successful detections of LyC leakage at z < 0.5 \citep[15 individual detections,][]{bergvall2006, leitet2013, borthakur2014, leitherer2016, izotov2016a, izotov2016b, izotov2018a, izotov2018b}, and at z\textasciitilde\ 2-3 \citep{vanzella2015, debarros2016, shapley2016, bian2017, steidel2018, fletcher2019,rivera2019}. This recent breakthrough has provided an ideal laboratory to explore the physical properties that favor the leakage of ionizing radiation in the LyC leakers.

Studies commonly refer to two major physical models to understand how LyC photons escape galaxies \citep[see e.g.][]{zackrisson2013}. In a {\it density-bounded} ISM, the \ion{H}{i} column density (\nh) surrounding the stellar populations is too low (< 10$^{17.9}$ cm$^{-2}$) to efficiently absorb all the ionizing photons passing through the neutral clouds, and the fraction of LyC radiation that escapes the ISM is proportional to the residual \nh\ in the galaxy. Conversely, in a {\it picket-fence} or {\it ionization bounded} system, the bulk of the stars is surrounded by an optically thick ISM, where the average \nh\ is high enough to efficiently absorb the LyC photons. In this model, ionizing photons escape through privileged paths through the ISM, referred to as holes or channels, which have no or little \ion{H}{i} column densities. This scenario is mainly characterized by the porosity of the neutral gas, which is defined as the fraction of all sightlines to all of the ensemble of far-UV continua sources that are "covered" by \ion{H}{i} gas with column densities above a certain value. One way to define this \ion{H}{i} "covering fraction" ($C_f$(\ion{H}{i})) is by using a set of \ion{H}{i} absorption lines in the far-UV that have a range of oscillator strengths and saturate above a certain column density (between \textasciitilde $10^{15}$ and $10^{16}$ cm$^{-2}$ for the range from Ly$\beta$ to Ly7). The range of oscillator strengths allows for simultaneous optical depth and covering fraction determination of the Lyman series lines. The \ion{H}{i} covering fraction measured using this approach is always a lower limit to the true geometric covering fraction of the neutral gas, because of potential kinematic effects \citep{jones2013, rivera2015, vasei2016}, or the presence of \ion{H}{i} residuals \citep{kakiichi2019}. Nevertheless, in this paper, we use this measure of $C_f$(\ion{H}{i}) as a proxy of the overall neutral gas porosity and study how that porosity relates to the escape of Ly$\alpha$ and LyC photons. 

Investigating which model represents the ISM of LCEs is also crucial to find indirect tracers of the leakage of ionizing photons. To do this, low-redshift observations remain the ideal approach. Indeed, far-UV spectroscopic observations are less likely to be contaminated by line-of-sight absorbers in the IGM, thus their LyC escape fraction can be measured directly from the flux at $\lambda$ < 912 \AA. Additionally, the study of their UV \ion{H}{i} and metal lines constrain their neutral gas properties and identify the processes driving the LyC leakage. 
Reliable LyC probes have already been identified by studying the \lya\ properties \citep[see e.g.][]{verhamme2017,izotov2018b,izotov2019}.

\lya\ also provides powerful insights on the distribution and kinematics of the neutral gas, and is closely related to the LyC properties. Indeed, the ionizing radiation arising from young massive stars creates \ion{H}{ii} regions surrounding the star-forming clusters, which produces \lya\ radiation due to the recombination of hydrogen atoms. \lya\ photons are resonantly scattered as they travel through the neutral gas, and the amount of scattering events, which is determined by the \ion{H}{i} gas distribution and column density, strongly impacts the shape of the \lya\ profile. Since \lya\ and LyC photons interact with the same neutral gas, the underlying physical mechanisms driving their leakage should be closely connected. 

Nevertheless, this is not trivial because the \lya\ transition is resonant. While LyC photons cannot escape from optically thick ISM, \lya\ photons pass through dense neutral clouds by being scattered out of the velocity range covered by the neutral gas. The theoretical connection between \lya\ and LyC has first been investigated using radiative transfer models \citep{verhamme2015, dijkstra2016}. These studies highlighted that the \lya\ spectral shape provides insights on the \ion{H}{i} column density and/or the existence of holes in the neutral gas spatial distribution. The presence of paths entirely cleared of \ion{H}{i} gas in the ISM should imprint a single \lya\ peak emission at the systemic, nevertheless, the presence of a double-peaked \lya\ profile with a narrow peak separation in all the confirmed LCEs suggested that they have density-bounded ISMs \citep{verhamme2017, izotov2018b}. However, a follow-up study revealed the presence of saturated Lyman series with non-unity covering fraction \citep{gazagnes2018}. This outcome favors an ionization bounded model with holes in an optically thick interstellar medium, hence at first sight incompatible with their double-peaked \lya\ profile. Additionally, \citet{steidel2018} reported a correlation between the EW(\lya) and the \ion{H}{i} covering fraction in stacks of z $\approx$ 3 galaxies, while \citet{mckinney2019} found a significant trend connecting the escape fraction of \lya\ photons and the \ion{Si}{ii} covering fraction in a sample of extreme Green Peas (GPs). Hence, the latest studies suggest that the ISM is likely a very complex environment, and improved \lya\ radiative transfer models are needed to constrain the origin of the  connection between the \lya\ spectral shape and the escape of LyC photons.

Promising insights were recently found by \lya\ simulation studies showing that a very clumpy distribution of neutral gas, or low-density channels produced by turbulence could favor the leakage of LyC photons and create a double peak \lya\ profile with low $v^{\rm sep}_{\rm \lya}$  \citep{gronke2016, gronke2017, kimm2019, kakiichi2019}. On the other hand, \citet{jaskot2019} recently proposed that different \lya\ markers probe different ISM properties; the peak separation could trace the presence of low-density gas, while the covering fraction relates to the gas porosity. Hence, further clarifications of the physical properties of the neutral gas of known LCEs are crucial to test the reliability of the \lya-LyC correlations, and to investigate the accuracy of indirect \fesclyc\footnote{In this paper, unless stated otherwise, \fesclyc\ refers to the escape of ionizing photons along the line of sight.} predictions that use the \ion{H}{i} covering fraction \citep{chisholm2018}.

In this work, we investigate the connection between the ISM porosity (characterised by the \ion{H}{i} covering fraction), the \lya\ properties, and the LyC escape fraction in an unique sample of 22 star-forming galaxies, including 13 confirmed LCEs, which have neutral hydrogen Lyman series and \lya\ observations. While we focus primarily on the presence of (dusty) low-density channels to explain the escape of LyC and \lya\ photons, we also explore the impact of the width of saturated \ion{H}{i} absorption lines on the spectral shape of the \lya\ profile.

This paper is organized as follows: Sect.~\ref{sect:dataobs} describes the observational data. Section \ref{sect:method} defines the methods used to measure the neutral gas and \lya\ properties of the galaxies in our sample. In Sect.~\ref{sect:results}, we compare and discuss the connection between the spatial distribution and kinematics of neutral gas on both the \lya\ properties and the escape of LyC photons. Section \ref{sect:disc} discusses how the porosity of the neutral gas triggers the LyC leakage, and how it can be used to provide a lower limit to the total escape fraction of LyC photons in high-z galaxies. Finally, we summarize the main conclusions from this work in Sect.~\ref{sect:conc}.

\section{Data}
\label{sect:dataobs}
\begin{table}
\caption{Properties of galaxies with Lyman series observations.} 
\label{table:sample}
\centering   
\begin{tabular}{lllllll}
\hline \hline
Galaxy name&  $z$ & \oh & \fesclycobs  \\
    &        & & \\  
   (1) & (2) & (3) & (4) \\  \hline
J1243+4646  & 0.4317  & 7.89\tablefootmark{a} &  0.726\tablefootmark{a}  \\ 
J1154+2443  & 0.3690  & 7.65\tablefootmark{b} &  0.460\tablefootmark{b}   \\ 
J1256+4509  & 0.3530  &7.87\tablefootmark{a} &  0.380\tablefootmark{a}  \\ 
J1152+3400  & 0.3419  &8.00\tablefootmark{c} &  0.132\tablefootmark{c}\\
J1011+1947  & 0.3322  &7.99\tablefootmark{a} &  0.114\tablefootmark{a}   \\ 
J1442-0209  & 0.2937  &7.93\tablefootmark{c} &  0.074\tablefootmark{c} \\
J0925+1409  & 0.3013  &7.91\tablefootmark{d} &  0.072\tablefootmark{d}  \\ 
J1503+3644  & 0.3537  &7.95\tablefootmark{c} &  0.058\tablefootmark{c}   \\ 
J1333+6246  & 0.3181  &7.76\tablefootmark{c} &  0.056\tablefootmark{c}  \\ 
J0901+2119  & 0.2993  &8.16\tablefootmark{a} &  0.027\tablefootmark{a}  \\
J1248+4259  & 0.3629  & 7.64\tablefootmark{a} &  0.022\tablefootmark{a}  \\
J0921+4509  & 0.23499 & 8.67\tablefootmark{e} &  0.010\tablefootmark{f}   \\ 
Tol1247-232 & 0.0488  &8.10\tablefootmark{g} &  < 0.004\tablefootmark{h}  \\
J0926+4427  & 0.18069 & 8.01\tablefootmark{i} &  -                      \\ 
J1429+0643  & 0.1736  & 8.20\tablefootmark{i} &   -                     \\ 
GP0303-0759 & 0.16488 & 7.86\tablefootmark{i}   &  -                     \\ 
GP1244+0216 & 0.23942 & 8.17\tablefootmark{i}   &  -                       \\ 
GP1054+5238 & 0.25264 & 8.10\tablefootmark{i}   &  -                       \\ 
GP0911+1831 & 0.26223 & 8.00\tablefootmark{i}   &   -                      \\
\sone\      &  2.92525 &-                    &   -  \\ 
\stwo\      &  2.76228 & $<8.5$\tablefootmark{j}  & - \\     
Cosmic Eye  &  3.07483 & 8.60\tablefootmark{k}& -  \\ \hline
\hline
\end{tabular}
\tablefoot{(1) Galaxy name; (2) redshift; (3) metallicities derived from oxygen optical emission lines; (4)  observed Lyman continuum escape fraction derived from SED fitting.  Dashes indicate that the quantities have not been measured. The reference studies for the metallicities and LyC escape fractions are listed below.}
\tablebib{(a) \citet{izotov2018b} (b) \citet{izotov2018a}; (c) \citet{izotov2016b};  (d) \citet{izotov2016a};  (e)  \citet{pettini2004};  (f) \citet{borthakur2014}; (g)  \citet{leitherer2016};   (h) \citet{chisholm2017leak}; (i) \citet{izotov2011}; (j) \citet{stark2008}; (k) \citet{wuyts}.}
\end{table}
    
In this work, we investigate the relation between the neutral gas covering fraction, \lya\ properties, and the escape of LyC photons in the sample of 22 star-forming galaxies listed in Table \ref{table:sample}. We selected these galaxies because they have publicly available rest-frame UV spectroscopy both for \lya\ and for the rest of the Lyman series (i.e between Lyman-$\beta$ at 1025 $\AA$ and the Lyman limit at 912 $\AA$). The latter can be observed with a spectral resolution R~$>$~1500 for galaxies at z~$>$~0.18 with the Cosmic Origins Spectrograph (COS) onboard the Hubble Space Telescope (HST) \citep{green2012}. The relation between the neutral gas properties and the escape of ionizing photons was already explored for 16 of the 22 galaxies \citep{gazagnes2018, chisholm2018}. This includes 13 low redshift galaxies (z < 0.4), 7 of which are confirmed LyC emitting galaxies; J0925+1409, J1503+3644, J1152+3400, J1333+6246, J1442-0209 from \citet{izotov2016a, izotov2016b}, J0921+4509 from \citet{borthakur2014} and Tol1247-232 from \citet{leitherer2016}. Four of them are Green Peas (GP) \citep{henry2015}, 2 are Lyman break analogs (LBA) \citep{heckman2011, heckman2015} and the 3 remaining galaxies are gravitationnally lensed galaxies at z $\approx$ 3 \citep[SGAS J122651.3+215220, SGAS J152745.1+065219, and the Cosmic Eye, ][]{stark2008, koester} from the Magellan Evolution of Galaxies Spectroscopic and Ultraviolet Reference Atlas \citep[\megasaura;][]{rigby}. Additionally, our sample includes 6 new galaxies that were not included in \citet{gazagnes2018} and are the recently discovered low redshift (0.2 < z < 0.5) LyC emitting galaxies J1154+2443, J1243+4646, J1256+4509, J1011+1947, J0901+2119, and J1248+4249 from \cite{izotov2018a, izotov2018b}; three of which have extreme LyC escape fractions of 38, 46 and 72.6 \%. \\

 Nineteen of the galaxies observed with HST/COS are at such redshift that at least the Ly$\beta$ line is observable with the COS G140L or G130M gratings. The resolving power (R) of the rest-frame UV spectra around the Lyman series is $\approx$ 1500 for all the confirmed leakers except that J0921+4509 which has a R $\approx$ 15000 G130M spectra. The 4 GPs and the 2 LBAs have observations of one or several \ion{H}{i} absorption lines with a spectral resolution of $\approx$ 15000. Additionally, all the \lya\ profiles were observed with the medium resolution grating G160M ($R \approx 16000$ at 1600 \AA). The data for the 13 leakers were reduced using CALCOSv2.21 and a custom method for faint COS spectra \citep{worseck2016}. The other COS/HST data were reduced with CALCOSv2.20.1 and the methods from \citet{wakker2015}.  
The three \megasaura\ galaxies have been observed with the MagE spectrograph on the Magellan Telescopes \citep{marshall2008}, and have moderate resolution spectroscopy ($R \sim 3000$) both for their Lyman series and \lya. They are the only galaxies in the \megasaura\ sample with a signal-to-noise ratio (S/N) sufficient (> 2) to constrain their neutral gas properties with the Lyman series. In this paper, we used the following short names for the two  sources: SGAS J122651.3+215220~=~\sone\  and SGAS J152745.1+065219~=~\stwo.

Table~\ref{table:sample} summarizes the galaxy properties of the sample. The metallicities have been derived from the optical [\ion{O}{iii}]~4366\AA\ oxygen emission lines, using the direct $T_e$ method, in all the low redshift galaxies. The metallicity of the Cosmic Eye has been measured by \citet{stark2008} using the R$_{23}$ index, and an upper limit has been derived from the [N II]/H$\alpha$ ratio for \stwo\ \citep[12+log(O/H) < 8.5;][]{wuyts}. The metallicity of \sone\ has not been measured because these lines are not accessible from the ground. Several different measurements of the escape fraction of ionizing photons from Tol1247-232 are reported in the literature: 4.2 $\pm$ 1.2 \% in \citet{leitherer2016}, < 0.4 \% in \cite{chisholm2017leak} and 1.5 $\pm$ 0.5 \% in \citet{puschnig2017}. We used the value derived in \citet{chisholm2017leak} since the measurement method is consistent with the one used for the other leakers.

\section{Method}
\label{sect:method}

We now describe the methods used to study the ISM and \lya\ properties of the 22 galaxies in our sample.

\subsection{Neutral gas properties}
\label{sect:neutprop}

\begin{table*}
\caption{Dust extinction and \ion{H}{i} properties derived from the Lyman series.}
\label{table:fitprop}
\centering   
\begin{tabular}{lllll}
\hline \hline
Galaxy name & \ebv & $C_f$($\ion{H}{I}$) & $v^{\rm shift}_{\rm \ion{H}{I}}$ &  $v^{\rm width}_{\rm \ion{H}{I}}$\\
 & [mag]  & &  [km s$^{-1}$] & [km s$^{-1}$] \\
 (1) & (2) & (3) & (4) & (5) \\ \hline 
J1243+4646 & 0.100 $\pm$ 0.021 & < 0.189 & - &  - \\ 
J1154+2443 & 0.118 $\pm$ 0.031 & 0.450 $\pm$ 0.086 & -289 $\pm$ 92 & 170 $\pm$ 86 \\ 
J1256+4509 & 0.076 $\pm$ 0.029 & 0.409 $\pm$ 0.079 & -48 $\pm$ 44 & 250 $\pm$ 50 \\ 
J1152+3400* & 0.144 $\pm$ 0.021 & 0.625 $\pm$ 0.054 & -346 $\pm$ 28 & 419 $\pm$ 60 \\ 
J1442-0209 & 0.140 $\pm$ 0.015 & 0.556 $\pm$ 0.038 & -261 $\pm$ 34 & 371 $\pm$ 53 \\ 
J0925+1409 & 0.164 $\pm$ 0.015 & 0.638 $\pm$ 0.086 & -214 $\pm$ 151 & 320 $\pm$ 60\\ 
J1011+1947 & 0.230 $\pm$ 0.084 & 0.708 $\pm$ 0.113 & -69 $\pm$ 32 & 285 $\pm$ 54\\ 
J1503+3644* & 0.217 $\pm$ 0.014 & 0.721 $\pm$ 0.055 & -79 $\pm$ 24 & 356 $\pm$ 49\\ 
J1333+6246 & 0.151 $\pm$ 0.043 & 0.804 $\pm$ 0.058 & -126 $\pm$ 48 & 280 $\pm$ 51 \\ 
J0901+2119 & 0.220 $\pm$ 0.057 & 0.637 $\pm$ 0.166 & -121 $\pm$ 84 & 300 $\pm$ 78 \\ 
J1248+4259 & 0.253 $\pm$ 0.073 & 0.954 $\pm$ 0.104 & 56 $\pm$ 49 & 232 $\pm$ 50 \\ 
J0921+4509 & 0.222 $\pm$ 0.015 & 0.761 $\pm$ 0.080 & -56 $\pm$ 13 & 440 $\pm$ 20\\ 
Tol1247-232* & 0.195 $\pm$ 0.028 & 0.518 $\pm$ 0.046 & 194 $\pm$ 41 & 453 $\pm$ 89 \\ 
J0926+4427*  & 0.175 $\pm$ 0.010 & 0.768 $\pm$ 0.034 & -199 $\pm$ 12 & 383 $\pm$ 46 \\ 
J1429+0643*  & 0.165 $\pm$ 0.020 & 0.931 $\pm$ 0.046 & -220 $\pm$ 36 & 420 $\pm$ 50 \\ 
GP0303-0759 & 0.121 $\pm$ 0.045 & 0.908 $\pm$ 0.207 & -266 $\pm$ 92 & 380 $\pm$ 50\\ 
GP1244+0216 & 0.290 $\pm$ 0.043 & 0.946 $\pm$ 0.123 & -78 $\pm$ 48 & 379 $\pm$ 49\\ 
GP1054+5238* & 0.253 $\pm$ 0.054 & 0.823 $\pm$ 0.101 & -161 $\pm$ 29 & 480 $\pm$ 29 \\ 
GP0911+1831 & 0.352 $\pm$ 0.038 & 0.752 $\pm$ 0.092 & -273 $\pm$ 40 & 374 $\pm$ 16\\
\sone\      & 0.201 $\pm$ 0.001 & 0.994 $\pm$ 0.009 & -264 $\pm$ 21 & 548 $\pm$ 29 \\ 
\stwo*      & 0.314 $\pm$ 0.002 & 0.990 $\pm$ 0.034 & -247 $\pm$ 25 & 480 $\pm$ 30 \\   
Cosmic Eye*  & 0.371 $\pm$ 0.006 & 0.990 $\pm$ 0.023 & 311 $\pm$ 16 & 467 $\pm$ 99 \\ 
\hline
\end{tabular}
\tablefoot{(1) Galaxy name; (2)  dust attenuation parameter ; (3)  \ion{H}{i} covering fraction; (4) \ion{H}{i} velocity shift;  (5) \ion{H}{i} velocity width of maximal absorption. * changes with respect to \citet{gazagnes2018}, see Sect.~\ref{sect:method} for details.}
\end{table*}

\begin{figure}
\centering
\includegraphics[width=\hsize]{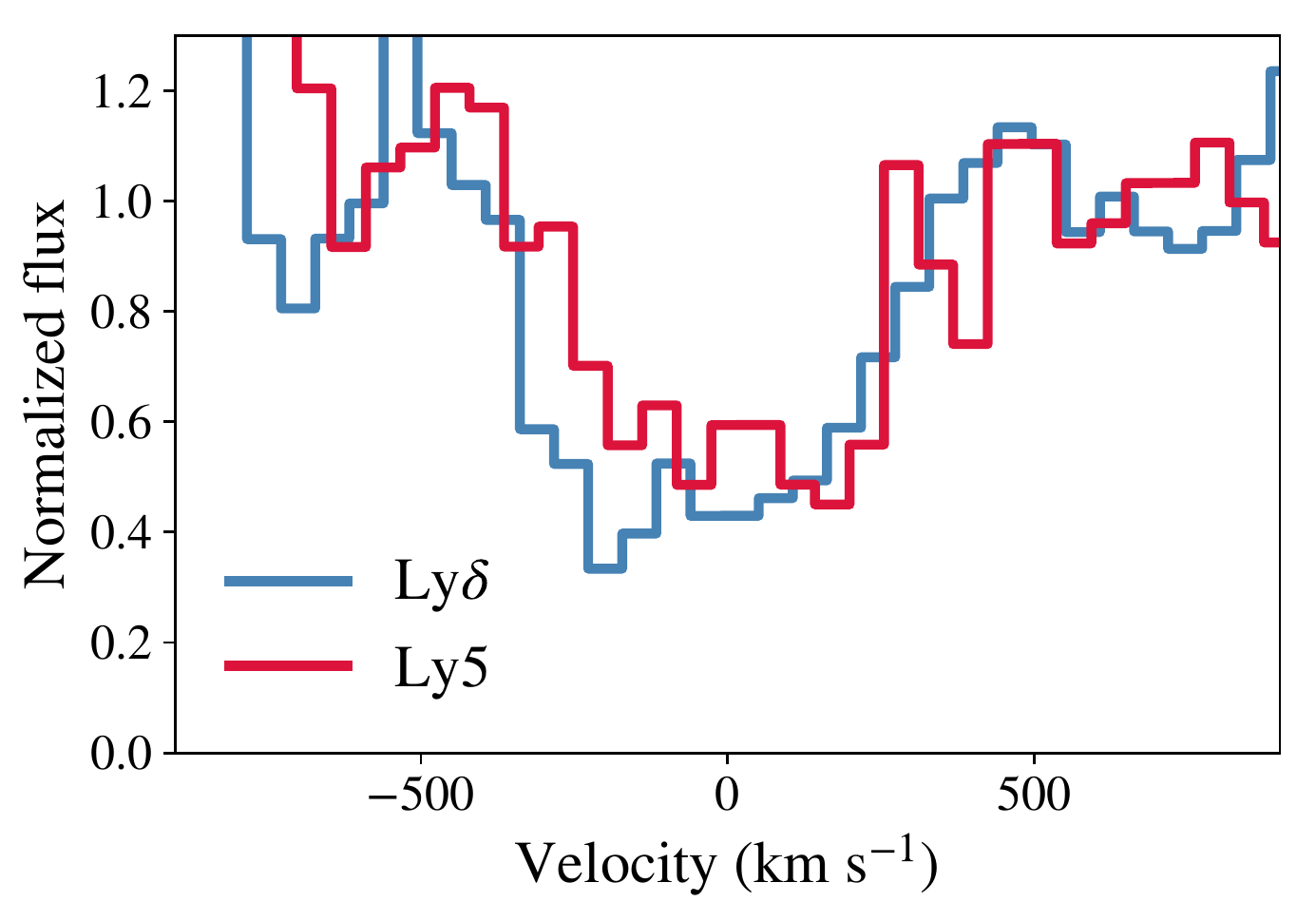}
\caption{Plot of the Ly$\delta$ (blue) and Ly5 (red) absorption lines in the galaxy J1256+4509. The flux has been normalized using the median of the observed spectra taken between 500 and 1000 km s$^{-1}$ from the absorption lines. The two \ion{H}{i} absorption lines have similar depth and width, which indicates that they are saturated.}
\label{fig:j1256ly}
\end{figure}

To measure the \ion{H}{i} velocity shift and covering fraction of the galaxies, we used the approach detailed in \citet{gazagnes2018}, and  recall here the main steps. We first correct the galaxy spectra for Milky Way extinction using the \citet{cardelli1989} extinction law, R(V) = 3.1, and the galactic \ebv\ reported in the NASA Extragalactic Database (NED\footnote{https://ned.ipac.caltech.edu}). We then fit the stellar continuum as in \citet{chisholm2019}, including dust extinction, metal and \ion{H}{i} absorption lines to consistently determine the UV attenuation in the galaxy, as well as the column densities and covering fractions of the individual ions. The stellar continuum model, $F^\star$, is a linear combination of 50 single-age fully theoretical stellar continuum models with 5 metallicities, 0.05, 0.2, 0.4, 1, and 2 Z$_\odot$, with ages of 1, 2, 3, 4, 5, 8, 10, 15, 20, and 40 Myr, drawn from the {\small STARBURST99} library \citep[S99;][]{leitherer1999}. This is numerically given as:

\begin{equation}
     \label{eq:stellarflux}
     F^\star = \Sigma_{i=1}^{50}{X_i \, F^{S99,\ {\rm Z_i}}_i}
\end{equation}

\noindent where $X_i$ and $F^{99, {\rm Z_i}}_i$ are respectively the linear coefficients and the {\small STARBURST99} theoretical stellar continuum models for a given age and metallicity. The S99 spectra were computed with the WM-Basic spectral library \citep{leitherer2010}, using a Kroupa initial mass function with a high and low mass exponent of 2.3 and 1.3 respectively, a high-mass cutoff of 100~M$_\odot$, and the stellar evolution tracks with high mass loss from \citet{meynet1994}. Additionally, the large amount of ionizing photons produced by young massive stars produces free-free, free-bound, and two-photon nebular continuum emission, can have a significant impact on the total continuum flux in young, low metallicity stellar populations \citep{steidel2016, Byler2018}. Following the procedure detailed in \citet{chisholm2019}, we created a nebular continuum for each single-age and metallicity stellar model using Cloudy v17.0 \citep{ferland2013, ferland2017}, assuming similar gas-phase metallicity and stellar metallicity, a volume hydrogen density of 100 cm$^{-3}$ and an ionization parameter log(U) = -2.5. The final output nebular continua were added to the stellar models and the final synthetic spectra have a spectral resolution R $\approx$ 2500, which is convolved to the resolution of the data.  We used the far-UV dust attenuation curve from \citet{reddy2016dustlaw} and an uniform dust screen model to account for the dust attenuation. 

Absorption lines from different ions were included using Voigt profiles defined by 4 free parameters: the velocity shift ($v^{\rm shift}$), the Doppler parameter ($b$), the column density ($N$), and the covering fraction ($C_f$). The linear combination of stellar continuum models, interstellar absorption lines, and dust attenuation produces the final fitted spectrum. 

As we are interested in the Lyman series, the spectral region that we fit is taken from 912 to 1050 $\AA$. We include redder portions of the spectrum, up to 1300 \AA, to further constrain the stellar model and dust attenuation. We aimed at including all the \ion{H}{i} absorption lines that are observed to improve the constraints on the \ion{H}{i} parameters. Nevertheless, in practice, the wavelength regime from 912 to 930 $\AA$ were excluded due to low S/N, and/or geocoronal emission. Additionally, Ly$\beta$ is not systematically fit since it is located close to a strong \ion{O}{VI} P-Cygni profile, which synthetic stellar models sometimes fail to reproduce (see the fits in Appendix~\ref{sect:fits}), and often decreases the fit quality when combined with bluer \ion{H}{i} absorption lines. The \ion{O}{i} absorption lines that directly blend with the Lyman series are always included, and their parameters are mostly constrained by the \ion{O}{i}~989 and 1039~$\angstrom$ lines. However, because of low S/N and/or low \no, the latter are not always resolved, such that we cannot accurately recover the \ion{O}{i} contribution in these galaxies. Finally, absorption lines from  \ion{O}{vi}, \ion{Si}{ii}, \ion{C}{ii}, \ion{C}{iii} or from the Milky Way are sometimes added, provided that they improve the fit around the Lyman series. For more details see \cite{gazagnes2018}.

Assuming a foreground dust attenuation, the final fitted model, $F_{\rm mod}(\lambda)$, can be expressed as
\begin{equation}
     F_{\rm mod}(\lambda)  = F^\star(\lambda)\times 10^{-0.4\ \ebv\ k_{\rm Reddy16} (\lambda)} \times \mu_{\rm ion}(\lambda)
     \label{eq:fittedflux}
\end{equation}
where $\mu_{\rm ion}(\lambda)$ represents the fitted profiles of absorption lines given by: 
\begin{equation}
\mu_{\rm ion}(\lambda) =
    \begin{cases}
      { 1-C_f({\rm ion}) + C_f({\rm ion})\times \exp^{-\tau_{ion}(\lambda)}}  & \text{if included} \\
      1 & \text{otherwise}
    \end{cases}
     \label{eq:abslines}
\end{equation}
  Equation \eqref{eq:fittedflux} assumes that all the photons escaping the ISM are affected by the same dust extinction. In practice, galaxies include several star-forming clumps, which might not be affected by the same dust attenuation. Thus, the recovered \ebv\ should be interpreted as the mean extinction in the galaxy.

The fitting method is based on an IDL routine that uses nonlinear least squares fitting, {\small MPFIT} \citep{markwardt2009}, and returns the best fit parameters and their statistical errors for each free parameter fitted. In this work, we focus mainly on the fitted \ion{H}{i} parameters: $v^{\rm shift}_{\rm \ion{H}{I}}$, \nh\ and $C_f$(\ion{H}{i}). As discussed in \citet{jones2013}, \citet{rivera2015} and \citet{vasei2016}, the interpretation of the measured $C_f$ must be taken with caution. This is because the kinematics of the absorbing gas impacts the observed depth of the absorption lines. Indeed, two dense \ion{H}{i} clouds with non-overlapping velocity distributions and each covering half of the galaxy will imprint \ion{H}{i} absorption lines with $C_f$(\ion{H}{i}) = 0.5. However, in this case, the total geometrical covering fraction is 1 because LyC photons are insensitive to kinematic effects (the \ion{H}{i} absorbs ionizing photons at all wavelengths below the Lyman limit). Hence, $C_f$(\ion{H}{i}) is always a lower limit to the covering fraction seen by the ionizing photons. Nevertheless, we showed in \citet{gazagnes2018} that it seems to be a good proxy to the true geometric covering fraction of the optically thick \ion{H}{i} clouds in the current sample.

 \begin{figure*} 
\centering \includegraphics[width=\hsize]{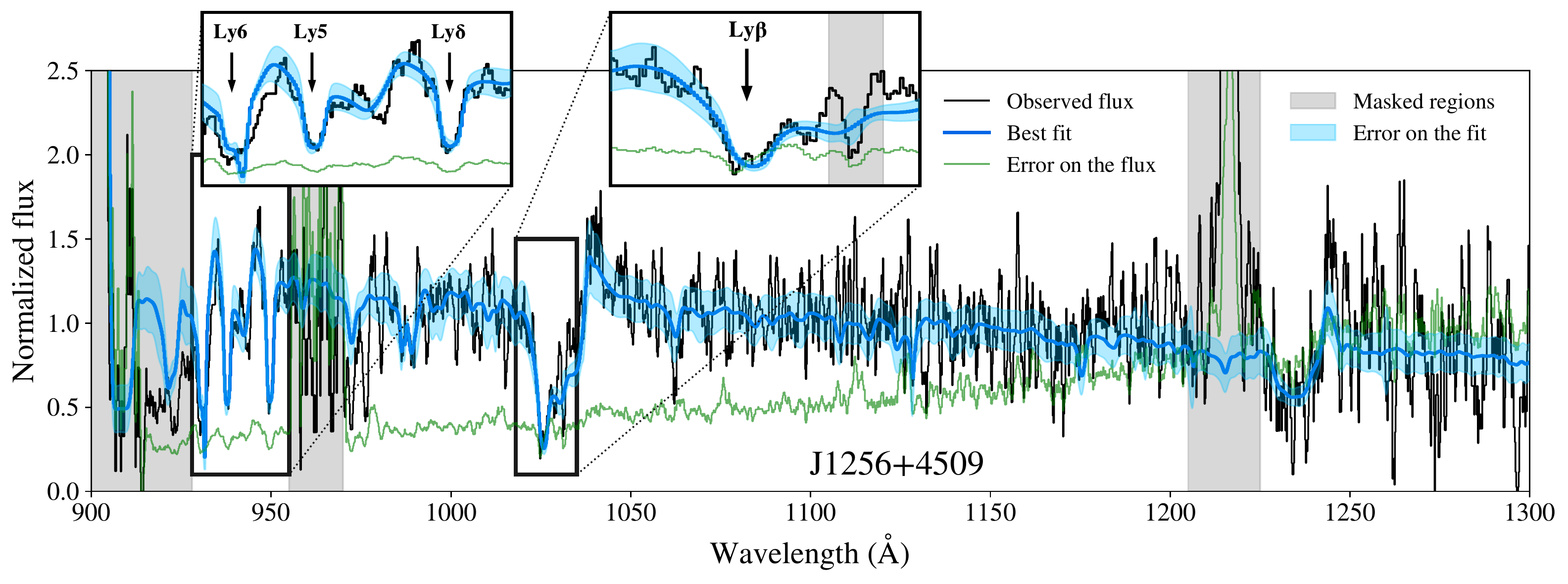}
 \caption{ Best fit (blue solid line) obtained for the galaxy J1256+4509. The black and green solid lines show the observed flux and the error on the flux, respectively. The top panels are zooms on the individual Lyman series lines fitted. The gray shaded areas are the principal wavelength regions masked during the fit, due to geo-coronal emission, low S/N or \lya\ emission. We additionally show in the top panels the masks for ISM/Milky Way absorption lines that are not included during the fitting procedure. For display purposes, these masks do not appear in the main panel. The blue shaded area represents the uncertainty on the fit, derived using a Monte-Carlo approach (see details in Sect.~\ref{sect:neutprop}). }
 \label{fig:fiitexample}
\end{figure*}

In \citet{gazagnes2018}, we used simulations to show that the \nh\ derived from the fits suffers from large uncertainties due to the degeneracy between the Doppler parameter $b$ and the column density when the absorption lines are saturated. The typical resolution and S/N of the observations is too low to constrain \nh\ directly from the Lyman series. Consequently, we neither report nor use the \nh\ and $b$ values derived using our fitting procedure. Alternatively, one can indirectly estimate \nh\ using an approach based on \no\ and the metallicity, \oh. Nevertheless, \ion{O}{i} is not detected in 6 galaxies, which might be due to low S/N, small \no\ or low covering fraction. Conversely, we established in \citet{gazagnes2018} that the \ion{H}{i} covering fraction can be inferred with a reasonable accuracy from Voigt fitting methods given the spectral resolution and S/N of the galaxies observed in our sample. A systematic error, relative to the resolution and S/N of a given spectrum, needs to be included in the final error term because the statistical error returned by {\small MPFIT} does not account for it \citep[see Sect 2 and Table 3 in][]{gazagnes2018}. The typical systematic error of the covering fraction derived from spectra with R = 1500 (typical resolution of the GL140 grating) and S/N $\approx$ 2 is 0.10.  Note that this typical systematic error is only valid if the Lyman series lines are saturated. When the \ion{H}{i} absorption lines are not saturated, the residual flux at the bottom of an absorption line is similarly impacted by the presence of low \nh\ and/or low $C_f$(\ion{H}{i}), such that an accurate constraint on these parameters would require higher S/N and resolution.

 Fig.~\ref{fig:fiitexample} shows the observed flux, the error on the flux, and the best fit obtained for the galaxy J1256+4509 using our fitting approach. It highlights the main regions masked during the fits, either due to geo-coronal emission, low S/N, \lya\ emission, or ISM or Milky Way absorption lines that are not fit. We estimate an uncertainty on the fit using a Monte-Carlo approach where the observed flux is modified by a Gaussian kernel centered on zero with standard deviation corresponding to the error on the flux. We performed 100 fit realizations and took the standard deviation as the error of the fit (represented by a blue shaded area in the figure). Figure~\ref{fig:fiitexample} shows that this uncertainty is roughly on the order of the error of the observed flux in the highest S/N regions. Interestingly, the error is low around the Lyman series lines, suggesting that the latter are robustly constrained and little affected by fluctuations in the fitted stellar continuum. This point is further discussed in Sect.~\ref{sect:dust}, where we emphasize the complexity of constraining the stellar continuum and dust extinction in galaxy spectra with low S/N. Section~\ref{sect:fits} presents the fits obtained for the 5 other leakers from \citet{izotov2018a, izotov2018b}, the 16 remaining fits can be found in \citet{gazagnes2018}.

Fitting the Lyman series using a Voigt profile assumes that the lines follow a single Gaussian velocity distribution, and this assumption might not be valid for absorption profiles arising from galactic outflows  \citep{heckman2000, pettini02, shapley2003, weiner09, chisholm2017mass}. Consequently, we used the non-parametric approach described in \citet{gazagnes2018} to measure the \ion{H}{i} covering fraction from the residual flux of the Lyman series lines after removing the stellar continuum.  This method does not presume a specific line profile or velocity distribution of the \ion{H}{i} gas. However, this assumes that the \ion{H}{i} absorption lines are saturated, i.e that $\nh \ga 10^{16} $~cm$^{-2}$ for the Lyman series lines that we fit (Ly$\beta$ to Ly6). In \citet{gazagnes2018}, we found evidence that the latter assumption is true for 13 galaxies which have \nh\ values > $10^{18.4} $~cm$^{-2}$ using the observed \no\ and the gas-phase metallicity. In the 9 remaining galaxies, we did not find a reliable measurement of \no\ \citep[for J0925+1409 and the 6 new leakers included from][]{izotov2018a, izotov2018b}, or the galaxy metallicity has not been measured (\sone, \stwo). Nevertheless, despite their different oscillator strengths, we observed a tendency for the observed Lyman series to have similar depths and shapes. This is illustrated in Fig.~\ref{fig:j1256ly} where the Ly$\delta$ and Ly5 absorption lines are plotted in velocity space for J1256+4509. Similar depths and widths are  robust indicators of saturated lines. No \ion{H}{i} absorption lines are detected in J1243+4646, the highest LyC escape fraction (see Fig.~\ref{fig:J1243}), likely due to either a very low \ion{H}{i} neutral gas column density or covering fraction. 

To measure the covering fraction from the residual flux, we used a Monte-Carlo approach: the observed flux is first divided by the stellar continuum (fit as  $F^\star$ in Eq.~\ref{eq:stellarflux}), and modified by a Gaussian kernel centered on zero with standard deviation corresponding to the error array. $C_f$(\ion{H}{i}) is derived from the median of 1 minus the residual flux in a velocity range that includes the deepest part of the absorption line. We repeated this procedure 1000 times and took the median and standard deviation of this distribution as the $C_f$ value and uncertainty for each \ion{H}{i} absorption line that is not polluted by Milky Way absorption lines or geocoronal emission. We additionally include the systematic errors in quadrature. We then obtain a final covering fraction by taking the error weighted mean of the $i$ observed Lyman series transitions. Table~\ref{table:$C_f$} lists the $C_f$ derived from the residual flux of each Lyman series transition in each galaxy, the resulting $C_f$(\ion{H}{i}) "Depth" and the measurement derived from the fitting method. The last column shows the final \ion{H}{i} covering fraction, derived from the error weighted mean between the values obtained from the two different approaches. For J1243+4646, which has no detected Lyman series, we still measure the median residual flux in a velocity range chosen where the flux is minimal. We consider the final value as an upper limit. We do not report a $C_f$(\ion{H}{i}) "Depth" for GP0303-0759 because its only \ion{H}{i} absorption line observed is contaminated by a Milky Way absorption line. 

 We note that the galaxies in our sample have a different number of observed Lyman Series lines. However, for galaxies with more than one observed \ion{H}{i} absorption line, the individual $C_f$(\ion{H}{i}) estimates using the depth of the absorption profiles are consistent at $\pm$ 1$\sigma$ with the value derived when all the lines are fitted simultaneously. Hence, we assume that the \ion{H}{i} parameter values derived in galaxies with a single Lyman series line does not suffer from significant systematic effects. Additionally, Table~\ref{table:$C_f$} shows that both approaches give comparable estimates and uncertainties, thus supporting the fact that both are robust techniques to measure the \ion{H}{i} covering fraction from saturated Lyman series.

Finally, using the same methodology, we measured the velocity width of each \ion{H}{I} absorption line, when it is not contaminated by foregrounds or Milky Way absorption, and has sufficient S/N so that the line profile is clearly observed. We estimate the velocity width as the velocity range where the absorption profile is at its maximum depth. The minimal and maximal velocities of this interval are chosen where the flux deviates by more than 20\% from the residual flux measurements reported in Table~\ref{table:$C_f$}. The same Monte-Carlo approach is used to derive the resulting $v^{\rm width}$ value and uncertainty for each absorption line, and we obtain the final $v^{\rm width}_{\ion{H}{i}}$ for each galaxy derived from the error weighted mean of the individual $v^{\rm width}$ (last column of Table~\ref{table:vel}).

The final covering fraction values, as well as the velocity shift of the line $v^{\rm shift}_{\rm \ion{H}{I}}$, the dust extinction \ebv\ obtained from the fitting method, and average velocity width of the Lyman series are reported in the Table~\ref{table:fitprop}. Note that in this work, the synthetic spectra used for the fitting are slightly different from \citet{gazagnes2018} where the final spectra were only 10 single-age stellar populations of a single metallicity and did not include the nebular continuum. Therefore, we re-measured all the properties in all the galaxies in the sample. While the covering fractions and velocity shift measurements all remained consistent at $\pm$ 1 $\sigma$, we find some small variations in the dust extinction (0.01 to 0.06) for 8 galaxies (marked with * in Table~\ref{table:fitprop}). This is expected because the incorporation of nebular continuum and the combination of 5 different metallicities can change the final fitted stellar spectral shape and therefore impact the \ebv\ inferred (see Sect~\ref{sect:dust}). Overall, the variations are relatively small and do not impact the results obtained in \citet{gazagnes2018} and \citet{chisholm2018}.

\subsection{Constraining the dust extinction}
\label{sect:dust}

\begin{table*}
\caption{Investigating the impact of different dust extinction assumptions in J1154+2443 and J1256+4509.}
\label{table:addfit}
\centering   
\begin{tabular}{lllllllllll}
\hline \hline
Galaxy name &  $f^{\rm obs}_{\rm esc}(\rm LyC)$& Law & \ebv & Age  & $C_f$(\ion{H}{i}) & $A$(LyC) & $A$(\lya) & WSS \\
 & &  &  & [Myr] & & &\\
 (1) & (2) & (3) & (4) & (5) & (6) & (7) & (8) & (9) \\ \hline 
\multirow{4}{*}{J1154+2443} & \multirow{4}{*}{0.46} &  \multirow{3}{*}{Reddy+16} &   0.118 $\pm$ 0.031 & 8.10 & 0.401 $\pm$ 0.152 & 1.51 & 1.20 & 733 \\ 
& &  & 0.00* & 15.07 & 0.379 $\pm$ 0.145 & 0.00 & 0.00 & 770 \\
& &  & 0.20* & 4.20 & 0.339 $\pm$ 0.183 & 2.57 & 2.04 & 743 \\[5pt]
& & SMC  & 0.034 $\pm$ 0.008 & 7.99 & 0.483 $\pm$ 0.390 & 0.87 & 0.55 & 733\\[15pt]

\multirow{4}{*}{J1256+4509} & \multirow{4}{*}{0.38} &\multirow{3}{*}{Reddy+16} & 0.076 $\pm$ 0.029 & 2.52 & 0.408 $\pm$ 0.125 & 0.98 & 0.78 & 1517\\ 
& &   &  0.00*  & 3.15 & 0.440 $\pm$ 0.121 & 0.00 & 0.00 & 1555 \\
& &   &  0.20* & 2.32 & 0.571 $\pm$ 0.183 & 2.57 & 2.04 & 1518 \\[5pt]
& & SMC &  0.038 $\pm$ 0.008  & 2.38 & 0.383 $\pm$ 0.133 & 0.98 & 0.62 & 1499 \\
\hline
\end{tabular}
\tablefoot{(1) Galaxy name; (2) Observed LyC escape fraction; (3) dust attenuation law; (4) dust extinction;  (5) light-weighted stellar continuum age from the fit, derived from the linear combination of the 50 single-age  STARBURST99 stellar continuum; (6) \ion{H}{i} covering fraction value and uncertainty returned by the fit; (7) Attenuation at 912 \AA; (8) Attenuation at 1216 \AA\ and (9) the value of the summed squared weighted residuals for the returned parameter values as given by MPFIT.\\
* Values fixed during the fitting procedure.}
\end{table*}

 \begin{figure*} 
\centering \includegraphics[width=0.9\hsize]{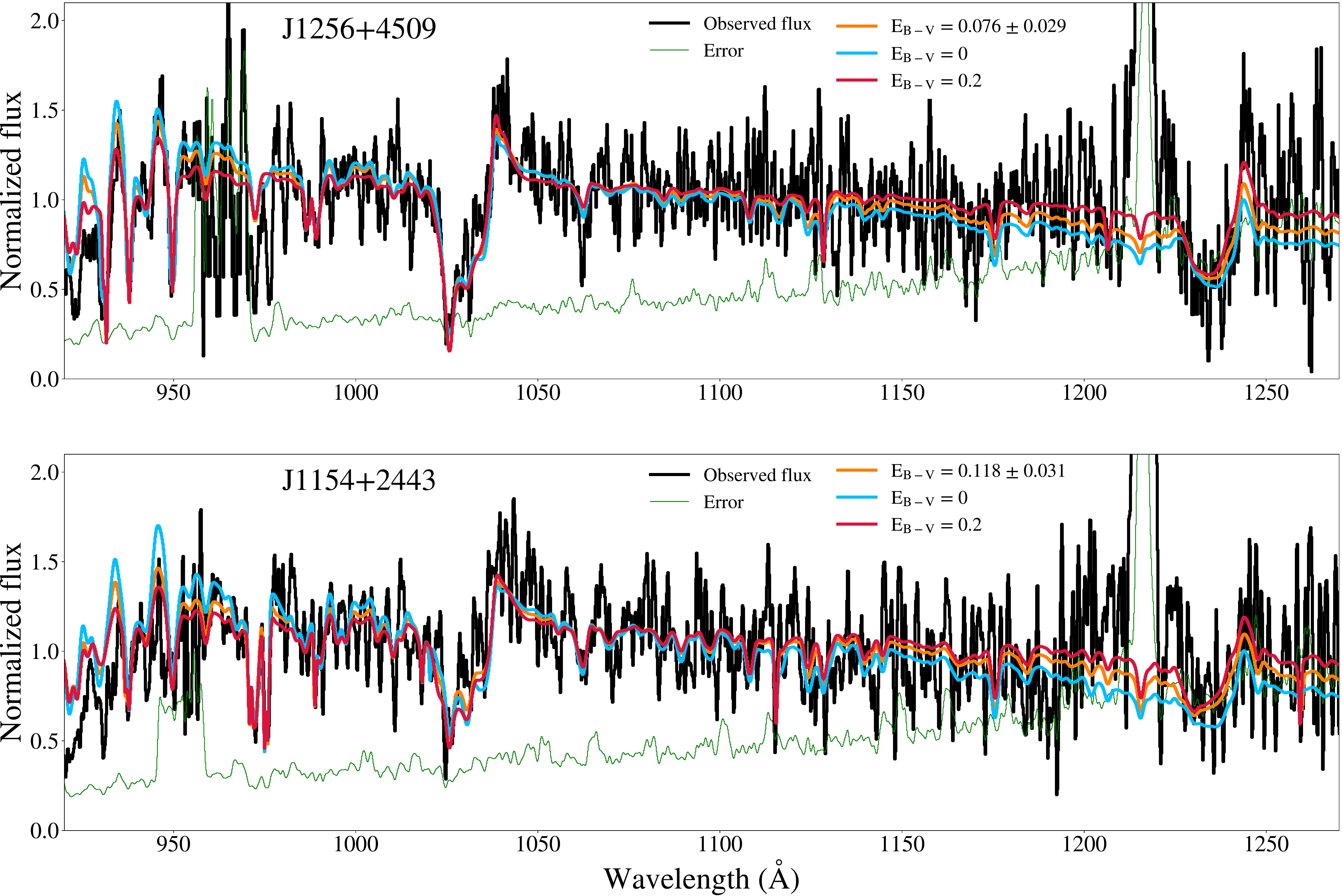}
 \caption{Top: the black curve is the observed flux in the galaxy J1256+4509, green is the error, orange is the fit obtained letting \ebv\ as a free parameter, blue and red lines are fits obtained when fixing \ebv\ to 0 and 0.2 respectively. Bottom: same for the galaxy J1154+2443. Both spectra have been normalized using the median flux between 1090 and 1100 \AA.  The wavelength regions with Ly$\alpha$ emission (between 1205 to 1225 \AA) or low S/N (< 1) are masked during the fitting procedure (see Fig.~\ref{fig:fiitexample}).}
 \label{fig:fixebv}
\end{figure*}

Accurately fitting the dust attenuation in the ISM of the galaxies is important to constrain its impact on the escaping radiation. Indeed, several dust models, and/or dust attenuation laws might lead to various physical interpretations of the impact of dust on the \lya\ or LyC radiations. \citet{gazagnes2018} carefully discussed the effects of dust models with a uniform dust screen (all photons are homogeneously attenuated) or with dust free holes (dust only lies in optically thick neutral regions). It was shown that these models result in different values of $C_f$(\ion{H}{i}) and \ebv, but lead to very similar estimates of the absolute escape fraction of LyC radiation.

Additionally, the choice of the dust extinction curve significant impacts the effect of dust obscuration on the wavelength region around and below the Lyman break. Typical dust attenuation curves  \citep[e.g from][or the  Small Magellanic Cloud (SMC) law]{calzetti2000} are unconstrained in the far-UV ($\lambda < 1250$ \AA), while measuring the impact of dust at these wavelengths is crucial when investigating the escape of ionizing photons from dusty galaxies. In this work, we choose the dust extinction law derived in \citet{reddy2016dustlaw}. Our choice is motivated by the fact that the authors used a large sample of 933 far-UV observations of Lyman Break Galaxies at z\textasciitilde 3, 121 having a deep spectroscopic coverage from 850 to 1300 \AA, to robustly constrain the shape of the dust attenuation curve between 950 and 1500 \AA. The authors report that the attenuation of LyC photons around 900 \AA\ is $\approx 2$ times lower than estimates derived from polynomial extrapolations of typical dust attenuation curves \citep{calzetti2000, reddy2015}. In \citet{gazagnes2018}, we investigated the effects of using different attenuation laws, such as the SMC attenuation law\footnote{Values have been taken from the IDL routine from J. Xavier Prochaska: \url{https://github.com/profxj/xidl/tree/master/Dust}}, which is significantly steeper than the \citet{reddy2016dustlaw} law, and showed that this had a relatively small impact on the derived values of the \ion{H}{i} covering fractions. In this work, we reconsider the effects of different dust curves using the examples of J1154+2443 and J1256+4509, two of the three largest leakers in our sample, and report in Table~\ref{table:addfit} the changes seen with respect to the stellar population age, the $C_f$(\ion{H}{i}), the dust extinction and attenuation at 912 \AA\ and 1216 \AA. We also report the value of the summed squared weighted residuals for the recovered parameter values, WSS, as returned by MPFIT, to demonstrate the differences in the quality of the fit in each case. 

The age of the stellar population and the \ion{H}{i} covering fraction are insensitive to the dust attenuation law used. This is because the presence of specific stellar features in the spectra fix the stellar populations during the fit \citep{chisholm2019}. However, we derive approximately two times lower attenuation  for the \lya\ and LyC radiation in J1154+2445 when using the SMC law. Thus, the choice of the dust law can impact the fitted attenuation, while the differences in the summed squared weighted residuals are too small to choose a dust extinction curve which significantly improves the final fit quality. 

Additionally, we further investigated the fluctuations of the parameters inferred when \ebv\ is fixed to 0 and 0.2 for a given dust extinction law. These results are reported in Table~\ref{table:addfit} and the obtained fits are shown in Fig.~\ref{fig:fixebv}. A $\pm$ 0.2 difference in \ebv\ leads to small variations (within the flux error) in the shape of the fitted spectra while having a significant impact on the attenuation at 912 and 1216 \AA, reducing the flux by 85\% in the model with \ebv\ = 0.2 compared to that in the model with \ebv\ = 0. However, the differences in the residuals are insignificant for selection of \ebv. This is likely because there is a strong degeneracy between the stellar population age and the dust extinction needed to model the observed flux. Indeed, younger populations have a steeper continuum towards the far-UV wavelengths, and require a larger dust extinction to reproduce the observed flux compared to older star populations. This is consistent with the results in Table~\ref{table:addfit}, since we obtained younger and older stellar populations when fixing \ebv\ to 0.2 and 0, respectively. 
The poorly constrained stellar population age can affect the reliability of \fesclycobs\ when the former is derived using the ratio of the observed flux at < 912 \AA\ over the estimated intrinsic emission of ionizing photons. 

Robustly constraining the stellar populations requires to identify specific markers of the presence of young or old star populations. \citet{izotov2016a, izotov2016b, izotov2018a, izotov2018b} report that all leakers have large EW(H$\beta$) (> 200 \AA), which should indicate young stellar populations \citep{stasiska1996}. However, in practice, exact age determinations from EW(H$\beta)$ are complicated, since it depends on the star formation history and the IMF of the stellar population. Additionally, the strength of the stellar features such as the \ion{O}{vi} and \ion{N}{v} P-Cygni profiles at 1020-1040 \AA\ and 1220-1240 \AA\ respectively is related to the stellar population properties. However, the relatively low S/N of our observations make it challenging to constrain the age of the stellar population. \citet{chisholm2019} recently emphasized that high S/N is required to properly constrain the stellar population of the galaxy, especially to accurately derive the escape of ionizing photons from SED fitting. Therefore, robustly constraining the dust attenuation in our sample would require deeper observations. 

Table~\ref{table:addfit} shows that the fitted $C_f$(\ion{H}{i}) is rather insensitive to variations of stellar age or dust extinction, and still provides reliable covering fraction estimate to investigate its impact on the \lya\ and LyC escape. 

\subsection{\lya\ properties}
\label{sect:lya}

\begin{table*}
\caption{\lya\ properties.}
\label{table:lyaprop}
\centering   
\begin{tabular}{lllllllllll}
\hline \hline
Galaxy name & $\frac{\rm F_{trough}}{\rm F_{\rm cont}}$ & EW(Ly$\alpha$) & $f_{\rm esc}^{\rm Ly\alpha}$ & $v^{\rm trough}_{\rm \lya}$ & $v^{\rm blue}_{\rm \lya}$ & $v^{\rm red}_{\rm \lya}$ & $v^{ \rm sep}_{\rm \lya}$\\
 & & [\AA] & & [km s$^{-1}$] & [km s$^{-1}$] & [km s$^{-1}$] & [km s$^{-1}$]\\
 (1) & (2) & (3) & (4) & (5) & (6) & (7) & (8) \\ \hline 
J1243+4646 &  20.37 $\pm$ 3.58 & 98 $\pm$ 3 & 0.52 $\pm$ 0.04 &  40 $\pm$ 20 & -40 $\pm$ 20 & 130 $\pm$ 40 & 160 $\pm$ 63\\ 
J1154+2443 & 11.70 $\pm$ 3.30 & 135 $\pm$ 4 &  0.61 $\pm$ 0.03 & 20 $\pm$ 20 & -80 $\pm$ 30 & 130 $\pm$ 20 & 240 $\pm$ 36\\ 
J1256+4509 &  5.83 $\pm$ 2.40 &  88 $\pm$ 3 &  0.32 $\pm$ 0.03 & 10 $\pm$ 40 & -90 $\pm$ 50 & 160 $\pm$ 20 & 260 $\pm$ 54\\ 
J1152+3400 & 3.75 $\pm$ 1.01  & 75 $\pm$ 6 & 0.34 $\pm$ 0.07 & 60 $\pm$ 30 & -120 $\pm$ 40 & 190 $\pm$ 30 & 320 $\pm$ 42\\ 
J1442-0209 & 4.68 $\pm$ 0.72 & 115 $\pm$ 6 &  0.54 $\pm$ 0.05 & -130 $\pm$ 20 & -250 $\pm$ 50 & 70 $\pm$ 30 & 320 $\pm$ 58\\ 
J0925+1409 &  3.10 $\pm$ 0.93  &  80 $\pm$ 5 &  0.29 $\pm$ 0.03 & -30 $\pm$ 30 & -160 $\pm$ 20 & 150 $\pm$ 20 & 310 $\pm$ 28\\ 
J1011+1947 & 7.66 $\pm$ 2.04 & 115 $\pm$ 4 &  0.18 $\pm$ 0.01 & -30 $\pm$ 20 & -130 $\pm$ 20 & 120 $\pm$ 10 & 260 $\pm$ 22\\ 
J1503+3644 & 0.95 $\pm$ 0.68 &98 $\pm$ 3 &  0.30 $\pm$ 0.04 & -100 $\pm$ 50 & -300 $\pm$ 30 & 140 $\pm$ 30 & 460 $\pm$ 58\\ 
J1333+6246 & 1.66 $\pm$ 0.78 & 73 $\pm$ 2 &  0.51 $\pm$ 0.09 & -150 $\pm$ 20 & -300 $\pm$ 40 & 70 $\pm$ 30 & 380 $\pm$ 58\\ 
J0901+2119 & 5.30 $\pm$ 2.40  & 170 $\pm$ 4 & 0.14 $\pm$ 0.01 & -80 $\pm$ 40 & -180 $\pm$ 40  & 140 $\pm$ 20 & 320 $\pm$ 45\\ 
J1248+4259 &  4.85 $\pm$ 2.25 & 258 $\pm$ 9 & 0.17 $\pm$ 0.01 & -10 $\pm$ 20 & -130 $\pm$ 30 & 140 $\pm$ 30 & 260 $\pm$ 50\\ 
J0921+4509 & 0.14 $\pm$ 0.10 &5 $\pm$ 3 & 0.01 $\pm$ 0.01 & -220 $\pm$ 50;  0 $\pm$ 50 & -460 $\pm$ 60 & 200 $\pm$ 50 & 660 $\pm$ 92\\ 
Tol1247-232  & 0.08 $\pm$ 0.02 & 29 $\pm$ 2 &  0.10 $\pm$ 0.02 & -100 $\pm$ 50 & -300 $\pm$ 10 & 150 $\pm$ 10 & 450 $\pm$ 14\\ 
J0926+4427  & 1.51 $\pm$ 0.10 & 59 $\pm$ 12 &  0.20 $\pm$ 0.06 & -80 $\pm$ 30 & -250 $\pm$ 50 & 160 $\pm$ 50& 410 $\pm$ 71\\ 
J1429+0643  & 0.82 $\pm$ 0.09  & 36 $\pm$ 3 &  0.10 $\pm$ 0.03 &  20 $\pm$ 30 & -220 $\pm$ 50 & 240 $\pm$ 40 & 460 $\pm$ 62\\ 
GP0303-0759 & 0.29 $\pm$ 0.20  & 13 $\pm$ 4 &  0.05 $\pm$ 0.01 & -90 $\pm$ 50 & -310 $\pm$ 20 & 150 $\pm$ 10 & 460 $\pm$ 22\\ 
GP1244+0216 &  0.23 $\pm$ 0.20 & 54 $\pm$ 8 & 0.07 $\pm$ 0.02  & -10 $\pm$ 50 & -260 $\pm$ 40 & 250 $\pm$ 20 & 510 $\pm$ 45\\ 
GP1054+5238 & 0.21 $\pm$ 0.13 & 14 $\pm$ 3 &  0.07 $\pm$ 0.02 & -10 $\pm$ 60 & -220 $\pm$ 50 & 190 $\pm$ 10 & 410 $\pm$ 51\\ 
GP0911+1831 & 1.31 $\pm$ 0.34 & 70 $\pm$ 12 &  0.16 $\pm$ 0.05 & -80 $\pm$ 30 & -280 $\pm$ 50 & 80 $\pm$ 20 & 360 $\pm$ 64\\
\sone\      & < 0.05  & -2 $\pm$ 1  &  < 0.01          &  -120 $\pm$ 40 & -            & 95 $\pm$ 10  &  -           \\ 
\stwo\     & 0.07 $\pm$ 0.21   &18 $\pm$ 1  &  < 0.01          &  -160 $\pm$ 20 & -            & 140 $\pm$ 20 &  -           \\   
Cosmic Eye & < 0.03   & -30 $\pm$ 2 & 0.00    &  -            &    -     &  -           &  -           \\ 
\hline
\end{tabular}
\tablefoot{(1) Galaxy name; (2) normalized flux at minimum of the \lya\ profile; (3) \lya\ restframe equivalent width; (4) \lya\ escape fraction; (5) \lya\ trough velocity (6) \lya\ blue peak velocity; (7) \lya\ red peak velocity; and (8) \lya\ peak velocity separation. \lya\ is seen in absorption in the Cosmic Eye, and only the \lya\ red peak is seen in emission in \sone\ and \stwo. J0921+4509 has two troughs in its \lya\ profile, see discussion in Section~\ref{sect:lya}. The blue and peak velocities reported in Col. (6) and (7) are measured with respect to the systemic velocity. The relative blue and red peak velocities discussed in Sect.~\ref{sect:$C_f$peak} and Fig.~\ref{fig:vel$C_f$} are obtained by subtracting the trough velocity in Col. (5) from the Cols. (6) and (7), respectively.  }
\end{table*}

Table~\ref{table:lyaprop} reports the \lya\ properties for all galaxies in our sample.  One galaxy has a \lya\ absorption profile, two have \lya\ seen both in absorption and emission with a single peak profile. The 19 remaining galaxies have \lya\ seen in emission, 18 exhibiting a double peak profile and one having a triple peak profile (J1243+4646). Eighteen of the 22 galaxies had their \lya\ profiles studied in the literature \citep{ henry2015,izotov2016a, izotov2016b, verhamme2017,puschnig2017, izotov2018a, izotov2018b, orlitova2018}.  We re-measured the \lya\ properties in all spectra to avoid inconsistencies due to different measurement methods. The \lya\ escape fractions were re-calculated coherently using the equation
\begin{equation}
    \fesc^{\lya} = \frac{\rm F(\lya)}{8.7\times \rm F_{\rm corr}(\rm H\alpha)},
    \label{eq:fesclya}
\end{equation}

\noindent where  F($\lya$) is the observed \lya\ flux, corrected for the Milky Way extinction, $\rm F_{\rm corr}(\rm H\alpha)$ is the H$\alpha$ flux corrected for both internal and Milky Way extinction, and 8.7 is the assumed ratio between the intrinsic \lya\ and H$\alpha$ flux assuming Case-B recombination with a temperature of 10$^4$ K and an electron density of $n_e$ = 350 cm$^{-3}$. We note that Case-B recombination assumes that the gas in the ISM is optically thick to radiation above 13.6 eV, thus may not be valid for galaxies where a substantial amount of ionizing photons escape. In galaxies with an optically thin ISM (Case-A recombination), the effective recombination coefficient for the Balmer hydrogen lines can increase by a factor \textasciitilde1.5 \citep{osterbrook1989}. In \citet{izotov2018a}, the authors derived an intrinsic \lya-H$\alpha$ flux ratio of $\approx 11.2$ in the galaxy J1154+2443 using \textsc{cloudy} models \citep{ferland2013, ferland2017}. We investigated in Appendix~\ref{app:sigma} the impact of using a different intrinsic   F($\lya$)/F(H$\alpha$) ratio on the observed \fesclya\ trends derived in this work. Overall, we show that the new significance levels of these trends differ by at most -0.5 $\sigma$, while all the correlations remain significant at least at the 2.5 $\sigma$ level. Thus we assume that fluctuations in the intrinsic \lya\ H$\alpha$ flux ratio should not substantially affect the results presented in this work.

For the low-redshift galaxies, we calculated \fesclya\ using the \lya\ and H$\alpha$ flux measurements, already corrected for internal and Milky Way extinction,  reported in the reference papers. While some of the sources in our sample have \lya\ emission outside the COS aperture \citep{henry2015}, we did not apply aperture correction because the size of the extended \lya\ emission varies from galaxy to galaxy. The final errors were derived by propagating the errors of the individual measurements. We derived an \lya\ escape fraction of 61\% for J1154+2443, which is inconsistent with the 98 \% reported in \citet{izotov2018a}. However, those authors corrected the \lya\ flux with the galaxy internal extinction, which is incompatible with Eq~\eqref{eq:fesclya}. For the three gravitationally lensed galaxies, we measured F($\lya$) using the Monte-Carlo approach described in the Section~\ref{sect:neutprop}, accounting for the lensing magnification factor. The observed flux at \lya\ in the Cosmic Eye is consistent with 0, hence we report $f_{\rm esc}^{\rm Ly\alpha}$ = 0. For \sone\ and \stwo, we used the H$\beta$ flux measurements reported in \citet{wuyts} and assumed an intrinsic $\rm F(\rm H\alpha)/\rm F(\rm H\beta)$ ratio of 2.85.

\begin{figure}[h] 
\begin{center}
\includegraphics[width=\textwidth/2]{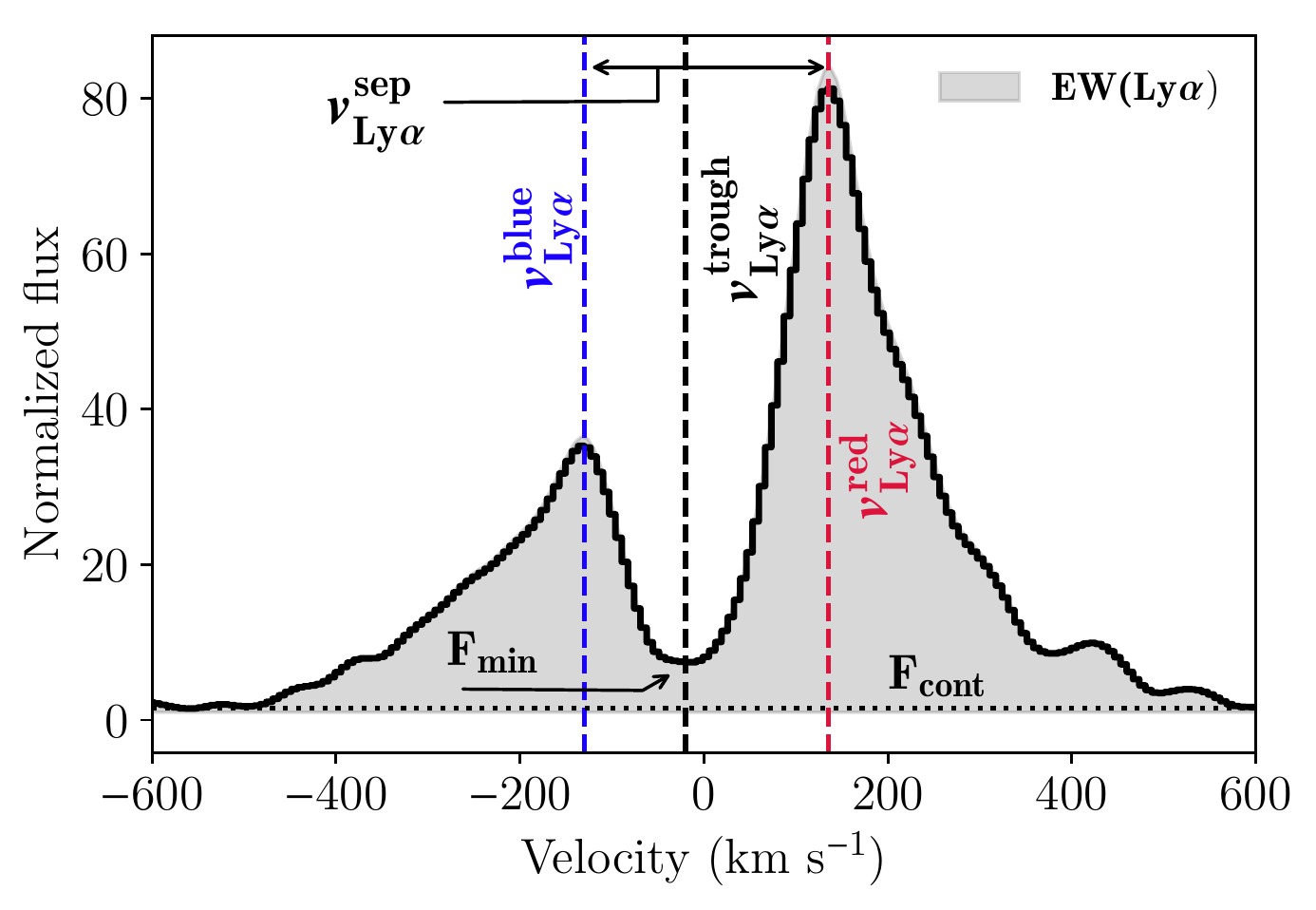}
 \caption{ The Ly$\alpha$ profile observed in the galaxy J1011+1947 with the characteristic measurements annotated. The observed flux has been smoothed by an arbitrary factor for display purposes, and normalized using the median of the flux between 1200 and 1210 \AA. The gray shaded area represents the integrated region to compute the \lya\ equivalent width.}
 \label{fig:lyaex}
 \end{center}
\end{figure}

Table~\ref{table:lyaprop} reports the \lya\ equivalent width, peak and trough velocities, and the normalized flux at the minimum of the profile. These different measurements are illustrated on the \lya\ profile of the galaxy J1011+1947 in Fig.~\ref{fig:lyaex}. All these properties have been measured using the Monte-Carlo method to have consistent values and uncertainties for the entire sample. The \lya\ equivalent widths were derived using the observed MW attenuation-corrected spectra, by integrating both the associated \lya\ emission and absorption such that the integral limits are chosen, by eye, where the flux meets the stellar continuum. The reported EW(\lya) values are respectively positive and negative for galaxies with a net emission or absorption \lya\ profile. The flux above and below the continuum level are respectively The trough velocities $v^{\rm trough}_{\rm \lya}$ are taken as the values when the intensity of the \lya\ flux reaches a local minimum. J0921+4509 and J1243+4509 have peculiar \lya\ profiles. The former has two distinct troughs between the red and blue peaks (see Fig.~\ref{fig:J0921vel}), and both their velocities are reported in Table~\ref{table:lyaprop}. The latter has two peaks bluer than the central trough \citep[see Fig. 7 in][]{izotov2018b}, and we only report the $v^{\rm blue}_{\rm \lya}$ of the peak of maximal intensity. The peak separation, $v^{\rm sep}_{\rm \lya}$, is defined as $v^{\rm red}_{\rm \lya}$ - $v^{\rm blue}_{\rm \lya}$, and the error is derived from the quadratic sum of both uncertainties. Additionally, we measure the flux at the minimum of the \lya\ profile as $\frac{\rm F_{trough}}{\rm F_{\rm cont}}$, where $\rm F_{trough}$ is the minimum flux measured at the \lya\ trough, and $\rm F_{\rm cont}$ is the median value of the stellar continuum estimated between 1160 and 1270 \AA, excluding all emission and absorption lines in that range. In this work, we assume that  the spectral resolution has a negligible impact on the derived $\frac{\rm F_{trough}}{\rm F_{\rm cont}}$ values, because the latter was measured in high-resolution \lya\ spectra (R $\approx$ 16000). However, the impact of low spectral resolution should be investigated to extend this analysis to samples with lower resolving power.
 Except for the \lya\ escape fraction in J1154+2443, all the values presented in Table~\ref{table:lyaprop} are consistent at $\pm$ 1 $\sigma$ with the previous measurements reported in the literature.  Appendix~\ref{app:lya} shows the \lya\ profiles for all galaxies in our sample.

\section{The ISM porosity enables the escape of \lya\ and LyC photons}
\label{sect:results}
In this section, we examine the connection between the neutral gas properties, the \lya\ properties and the escape of LyC photons.

\subsection{The scattering of \lya\ photons in a porous ISM}
\label{sect:$C_f$peak}

\begin{figure*}[h] 
\begin{center}
\includegraphics[scale=1]{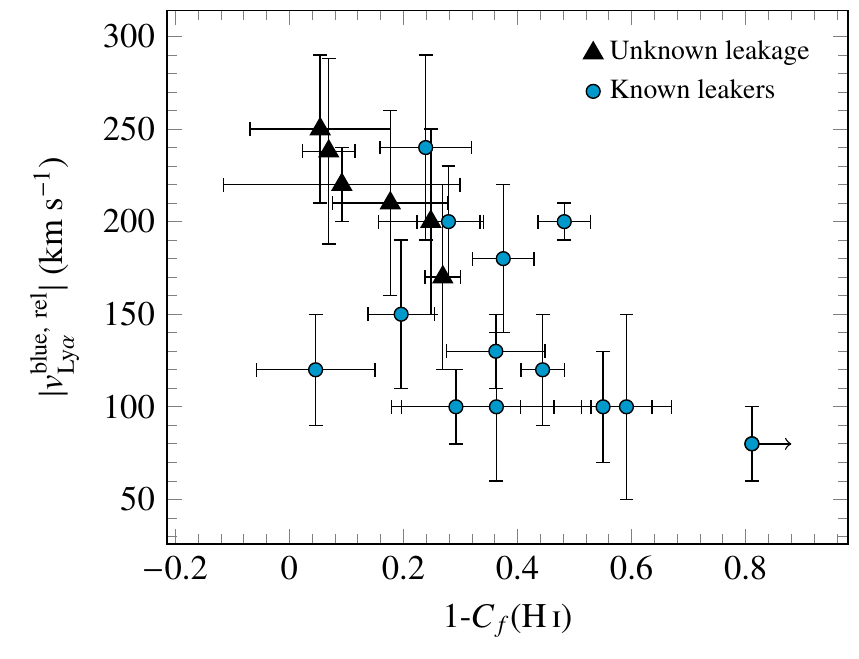}
\includegraphics[scale=1]{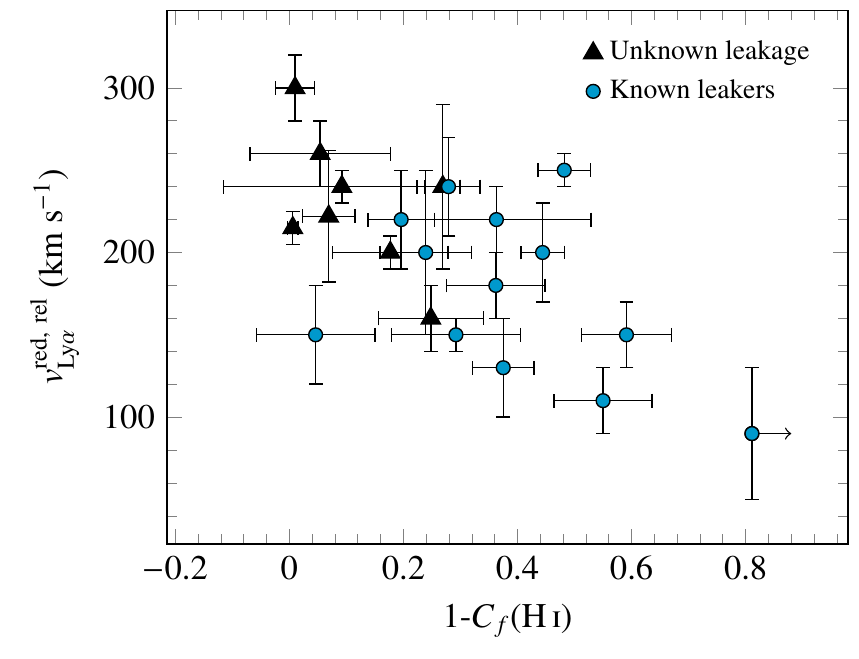}
 \caption{Left: relation between the relative \lya\ blue peak velocity ($\lvert v^{\rm blue, rel}_{\rm \lya}\rvert$ = $\lvert v^{\rm blue}_{\rm \lya}$ - $v^{\rm trough}_{\rm \lya}\rvert$) and 1-$C_f$(\ion{H}{i}).  Right: relation between the relative \lya\ red peak velocity  ($v^{\rm red, rel}_{\rm \lya}$ = $v^{\rm red}_{\rm \lya}$ - $v^{\rm trough}_{\rm \lya}$) and 1-$C_f$(\ion{H}{i}). Our sample is represented by black triangles and blue circles for galaxies with unknown and known leakage, respectively. J0921+4509 has two main troughs, and we defined its relative peak velocities with respect to the respective closest local trough (see text in Sect.~\ref{sect:$C_f$peak}). }
 \label{fig:vel$C_f$}
 \end{center}
\end{figure*}

\begin{figure} 
\includegraphics[scale=1]{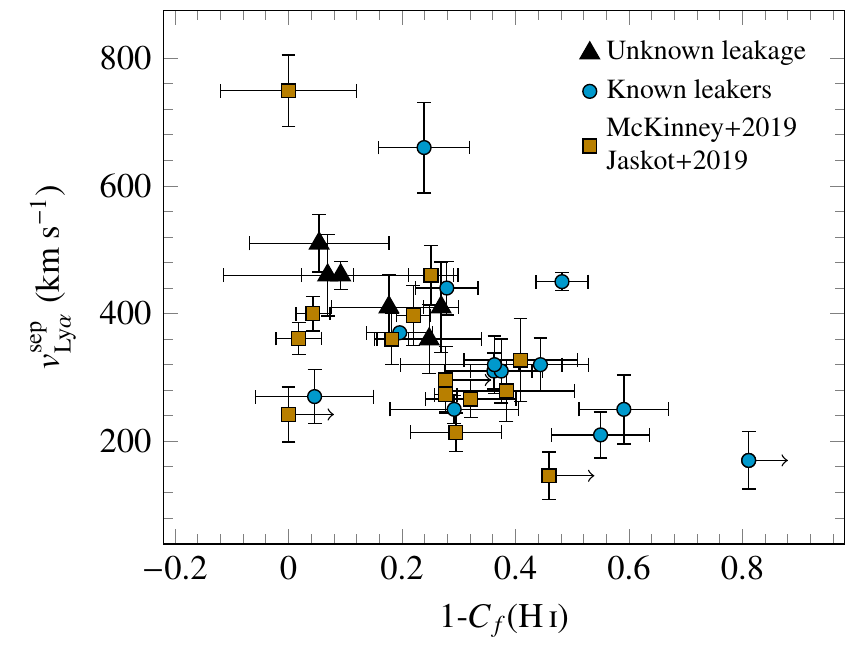}
 \caption{Relation between the \lya\ peak velocity separation ($v^{\rm sep}_{\rm \lya}$) versus 1-$C_f$(\ion{H}{i}). Our sample is represented by black triangles and blue circles for galaxies with unknown and known leakage, respectively. The sample from \citet{mckinney2019} and \cite{jaskot2019} is shown with orange squares. Note that the latter studies only report $C_f$(\ion{Si}{ii}) measurements, from which we derived the corresponding $C_f$(\ion{H}{i}) using the empirical $C_f$(\ion{Si}{ii})-$C_f$(\ion{H}{i}) relation found in \citet{gazagnes2018}. }
 \label{fig:velsep$C_f$}
\end{figure}

The \lya\ transition is a resonant line and each interaction between \lya\ photons and hydrogen atoms shifts the photon’s frequency depending on the velocity of the \ion{H}{i} gas. Therefore, the emergent \lya\ profile provides insights on the neutral hydrogen spatial and velocity distribution. In this section, we investigate the connection between the \ion{H}{i} covering fraction and the scattering of \lya\ photons, depicted by the blue and red velocity shift of the double peak \lya\ profiles. We choose to consider the \lya\ emission velocity relative to the \lya\ absorption trough velocity as $v^{\rm peak, rel}_{\rm \lya}$ = $v^{\rm peak}_{\rm \lya}$ - $v^{\rm trough}_{\rm \lya}$. Hence, $v^{\rm blue, rel}_{\rm \lya}$ and $v^{\rm red, rel}_{\rm \lya}$ are derived by subtracting the Col.~(5) from the Cols.~(6) and (7) in Table~\ref{table:lyaprop}, respectively. This alternative approach provides insights about the velocity of the last scattering of the blue and red shifted \lya\ photons relative to the velocity of the predominant absorption, and also accounts for potential errors in the systemic redshifts \citep[see the discussion in][]{orlitova2018}. J0921+4509 is a peculiar case with two local minima at different velocities (-220 and 0 km s$^{-1}$, see Fig~\ref{fig:J0921vel}). We define its relative peak velocities with respect to the closest absorption troughs, such that $v^{\rm blue, rel}_{\rm \lya}$~=~(-460)~-~(-220)~=~-240 km s$^{-1}$ and $v^{\rm red, rel}_{\rm \lya}$~=~(200)~-~(0)~=~200 km s$^{-1}$.

Figure~\ref{fig:vel$C_f$} shows the relation between the $v^{\rm blue, rel}_{\rm \lya}$ and $v^{\rm red, rel}_{\rm \lya}$ and the neutral gas covering fraction. We find a strong correlation between the blue and red relative velocities of the \lya\ photons with respect to the \ion{H}{i} gas covering fraction (3 $\sigma$  significance, p-value of 0.0026 and 0.0024 respectively). Additionally, we show in Figure~\ref{fig:velsep$C_f$} the connection between the \lya\ peak velocity separation and the \ion{H}{i} covering fraction. We included the recent analysis of the neutral gas properties of 13 low-z GPs with HST-COS observations (GO-14080, PI Jaskot) studied in \citet{mckinney2019} and \citet{jaskot2019}. In the latter study, the authors measured the $C_f$(\ion{Si}{ii}) of these galaxies, from which we derived the corresponding $C_f$(\ion{H}{i}) using the empirical relation $C_f$(\ion{Si}{ii})-$C_f$(\ion{H}{i}) found in \citet{gazagnes2018}. We report a correlation at the 3-$\sigma$ level\footnote{In all this work, for galaxies only having upper limit on $C_f$(\ion{H}{i}), we adopt the upper limit when deriving the significance level of the $C_f$(\ion{H}{i}) correlations.  In Appendix~\ref{app:sigma}, we investigate how the significance level of the reported trends changes for different assumptions, such as fixing $C_f$(\ion{H}{i}) to the lower limit (0), or excluding one or several observations.} (p-value of 0.0018) between $v^{\rm sep}_{\rm \lya}$ and 1-$C_f$(\ion{H}{i}). The trends reported in Figs.~\ref{fig:vel$C_f$} and \ref{fig:velsep$C_f$} indicate that \lya\ photons experience fewer scattering events in galaxies having low \ion{H}{i} covering fraction. Interestingly, this somehow differs from \citet{jaskot2019} and \citet{mckinney2019} who found a weak correlation between $v^{\rm sep}_{\rm \lya}$ and $C_f$(\ion{Si}{ii}). This difference might be explained by the fact that \ion{Si}{ii} and \ion{H}{i} may not trace exactly the same gas. The \ion{Si}{ii} and \ion{H}{i} ionization potential are similar, but not identical. Hence, the \ion{Si}{ii} covering fraction is related to, but not equal to $C_f$(\ion{H}{i}) \citep{reddy2016stack, gazagnes2018}. Additionally, the \ion{Si}{ii} covering fractions measurements might be impacted by scattering and fluorescent emission in-filling \citep[][Mauerhofer et al. in prep]{prochaska2011, scarlata2015}. 

The scaling relation between $v^{\rm sep}_{\rm \lya}$ and $C_f$(\ion{H}{i}) contrasts with the theoretical \lya\ studies of \citet{verhamme2015} and \citet{dijkstra2016}. The latter radiative transfer simulations showed that $v^{\rm sep}_{\rm \lya}$ should scale with decreasing \nh\ in the ISM, but the presence of paths cleared of \ion{H}{i} gas should imprint a single peak profile at the systemic velocity. Nevertheless, the latter analysis assumes that the escape channels are entirely cleared of gas. Other studies showed that the presence of a clumpy ISM \citep{gronke2016} and/or the presence of \ion{H}{i} residuals in the channels \citep{kakiichi2019} could lead to a non-unity neutral gas covering fraction, while imprinting a double peak profile. Using the \ion{O}{i} column densities and reported metallicities, we find that 13 of 22 galaxies in our sample have \nh\ larger than 10$^{18}$ cm$^{-2}$, while \citet{mckinney2019} found similar results in their sample using both \ion{O}{i} and \ion{Si}{ii}. Hence, in these galaxies, the porosity of the ISM (low $C_f$(\ion{H}{i})) should be physically interpreted as the existence of channels with low column densities in an optically thick \ion{H}{i} environment. To refer to this bi-modal distribution of \ion{H}{i}, we introduce the notations \nhchan\ and \nhcloud\ to denote the column densities within the channels and within the clouds, respectively. The scaling relation between $v^{\rm sep}_{\rm \lya}$ and  $C_f$(\ion{H}{i}) suggests that galaxies that have more of these channels also have lower \nhchan. Hence, the presence of low $C_f$(\ion{H}{i}) could indirectly trace the abundance of \ion{H}{i} in the lowest column density regions of the ISM.

The tight connection between the velocity shift of the peaks of the \lya\ emission and $C_f$(\ion{H}{i}) additionally indicates that the porosity of the ISM impacts the shape of the \lya\ profile. Indeed, in this work, we investigate the relative \lya\ peak velocities which can be interpreted as the velocity shift of the escaping \lya\ photons with respect to the velocity of the predominant \ion{H}{i} absorption \citep[similarly to][]{orlitova2018}. In the literature, \lya\ peak velocities are usually considered with respect to the systemic velocity. Interestingly, we do not find any correlation between the red peak velocity relative to the systemic velocity and the \ion{H}{i} covering fraction and the significance level of the correlation between the blue peak velocity relative to the systemic velocity and $C_f$(\ion{H}{i}) is lower (2 $\sigma$). This suggests that the shift of both the blue and red-shifted \lya\ photons relates to the properties of the main \ion{H}{i} absorption, which traces the bulk of the \ion{H}{i} gas along the line of sight (LOS). The velocity of the last scattering of \lya\ photons likely correlates with the velocity coverage of the thick \ion{H}{i} gas, and with the presence of low-density channels in the ISM. This trend is somewhat surprising for the red \lya\ emission. Indeed, the red peak of the \lya\ profile corresponds to the back-scattered \lya\ photons, which were first emitted in the direction opposite to the observer. In theory, they probe the properties of the gas in the back side of the galaxy ISM, and are expected to have a weaker connection to LOS-dependent properties, such as $C_f$(\ion{H}{i}), than blue-shifted photons. Figure~\ref{fig:vel$C_f$} shows that this is not the case. This could be because galaxies that have lower $C_f$(\ion{H}{i}) also have lower \ion{H}{i} column densities in the ISM, such that $v^{\rm red, rel}_{\rm \lya}$ indirectly probes the average \nh\ of the neutral gas.

The strong correlation between the peak velocities and the presence of low column density channels could in principle be used to indirectly detect the leakage of \lya\ and LyC photons through these sightlines. \citet{kakiichi2019} suggested that, in an ionization-bounded ISM, the \lya\ peak velocity separation could probe \nhchan. A similar idea was discussed in \citet{jaskot2019} where the authors proposed $v^{\rm sep}_{\rm \lya}$ as a probe of the residual \nh\ along the paths of "least resistance" through which \lya\ photons escape. These assumptions would explain the strong correlations between low peak velocity separations and \fesclycobs\ \citep{verhamme2017,izotov2018b, izotov2019}. Nevertheless, we further show in Sect.~\ref{sect:ismvel} that both low and high column densities of gas impact the \lya\ peak separation, such that a single parameter does not fully determine \fesclyc.

\subsection{Narrower \ion{H}{i} absorption lines lead to larger observed \lya\ emission}
\label{sect:ismvel}

\begin{figure*}
\includegraphics[scale=1]{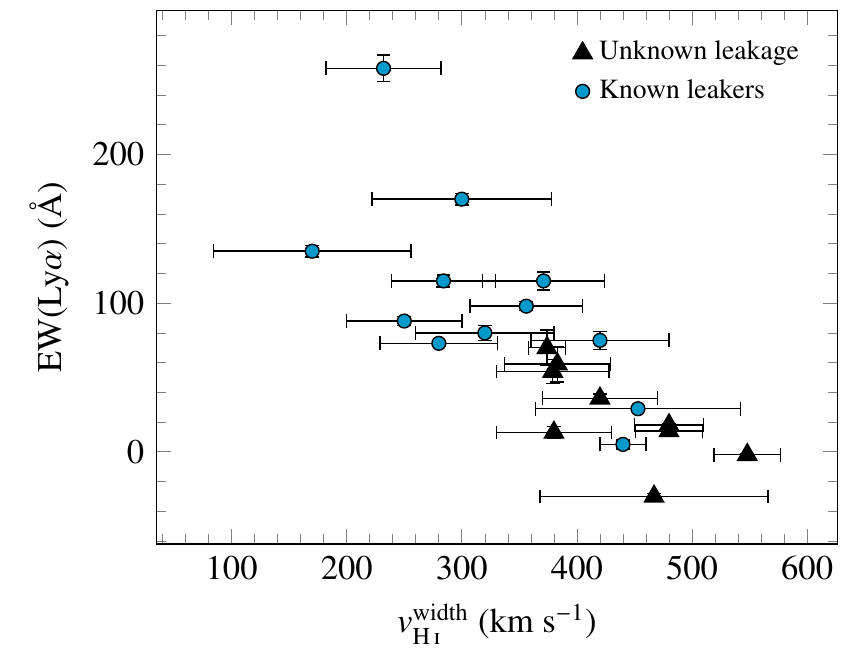}
\includegraphics[scale=1]{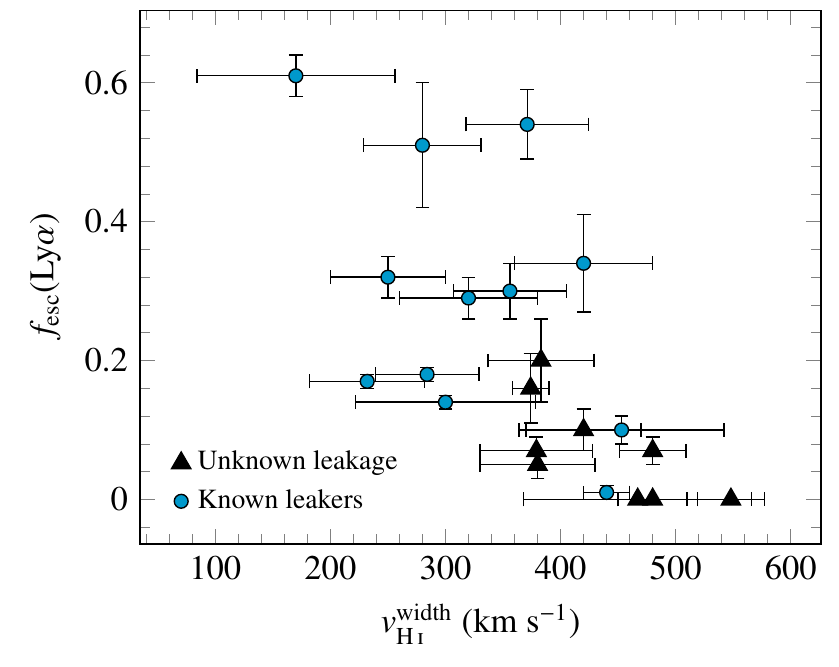}
 \caption{\lya\ equivalent width (left) and escape fraction (right) versus the width of the \ion{H}{I} maximal absorption. Galaxies with a lower velocity width of the Lyman series have higher EW(\lya) and \fesclya. }
 \label{fig:vwidth}
\end{figure*}

\begin{figure} 
\includegraphics[width=\hsize]{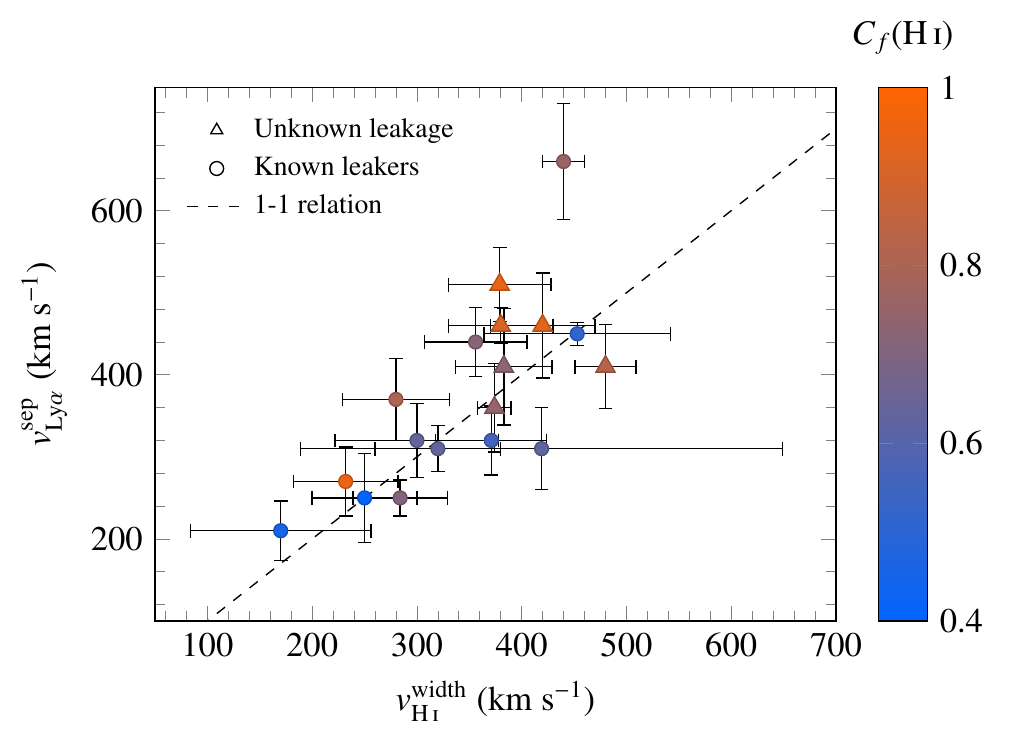}
 \caption{\lya\ peak velocity separation versus the \ion{H}{I} velocity width of maximal absorption. The color bar represents the \ion{H}{i} covering fraction measurement of each galaxy. Overall, the \lya\ peak velocity separation scales with the width of the maximal absorption of the \ion{H}{i} gas, and galaxies with larger $C_f$(\ion{H}{i}) have larger $v^{\rm sep}_{\rm \lya}$ than $v^{\rm width}_{\rm \ion{H}{i}}$, such that the \lya\ emission peaks at velocities outside of the \ion{H}{i} absorption. }
 \label{fig:vwidthvsep}
\end{figure}

\begin{figure*}  
\includegraphics[width=\hsize/2]{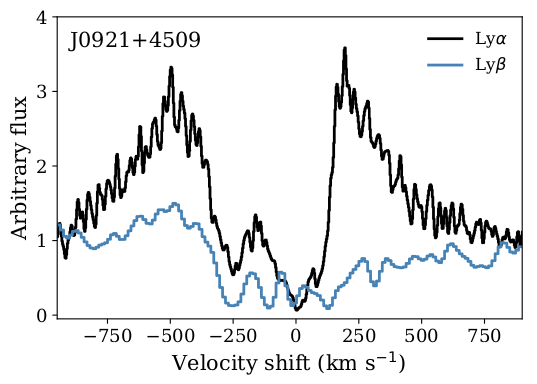}
\includegraphics[width=\hsize/2]{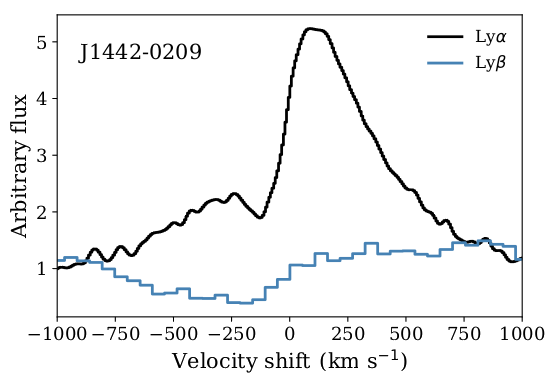}
 \caption{Left: plot of the \lya\ emission line and Ly$\beta$ absorption line in J0921+4509 ($C_f(\ion{H}{i}$) = 0.761). Right: plot of the \lya\ emission line and Ly$\beta$ absorption line in J1442-0209 ($C_f(\ion{H}{i}$) = 0.556). The \lya\ spectra have been smoothed to half the resolution of the observations, and scaled down with a power law for display purposes. The presence of \lya\ emission on top of the optically thick \ion{H}{i} absorption line suggests that \lya\ photons can escape at lower velocities through low-density channels.  Similar plots for all the galaxies in our sample are in the Appendix~\ref{app:lya}.}
 \label{fig:J0921vel}
\end{figure*}

 Because the \lya\  is a resonant transition, the \lya\ photons are strongly impacted by the neutral gas kinematics. \citet{rivera2015} outlined the role of the velocity coverage of the \ion{H}{i} gas as an important factor governing the observed emission of \lya\ photons. In Fig.~\ref{fig:vwidth}, we compare the \lya\ equivalent width (left panel) and the \lya\ escape fraction (right panel) with the \ion{H}{i} velocity width of the Lyman series in our sample. The latter is defined as the velocity interval where the \ion{H}{i} absorption profile is at its maximum depth. We find strong anti-correlations both with EW(\lya) (4 $\sigma$, p-value of 0.000025) and  with \fesclya\ (3 $\sigma$, p-value of 0.0007) that confirm that the velocity width of the optically thick \ion{H}{i} absorption significantly impacts the emission and escape of \lya\ photons. These results suggest that fewer \lya\ photons escape from galaxies with broad \ion{H}{i} absorption lines, because \lya\ photons are more likely to encounter optically thick neutral gas and hence have a higher probability to be destroyed by dust grains. However, disentangling the physical origin of a large $v^{\rm width}_{\rm \ion{H}{i}}$ is not trivial. This can either result from large \nh, or from a broad velocity range of the absorbing gas. Both cases should similarly lower the escape and observed emission of \lya\ photons because for \lya\ photons to escape, it needs to be scattered to velocities where the gas is transparent. Nevertheless, this is different for LyC photons, since they are sensitive to the \nh\ in the ISM, but not to the kinematics of the \ion{H}{i} gas.
 
Fig.~\ref{fig:vwidthvsep} provides additional insights about the origin and impact of broad \ion{H}{i} absorption lines on the escape of \lya\ and LyC photons. It shows the linear relation (at the 3 $\sigma$ significance, p-value of 0.0007) between the \ion{H}{i} velocity width of the maximal absorption and the \lya\ peak velocity separation. $v^{\rm sep}_{\rm \lya}$ is linearly consistent at $\pm 1 \sigma$ with $v^{\rm width}_{\rm \ion{H}{i}}$  for 17 galaxies (\textasciitilde 80\%) in our sample. Because $v^{\rm sep}_{\rm \lya}$ scales with \nh\ \citep{verhamme2015, dijkstra2016}, this trend could suggest that the observed width of the \ion{H}{i} absorption lines indirectly probes the \ion{H}{i} column densities in the dense regions (\nhcloud) which are imprinted from the saturated Lyman series. Additionally, recent observational studies have highlighted the connection between low peak velocity separation and the escape of ionizing photons \citep{verhamme2017,izotov2018b,izotov2019}. Interestingly, we find that J1154+2443 and J1256+4509, the two largest leakers (38 and 46\%) with Lyman series observations, have the lowest $v^{\rm width}_{\rm \ion{H}{i}}$ in our sample. Similarly, we do not detect \ion{H}{i} in the largest \fesclycobs\ LCE (J1243+4646). Hence, in these galaxies, the thick neutral clouds in the ISM might have \nhcloud\ low enough to let a significant fraction of ionizing photons escape. Note that this hypothesis is not incompatible with the presence of saturated \ion{H}{i} absorption lines, because the range of Lyman series lines we study in this work (Ly$\beta$ to Ly6) saturates at \nh\  larger than \textasciitilde10$^{16}$ cm$^{-2}$. Additionally, we do not report any significant trend between \fesclycobs\ and $v^{\rm width}_{\rm \ion{H}{i}}$ in the low LyC leakers (\fesclycobs\ $< 13\%$). Hence, for these galaxies, the \nhcloud\ is likely too large and efficiently absorbs all the ionizing radiation that passes through the clouds. This is consistent with \citet{gazagnes2018} where we estimated \nh\ larger than 10$^{18}$ cm$^{-2}$ using \no\ for 6 of the 10 low LyC leakers with observed \ion{O}{i} absorption lines. Hence, this result could emphasize that the leakage of LyC photons in the largest leakers is a combination of ionization and density bounded mechanisms, highlighted by the presence of both low $C_f$(\ion{H}{i}) and low $v^{\rm width}_{\rm \ion{H}{i}}$. However, this outcome should be taken with caution, because $v^{\rm width}_{\rm \ion{H}{i}}$ is highly degenerate with the velocity distribution of the \ion{H}{i} gas. We discuss further this point in Sect.~\ref{sect:cfpred}.

In Sect.~\ref{sect:$C_f$peak}, we showed the tight correlation between $v^{\rm sep}_{\rm \lya}$ and the \ion{H}{i} covering fraction. Fig.~\ref{fig:vwidthvsep} additionally shows the impact of low $C_f$(\ion{H}{i}) on the relation between $v^{\rm sep}_{\rm \lya}$ and $v^{\rm width}_{\rm \ion{H}{i}}$. Overall, galaxies with large $C_f$(\ion{H}{i}) have $v^{\rm sep}_{\rm \lya}$ larger than $v^{\rm width}_{\rm \ion{H}{i}}$, such that the \lya\ emission peaks at velocities beyond the \ion{H}{i} absorption. This is illustrated in the left panel of Fig.~\ref{fig:J0921vel}, where we superpose the \lya\ profile and the Ly$\beta$  line for the galaxy J0921+4509. J0921+4509 has a relatively high covering fraction, ($C_f$(\ion{H}{i}) $\approx$ 0.8) and the peaks of its \lya\ emission are located on the edges of the optically thick \ion{H}{i} absorption. On the other hand, Fig.~\ref{fig:vwidthvsep} shows that galaxies with a lower  $C_f$(\ion{H}{i}) have $v^{\rm sep}_{\rm \lya}$ $\leq$ $v^{\rm width}_{\rm \ion{H}{i}}$. This is because decreasing $C_f$(\ion{H}{i}) increases the probability of \lya\ photons to find an escape road through low-density channels and thus lowers the amount of scattering events for these photons (see Sect.~\ref{sect:$C_f$peak}). The right panel of Fig.~\ref{fig:J0921vel} shows the superposition of \lya\ and Ly$\beta$ for J1442-0209 which has a lower $C_f$(\ion{H}{i}) ($\approx$ 0.6). The blue peak falls directly on top of the optically thick part of the Ly$\beta$ absorption line, which implies that a dominant proportion of \lya\ photons on the blue portion of the line escape at velocities where the neutral gas is optically thick. Thus, the presence of \lya\ emission on the top of the \ion{H}{i} absorption is evidence of the existence of low-density channels (low $C_f(\ion{H}{i})$), and strongly indicates whether \lya\ photons are able to escape from the galaxy. This also supports a plausible origin for the role of the blue \lya\ peak in identifying the LyC escape \citep{henry2015, orlitova2018}, because LyC photons should escape through the same low-density paths along the line of sight. 
 
In Sect.~\ref{sect:$C_f$peak}, we found a tight correlation between $v^{\rm red, rel}_{\rm \lya}$  and $C_f$(\ion{H}{i}) indicating that the back-scattered emission is similarly affected by the presence of low-density channels. Nevertheless, Fig.~\ref{fig:J0921vel} shows that the red emission peaks at velocities outside of the optically thick \ion{H}{i}  absorption. More generally, we found that this is the case for all galaxies in our sample (Appendix~\ref{app:lya} shows the combinations of the \lya\ and Lyman series profiles). In Sect.~\ref{sect:$C_f$peak}, we suggested that the relation between the velocity shift of the red peak of the \lya\ profile and $C_f$(\ion{H}{i}) could arise from the presence of both lower \nhchan\ and lower \nhcloud\ in the galaxies with the lowest neutral gas covering fraction. Hence, this may suggest that there exists a connection between low \ion{H}{i} velocity width of maximal absorption (related to \nhcloud) and low \ion{H}{i} covering fraction (related to \nhchan). We report a moderate 2-$\sigma$ correlation between $C_f$(\ion{H}{i}) and $v^{\rm width}_{\rm \ion{H}{i}}$, hence, we can not significantly conclude that such relation exists. 

 Overall, the strong correlations between the peak velocity separation and the neutral gas covering fraction (Sect.~\ref{sect:$C_f$peak}), or the width of the saturated Lyman series, suggests that $v^{\rm sep}_{\rm \lya}$ is sensitive both to the presence of low column density paths and to the properties of the dense \ion{H}{i} regions in the ISM. As mentioned above, this might be because the blue peak is more sensitive to the existence of channels through which \lya\ and LyC photons more easily escape, while the red peak probes the overall abundance of \nh\ in the ISM. Hence, both lower \ion{H}{i} covering fraction, and lower \nhcloud\ similarly impact the peak separation of the \lya\ profile because the latter is sensitive to the "total" \ion{H}{i} properties of the ISM of galaxies. While this outcome is somewhat surprising, it suggests that the peak velocity separation is a robust probe of the presence of low column density paths towards the observer, and a key observable to investigate the escape of \lya\ or LyC photons in low redshift LCE candidates \citep{verhamme2017, izotov2018b, izotov2019}. Nevertheless, these outcomes also suggest that measuring a low $v^{\rm sep}_{\rm \lya}$ is not enough to disentangle the dominant leakage mechanisms, because both the ISM porosity and the width of the saturated \ion{H}{i} absorption lines impact the peak separation velocity. We discuss this point further in Sect.~\ref{sect:physpic}. 

At higher redshift,  where observing the Lyman Series is highly unlikely because of the presence of line-of-sight neutral gas in the IGM, the ISM porosity can be traced with $C_f$(\ion{Si}{ii}), which probes the presence of low-density channels when the Lyman series is not observable \citep{gazagnes2018}. Additionally, \citet{chisholm2016} found tight correlations between the velocity widths of different ionic transitions, and suggest that the velocity width of \ion{Si}{ii} is related to $v^{\rm width}_{\rm \ion{H}{i}}$. The combination of both could be used to estimate the \lya\ peak separation and probe the \lya\ and LyC escape of photons at higher redshift. We discuss this further in Sect.~\ref{sect:discpred}.

\subsection{The \lya\ equivalent width scales with low $C_f$(\ion{H}{i})}
\label{sect:ew}
\begin{figure} 
\includegraphics[width=\hsize]{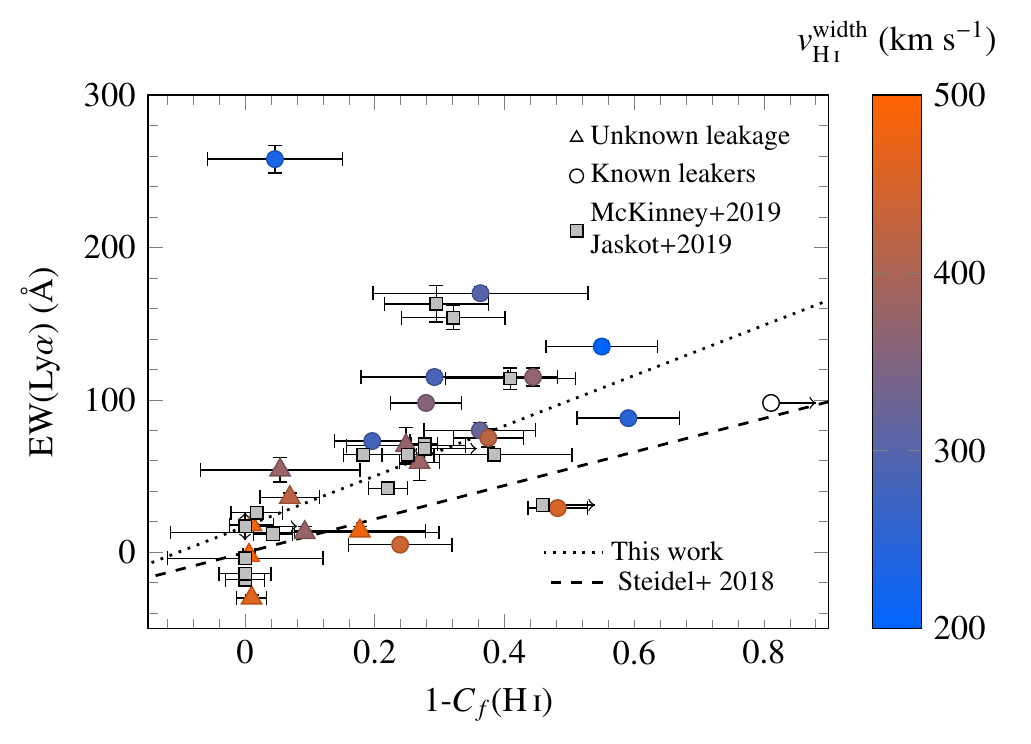}
 \caption{ \lya\ equivalent width versus the porosity of the neutral gas as given by 1 - $C_f(\ion{H}{i})$. The color bar represents the \ion{H}{i} velocity interval where the Lyman series is at its maximum depth. Our sample is represented by triangles and circles for galaxies with unknown and known leakage respectively, and the sample from \citet{mckinney2019} and \cite{jaskot2019} is shown with gray squares as they do not have \ion{H}{i} velocity width measurements. The dashed line shows the empirical relation found in \citet{steidel2018}, while the dotted line is the fit to the data, excluding J1248+4259 (see Eq~\eqref{eq:$C_f$ew}). This figure suggests that \lya\ emission strongly scales with the covering fraction of the channels and with the width of the saturated \ion{H}{i} absorption lines. }
 \label{fig:$C_f$vsew}
\end{figure}

The EW(\lya) and neutral gas properties of Lyman $\alpha$ Emitters have been investigated in numerous studies \citep{rivera2015, verhamme2017, chisholm2017leak, steidel2018, orlitova2018, du2018, mckinney2019, trainor2019, jaskot2019,runnholm2020}. Figure~\ref{fig:$C_f$vsew} explores the connection between the EW(\lya), the neutral gas coverage and the velocity width of the maximal absorption of the \ion{H}{i} lines.  \citet{jaskot2019} found a positive correlation between the escape of \lya\ photons and the \ion{Si}{ii} covering fraction in highly ionized GPs. Their sample, as well as J1243+4646, appear in grey in Fig.~\ref{fig:$C_f$vsew} as they do not have measured $v^{\rm width}_{\rm \ion{H}{i}}$.
We report a 2.5 $\sigma$ significance relation (p-value of 0.004), between EW(\lya) and 1-$C_f(\ion{H}{i})$. The trend is scattered, and Fig.~\ref{fig:$C_f$vsew} shows that most galaxies with high EW(\lya) have both low $C_f(\ion{H}{i})$ and low $v^{\rm width}_{\rm \ion{H}{i}}$. This supports the picture where both the presence of channels with low \nhchan\ and narrower \ion{H}{i} absorption lines increase the observed emission of \lya\ photons. Nevertheless, the interpretation of the \lya\ EW is not trivial because EW(\lya) is also degenerate with varying galaxy properties, such as burst age, star formation history or metallicity. 

One can note the peculiar case of the galaxy J1248+4259, with the largest EW(\lya) in both samples (258 \AA), but a relatively low 1-$C_f$(\ion{H}{i}) (0.046). Surprisingly, J1248+4259 has one of the weakest \ion{O}{VI} and \ion{N}{V} P-Cygni profiles in our sample (see Fig.~\ref{fig:J1243}), which suggest the presence of an older population of stars, at first sight incompatible with the presence of large EW(\lya) \citep{schaerer2003}. On the other hand, weak \ion{O}{VI} and \ion{N}{V} P-Cygni profiles could be explained by a young stellar population, but without forming the most massive stars. We also report a low $v^{\rm width}_{\rm \ion{H}{i}}$ for J1248+4259, which could partly explain the large EW(\lya) measured. Indeed, as seen in Sect.~\ref{sect:ismvel}, low $v^{\rm width}_{\rm \ion{H}{i}}$  either trace a narrow \ion{H}{i} velocity distribution or low \nh, both enhancing the observed emission of \lya\ photons. However, the large EW(\lya) in J1248+4259 likely results from a combination of several galaxy properties, such that disentangling its origin is not trivial.

Consequently, we exclude J1248+4259 from the sample and derive the equation that linearly relates EW(\lya) and 1-$C_f(\ion{H}{i})$ as (4 $\sigma$ significance level, p-value < 0.00002):
\begin{equation}
    EW(Ly\alpha) = 165 \times (1-C_f(\ion{H}{i})) + 17.
    \label{eq:$C_f$ew}
\end{equation}

\noindent  We include this relation in Figure~\ref{fig:$C_f$vsew} (dotted line), and compare it with the one obtained by \citet{steidel2018} (dashed line), where the authors investigated the connection between the neutral gas properties and the escape of \lya\ and LyC photons in composite spectra of z\textasciitilde 3 galaxies observed with the Low Resolution Imaging Spectrometer \citep[LRIS;][]{oke1995, steidel2004}  on the Keck I telescope. Both linear relations have similar slopes, but the relation from \citet{steidel2018} slightly underestimates the EW(\lya) from our sample. One reason for this is that \citet{steidel2018} sample is drawn from composite spectra with EW(\lya) < 50 $\AA$, while \textasciitilde60\% of both our and the \citet{jaskot2019, mckinney2019} samples have EW(\lya) larger than 50 \AA. Besides, the \lya\ equivalent width is also known to relate to the aperture size of the instrument \citep{steidel2011,hayes2014, wisotzki2016}, and therefore complicates direct comparison between samples observed with different instruments. Finally, the higher EW(\lya) in our sample might be biased by different galaxy properties (i.e. stellar masses, velocity widths, etc) and because our sample was selected to be only galaxies with intense star formation and young stellar populations. 

Although the physical interpretation of high EW(\lya) is complex, the tight connection between EW(\lya) and 1-$C_f(\ion{H}{i})$ is consistent with the findings from Sect.~\ref{sect:$C_f$peak} and \ref{sect:ismvel} and confirms the strong impact of low neutral gas coverage on the observed emission of \lya\ photons.

\subsection{The escape of \lya\ and LyC photons is larger in galaxies with lower $C_f$(\ion{H}{i}) and \ebv}
\label{sect:$C_f$pred}

\begin{figure*} 
\includegraphics[width=0.5\textwidth]{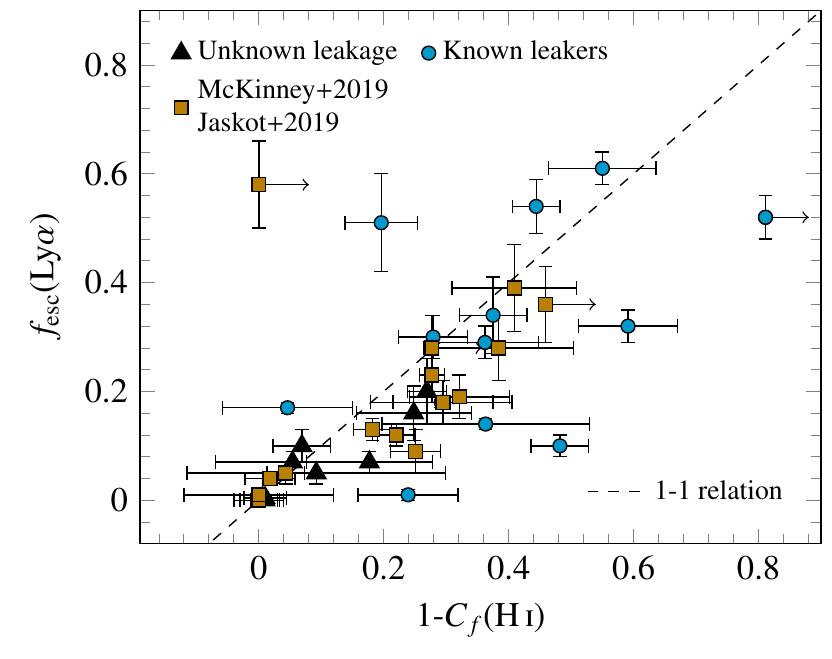}
\includegraphics[width=0.5\textwidth]{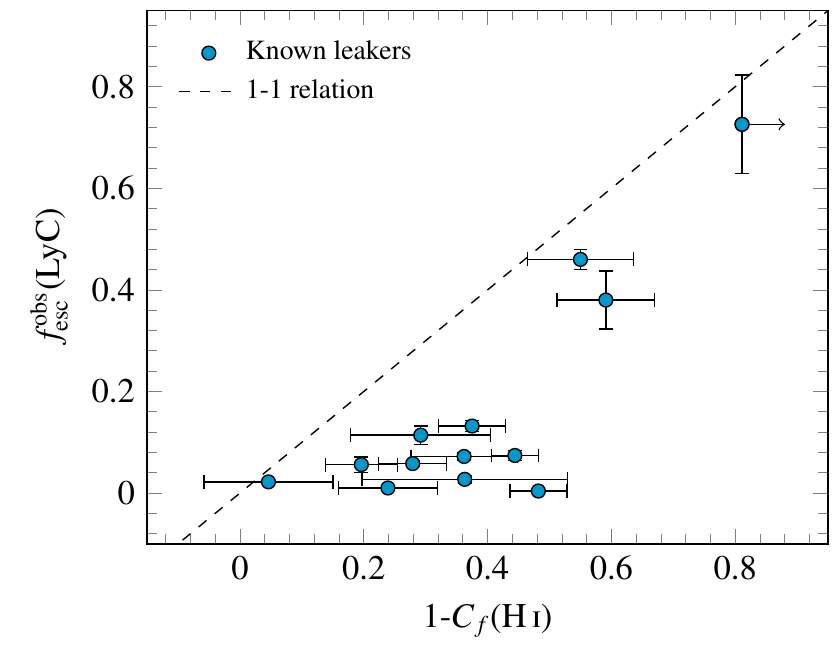}
 \caption{Observed \fesc(\lya) (left) and \fesclycobs\ (right) versus the 1-$C_f$(\ion{H}{i}). There is a strong correlation between the escape of \lya\ and LyC photons and the porosity of the \ion{H}{i} gas (4 $\sigma$ and 3 $\sigma$ level, respectively). The right panel shows that 1-$C_f$(\ion{H}{i}) is always an upper limit to the observed escape fraction of ionizing photons. }
 \label{fig:fesc$C_f$}
\end{figure*}

\begin{figure*} 
\includegraphics[width=0.5\textwidth]{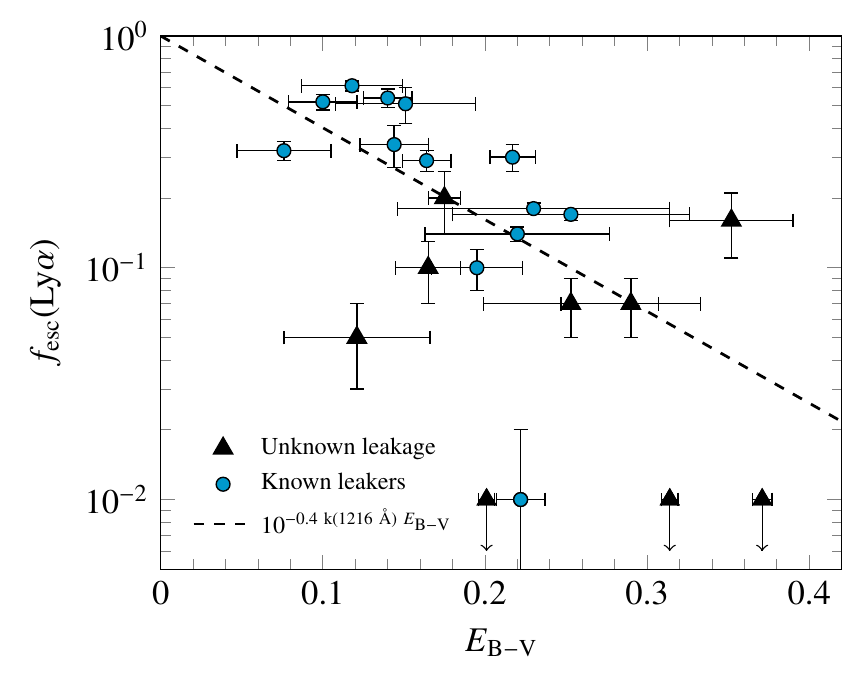}
\includegraphics[width=0.5\textwidth]{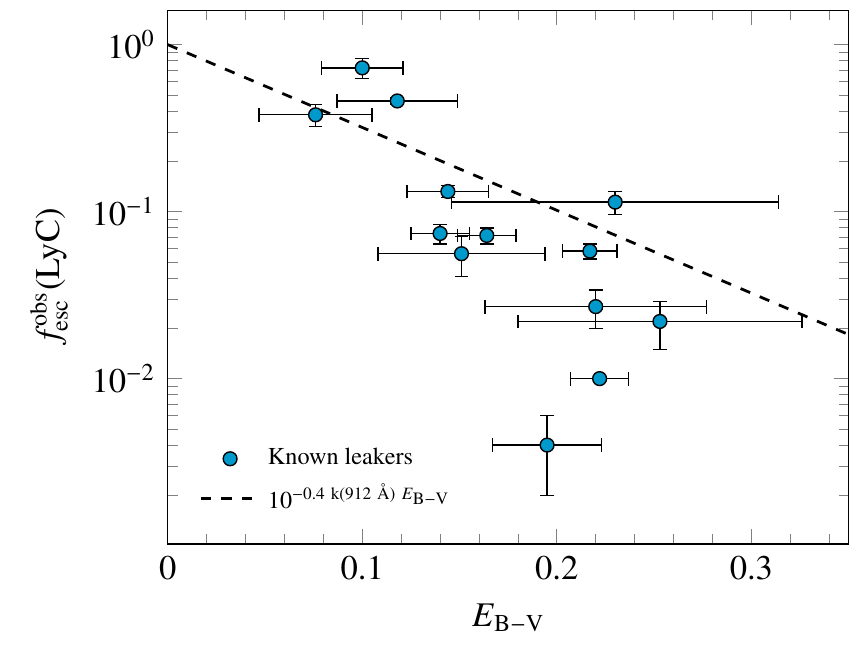}
 \caption{Observed \fesc(\lya) (left) and \fesclycobs\ (right) versus the dust extinction of the stellar continuum with logarithmic y-axes. We included the shape of the dust attenuation curve at the wavelength of interest (either \lya\ or LyC) with a dashed line.   The correlation between the escape fractions of \lya\ and LyC photons and the respective dust attenuation curve is significant at 3 $\sigma$.}
 \label{fig:fescebv}
\end{figure*}

Investigating the connection between the \ion{H}{i} column density, \ion{H}{i} covering fraction, and the escape fraction of \lya\ and LyC photons should allow us to understand the \ion{H}{i} geometry that controls and regulates the leakage of these photons out of the ISM. Recently, we found saturated Lyman series with non-unity covering fraction in 9 LCEs included in this work,  highlighting the existence of both low and high column density paths in galaxies that leak ionizing photons \citep{gazagnes2018}. Additionally, we found that 6 LCEs have \nh\ measurements supporting the presence of optically thick regions (> 10$^{18}$ cm$^{-2}$) in the ISM, which is incompatible with a density bounded scenario. Several studies similarly highlighted the close connection between $C_f$(\ion{H}{i}) and/or $C_f$(\ion{Si}{ii}) and the \lya\ escape fraction and equivalent width \citep{rivera2015, steidel2018, mckinney2019, jaskot2019}.

In Fig.~\ref{fig:fesc$C_f$} we analyze the dependence between the covering fraction and the observed escape fractions of \lya\ and LyC photons. We find a 4 $\sigma$ correlation between 1-$C_f$(\ion{H}{i}) and \fesc(\lya) (p-value < 0.00002) and a 3 $\sigma$ significance level for the trend that connects  1-$C_f$(\ion{H}{i}) to \fesclycobs\ (p-value of 0.0005). Further, the left panel of Fig.~\ref{fig:fesc$C_f$} shows that most galaxies have a \fesclya\ lower than or equal to $1-C_f$(\ion{H}{i}).
For \fesclycobs, the trend is slightly more scattered, but emphasizes that 1-$C_f$(\ion{H}{i}) always overestimates the observed escape fraction of ionizing photons. This observational result is consistent with theoretical studies which suggest that $C_f$(\ion{H}{i}) is always an upper limit to the escape of ionizing photons  \citep{kakiichi2019}. Indeed, several reasons explain why the ionizing escape fraction should deviate from 1-$C_f$(\ion{H}{i}): the presence of \ion{H}{i} residuals in the channels, the necessity to account for the dust attenuation \citep{chisholm2018}, or because the covering fraction measured is always a lower limit to the true geometrical neutral gas covering \citep[see discussion in][Mauerhofer et al, in prep]{rivera2015,vasei2016, gazagnes2018}. While these effects also impact the escape fraction of \lya\ photons, \lya\ photons have a higher chance to escape by being scattered and either shifted out of the velocity range of the optically thick neutral gas, or re-emitted in the direction of a low column density sightline. Hence, the kinematics of the \ion{H}{i} gas can affect the escape of \lya\ photons, while it does not impact the escape of LyC photons. This could explain why the neutral gas coverage is more correlated with the escape of \lya\ photons. Further LyC observations are needed to confirm this trend. Nonetheless, these results confirm that the common origin of the leakage of \lya\ and LyC photons is the ISM porosity.

Additionally, we also investigated the impact of dust extinction on the escape fraction of \lya\ and LyC photons. Several studies have shown that dust readily destroys \lya\ photons \citep{neufeld1991,hayes2011, verhamme2015, verhamme2017,du2018, kimm2019, jaskot2019}, while \citet{chisholm2018} showed that the dust attenuation is a crucial ingredient for the escape of ionizing photons. On the other hand, recent \lya-LyC simulations from \citet{kimm2019} found that the presence of dust grains should not significantly affect the leakage of LyC photons. Hence, the true impact of dust on the ionizing radiation is still poorly understood. 

In Fig.~\ref{fig:fescebv}, we compare the escape fractions of \lya\ and LyC photons with the \ebv\ obtained from the stellar continuum UV fits. Both panels highlight that galaxies with large \fesclyc\ and \fesclya\ have lower dust extinction. Interestingly, the three largest LyC leakers have the lowest \ebv, suggesting that low amount of dust favors the escape of LyC photons. We superposed on the same figure the theoretical attenuation curve (dashed lines) given by 10$^{-0.4A(\lambda)}$, where $A(\lambda)$ is derived using the dust extinction law from \citet{reddy2016dustlaw} at $\lambda =$ 1216 \AA\ for \lya\ photons and 912 \AA\ for LyC photons. We find a 3 $\sigma$ significance level correlation between the dust attenuation and the escape fractions of LyC and \lya\ photons (p-value of 0.0014 and 0.0011, respectively). Hence, this indicates that the presence of interstellar dust grains in the ISM has a significant impact the escape of both types of emission. However, these results need to be taken with caution. 

Figure~\ref{fig:fescebv} shows that a few galaxies have their \lya\ or LyC escape fractions above the dust attenuation curve. In theory, in galaxies having no \ion{H}{i} and a screen layer of dust, the fraction of LyC and \lya\ radiation escaping should be directly given by 10$^{-0.4A(\lambda)}$. Hence, no galaxies should have \fesclycobs\ or \fesclya\ larger than the dust attenuation curve. Several reasons for this discrepancy were considered in Section~\ref{sect:dust}, where we highlighted the complex task of measuring \ebv, and showed that different dust extinction laws can lead in significant differences in the fitted attenuation at 1216 and 912 \AA. Additionally, in this work, we use the \citet{reddy2016dustlaw} dust attenuation curve, which is derived in the far-UV observations ($\lambda = 950 -1500\ \AA$). Nevertheless,  the physical impact of dust extinction on the ionizing flux (< 912 \AA) is still unconstrained. Finally, the dust geometry model can also impact these results. A clumpy distribution of \ion{H}{i} and dust could theoretically have a shielding effect such that \lya\ photons have a lower probability to encounter dust \citep{neufeld1991, gronke2016}, and enhance the observed emission of \lya\ photons \citep{finkelstein2008}. \citet{scarlata2009} found that such clumpy distribution better reproduces the observed \lya/H$\alpha$ and H$\alpha$/H$\beta$ ratio in 31 low-z \lya\ emitters. Hence, several factors can lead to different estimations of the dust impact, and robustly constraining the latter requires deeper observations. Furthermore, both the \fesclycobs\ and \fesclya\ depend on accurately determining the intrinsic emission, which may be mis-estimated by our assumed models. 

\begin{figure} [t]  
\includegraphics[width=\hsize]{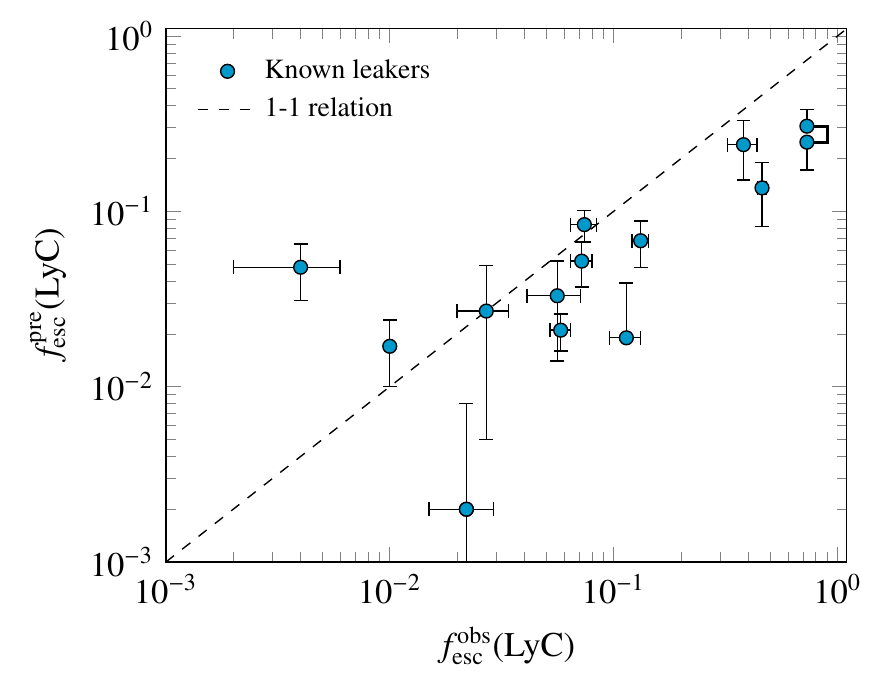}
 \caption{Predicted LyC escape fraction $f^{\rm pre}_{\rm esc}(\rm LyC)$, using the \ion{H}{i} covering fraction and dust extinction,  versus the observed \fesclyc\ plotted on a logarithmic scale. Dashed lines show the one-to-one relation. Two different predictions have been derived for J1243+4646 using $C_f$(\ion{H}{i})~=~0.18 (upper limit), and $C_f$(\ion{H}{i})~=~0, and are shown on the upper right part of the Figure. The error bars on $f^{\rm pre}_{\rm esc}(\rm LyC)$ are obtained by propagating the uncertainties from $C_f$(\ion{H}{i}) and \ebv. This Figure extends the work done in figure 1 of \citet{chisholm2018} by including the 6 new LyC detections from \citet{izotov2018a,izotov2018b}. }
 \label{fig:$C_f$pred}
\end{figure}

\subsection{Predicting the LyC escape fraction from absorption line measurements}
\label{sect:cfpred}

The outcomes of Figs.~\ref{fig:fesc$C_f$} and \ref{fig:fescebv} favor an ionization-bounded scenario, where dusty channels with low \nh\ act as the dominant escape roads for \lya\ and LyC photons, and support the conclusions drawn in \citet{gazagnes2018} and \citet{chisholm2018}. The latter work additionally proposed a novel approach to accurately recover the observed escape fractions based on the $C_f$(\ion{H}{i}) and dust extinction. This indirect prediction method will be particularly useful at higher redshift where the flux at < 912 \AA\ can not be directly observed. Indeed, the upcoming James Space Webb Telescope (JWST) will have a sufficient resolution to observe low-ionization interstellar (LIS) absorption lines and measure $C_f$(\ion{Si}{ii}), which can be then used to infer $C_f$(\ion{H}{i}) \citep{gazagnes2018}. The analysis done in \citet{chisholm2018} included 9 LCEs\footnote{Two of them, Tol0440-381 and Mrk54 \citep{leitherer2016}, are not included in this work because they do not have \lya\ observations} with \fesclycobs\ between 1 and 13 \%, and we test here its reliability using the six new leakers from \citet{izotov2018a,izotov2018b}, three of them with large LyC escape fractions ($\geq$ 38\%). 

This is shown in Fig.~\ref{fig:$C_f$pred}, where we plot the predicted escape fraction of LyC photons ($f^{\rm pre}_{\rm esc}(\rm LyC)$), computed as $(1 - C_f(\ion{H}{i}))\times 10^{-0.4A(\lambda = 912\ \AA)}$ \citep[see][]{chisholm2018}, versus the observed escape fraction of LyC photons ($f^{\rm obs}_{\rm esc}(\rm LyC)$), derived using the ratio of the observed flux at \textasciitilde 900 \AA\ over the modelled flux obtained from SED fitting for the 13 confirmed LCEs in our sample. The recovered LyC escape fractions accurately match at $\pm 3 \sigma$ with $f^{\rm obs}_{\rm esc}(\rm LyC)$ for 10 leakers, underestimates it  for the 2 largest leakers, and overestimates it in Tol1247-232. For the latter, the bias is likely explained by a possible contamination from geocoronal emission as discussed in \citet{chisholm2017leak,chisholm2018}. We mentioned in Sect.~\ref{sect:dataobs} that several studies using different measurement methods have reported LyC escape fractions varying from 0.004 to 0.042 for  Tol1247-232 \citep{leitherer2016, chisholm2017leak, puschnig2017}, which gives an interval consistent with the predicted 4.8 $\pm$ 1.7 \%. For the two largest leakers, several factors can explain the inconsistency between \fesclycobs\ and $f^{\rm pre}_{\rm esc}(\rm LyC)$. The measurement of \fesclycobs\ can suffer from poor constraints on the estimation of intrinsic emission of LyC photons \citep{chisholm2019}, and different \fesclycobs\ measurement approaches can lead to significant variations in some cases \citep[for example J1011+1947 has \fesclycobs\ = 11.4 $\pm$ 1.8\% when using an approach based on SED fitting, and \fesclycobs\ = 6.2 $\pm$ 0.7 \% when using the H$\beta$ flux density;][]{izotov2018b}. Similarly, reliable measurements of the dust attenuation are required to accurately estimate $f^{\rm pre}_{\rm esc}(\rm LyC)$. We showed in Sect.~\ref{sect:dust} that a 0.1 difference in \ebv\ is a factor 2 different in attenuation, and hence in $f^{\rm pre}_{\rm esc}(\rm LyC)$. Nevertheless, robustly constraining the stellar continuum and dust extinction is highly complicated in galaxies with low S/N spectra (see Sect.~\ref{sect:dust}).

Additionally, we showed in \citet{gazagnes2018, chisholm2018} that the dust geometry should not impact the derived $f^{\rm pre}_{\rm esc}(\rm LyC)$ because only the combination of $C_f$(\ion{H}{i}) and \ebv\ changes to match the observed data. We re-fitted J1154+2443 and J1256+4509 (\fesclycobs\ of 0.46 and 0.38, respectively) using a model where the dust lies only in the dense \ion{H}{i} clouds to check if this is still the case in large leakers. We found  $f^{\rm pre}_{\rm esc}(\rm LyC)$ = 0.074 $\pm$ 0.048 and 0.242 $\pm$ 0.095 for J1154+2443 and J1256+4509 respectively, which are consistent with the predictions using the screen model (0.136 $\pm$ 0.054 and 0.240 $\pm$ 0.089). Hence, assuming a different dust model does not improve the \fesclyc\ estimation of galaxies with high \fesclycobs.

Finally, indirectly predicting the ionizing escape in very strong leakers might be more complicated because their escape fraction could be a combination of the ionization and density bounded models, as suggested in \citet{kakiichi2019}. The authors showed that the leakage mechanisms in galaxies with a turbulent ISM dynamically evolves through time, such that the correlation between the leakage of LyC photons and the \ion{H}{i} covering fraction becomes weaker as more and more channels,  with lower \nhchan, form in the ISM. This is because these regions finally dominate the ISM such that the leakage is mostly regulated by the density-bounded scenario. While constraining the \nh\ requires large S/N \citep{gazagnes2018}, the shape of the \lya\ profile can provide insights of the dominant escape mechanisms. \citet{kakiichi2019} show that galaxies with a density bounded ISM have a lower red peak asymmetry ($A_f$) \citep[defined by Eq (33) in][]{kakiichi2019} than galaxies with an ionization bounded ISM. In particular, they find that the three largest leakers in our sample have a red peak asymmetry consistent with galaxies with a density bounded ISM. Additionally, we showed in Sect.~\ref{sect:ismvel} that two of these leakers had the lowest \ion{H}{i} velocity width of the maximal \ion{H}{i} absorption, while we do not observe \ion{H}{i} absorption lines in the largest LCE. This supports the fact that galaxies with large \fesclycobs\ must also have lower \ion{H}{i} column densities within the clouds that have the highest optical depth. In fact, Ramambason et al.\ (in preparation) find evidence from the optical emission lines pointing towards a density bounded ISM for these galaxies. This could explain why our approach based on $C_f$(\ion{H}{i}) underestimates the LyC escape fraction in these extreme leakers. 
In any case, Fig.~\ref{fig:$C_f$pred} shows that the UV absorption lines can be used to determine a reliable lower limit of the escape of LyC photons in LCEs candidates. 

Overall, our results suggest that the \ion{H}{i} porosity and dust attenuation play a crucial role in the escape of \lya\ and LyC photons and \fesclyc\ predictions based on $C_f$(\ion{H}{i}) and \ebv\ can provide a lower limit on the \fesclyc\ along the line of sight, despite the potentially complex leakage mechanisms in LCEs with high LyC escape fraction. This indirect approach might be valuable for high-redshift observations, in order to compare the range of \fesclyc\ of galaxies during the Reionization era and hence whether star-forming galaxies reionized the Universe. We discuss this point further in Sect.~\ref{sect:discpred}.

\subsection{The flux at the \lya\ profile minimum correlates with the LyC escape fraction}
\label{sect:fmin}

\begin{figure} [t]  
\includegraphics[scale =1]{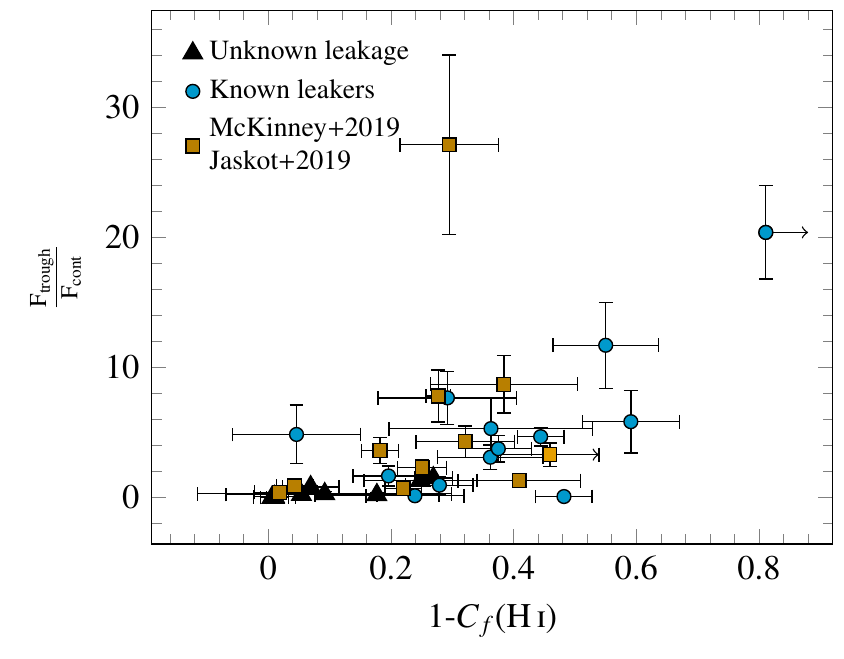}
 \caption{The normalized \lya\ flux at the minimum of the \lya\ profile $\frac{\rm F_{trough}}{\rm F_{\rm cont}}$ versus 1-$C_f$(\ion{H}{i}). The trend is scattered, but shows that galaxies with lower \ion{H}{i} covering fraction have higher residual flux at the \lya\ profile minimum. }
 \label{fig:fminfcont$C_f$}
\end{figure}

Finally, we find that the presence of \lya\ emission at the profile trough is a valuable marker of peculiar neutral gas properties. Indeed, it suggests that a fraction of \lya\ photons can escape unaffected where the optical depth of the \ion{H}{i} gas should be maximal. In particular, a \lya\ profile with a single peak at the systemic velocity should emerge from any patchy ISM with fully cleared holes (no \ion{H}{i} residuals), because all the \lya\ photons should find these holes after a negligible amount of scattering events \citep{ verhamme2015,dijkstra2016}. None of the galaxies in our sample with $C_f$(\ion{H}{i}) $<$ 1 exhibit such \lya\ spectral shape, including J1243+4646, which has no detected \ion{H}{i} absorption (all \lya\ profiles are shown in Appendix~\ref{app:lya}). However, all LCEs have net flux at the \lya\ trough, suggesting that some \lya\ photons escaped without being scattered by the \ion{H}{i} gas. This peculiarity has been observed in simulations in the presence of a very dense clumpy neutral gas distribution \citep{gronke2016, gronke2017}, or when the \ion{H}{i} column density is low enough such that \lya\ photons escape with fewer scattering events \citep{kakiichi2019}. In Sects.~\ref{sect:$C_f$peak} and \ref{sect:ismvel}, we argued that $C_f$(\ion{H}{i}) probes both the fraction of sightlines between the galaxy and the observer covered by low column density \ion{H}{i} gas, such that \lya\ and LyC photons passing through these escape roads have a lower probability to interact with the neutral gas. Hence, a larger $\frac{\rm F_{trough}}{\rm F_{\rm cont}}$ should correlate with the covering fraction of the low \nh\ channels  (1-$C_f$(\ion{H}{i})), and with a larger escape fraction of ionizing photons. 

Figure~\ref{fig:fminfcont$C_f$} explores the connection between $\frac{\rm F_{trough}}{\rm F_{\rm cont}}$ and 1-$C_f$(\ion{H}{i}) in our sample and in the GP sample studied by \citet{mckinney2019} and \citet{jaskot2019}. The trend is slightly scattered, but the correlation is significant at the 3 $\sigma$ (p-value of 0.0014). Hence, the presence of more flux at the \lya\ trough is related to a lower \ion{H}{i} covering fraction. In a highly porous ISM, higher amounts of \lya\ photons can escape at velocities where the \ion{H}{i} optical depth is large because they find low column density sightlines (see Section~\ref{sect:ismvel}). Note that, similar to \citet{orlitova2018} and \citet{jaskot2019}, we report a strong correlation (3.5 $\sigma$) between the \lya\ peak velocity separation and the strength of the flux at the \lya\ profile minimum (not shown). Hence, $v^{\rm sep}_{\rm \lya}$, $\frac{\rm F_{trough}}{\rm F_{\rm cont}}$ and $C_f$(\ion{H}{i}) likely provide similar insights about the presence of channels with low \nhchan\ in the ISM. 

\begin{figure} [t]  
\includegraphics[scale =1]{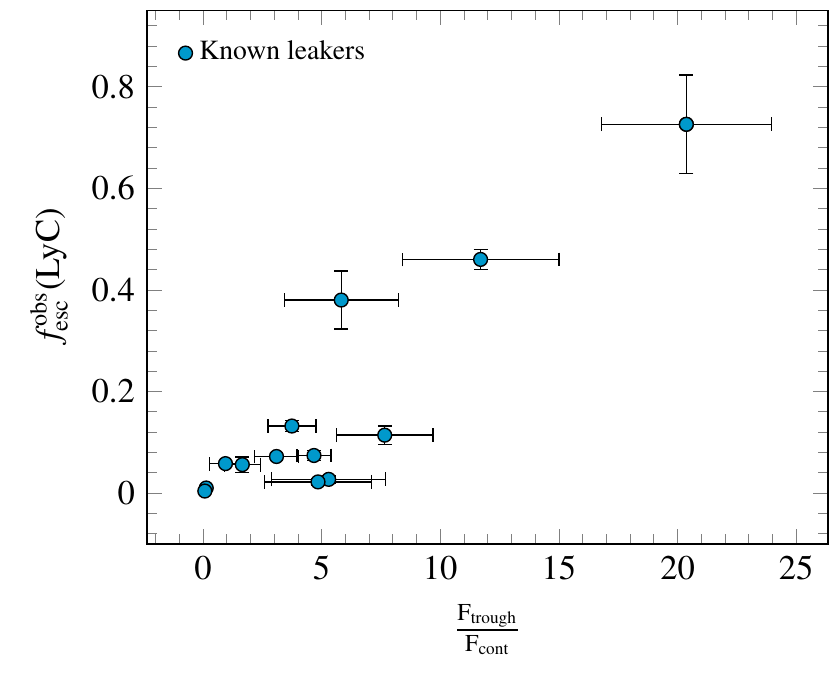}
 \caption{Observed escape fraction of LyC photons versus $\frac{\rm F_{trough}}{\rm F_{\rm cont}}$. Galaxies with higher residual flux at the \lya\ profile minimum have a higher escape fraction of ionizing photons. Note that the \lya\ trough depth is a resolution dependent measurement and
 requires a sufficient/high spectral resolution.}
 \label{fig:fminfcontfesc}
\end{figure}

Additionally, we investigate in Fig.~\ref{fig:fminfcontfesc} the connection between $\frac{\rm F_{trough}}{\rm F_{\rm cont}}$ and the escape of LyC photons. This Figure clearly highlights that all LCEs have net emission at the minimum profile, and larger \fesclycobs\ scales with larger $\frac{\rm F_{trough}}{\rm F_{\rm cont}}$ (at the 3.5 $\sigma$ level, p-value of 0.000022). Hence, similar to $v^{\rm sep}_{\rm \lya}$ \citep{verhamme2017, izotov2018b}, $\frac{\rm F_{trough}}{\rm F_{\rm cont}}$ is likely a reliable indicator of the presence of ISM properties that favor the leakage of ionizing photons. We derive the linear relation that relates \fesclyc\ to $\frac{\rm F_{trough}}{\rm F_{\rm cont}}$ in our sample (with $R \sim 15000$ observations from COS):

\begin{equation}
    \fesclyc = (0.032\ \pm 0.006) \times \frac{\rm F_{trough}}{\rm F_{\rm cont}} - 0.032\ \pm 0.053
    \label{eq:fminfesc}
\end{equation}
In \citet{gazagnes2018}, we posit that J0926+4427, J1429+0643, and GP1054+5238 are potential LCE candidates, and estimated \fesclyc\ of $\approx$ 1\% using their $C_f$(\ion{H}{i}) and dust extinction measurements. Eq.~\eqref{eq:fminfesc} suggests that follow-up observations could measure LyC escape fractions of $<$ 1\% in GP1054+5238 and J1429+0643 ($\frac{\rm F_{trough}}{\rm F_{\rm cont}}$ of 0.21 and 0.82 respectively), and 1.6\% in J0926+4427 ($\frac{\rm F_{trough}}{\rm F_{\rm cont}}$ = 1.51). The largest measured flux at minimum of \lya\ profile in the GPs sample analyzed by \citet{mckinney2019} and \citet{jaskot2019} is J1608+3528 with $\frac{\rm F_{trough}}{\rm F_{\rm cont}} = 27.1$, which corresponds to a very large \fesclyc\ of $\approx$ 83\% according to Eq.~\eqref{eq:fminfesc}. Additionally, J1608+3528 also has a narrow peak velocity separation  \citep[214 +- 30 km s$^{-1}$,][]{mckinney2019}, which strongly supports the presence of LyC leakage. 

Overall, Eq.~\eqref{eq:fminfesc} suggests that galaxies with $\frac{\rm F_{trough}}{\rm F_{\rm cont}}$ > 1 have \fesclyc\ > 1\%. Nevertheless, the reliability of Eq.~\eqref{eq:fminfesc} should be taken with caution because this linear relation is largely constrained by the large \fesclyc\ LCEs, while the $\frac{\rm F_{trough}}{\rm F_{\rm cont}}$ values of the low \fesclyc\ LCEs are largely scattered. More observations, with sufficiently high spectral resolution, are needed to confirm its reliability.

\section{Discussion}
\label{sect:disc}

\subsection{The ISM porosity enables the escape of \lya\ and LyC photons}
\label{sect:physpic}

\begin{figure*} [t]  
\includegraphics[width =\textwidth]{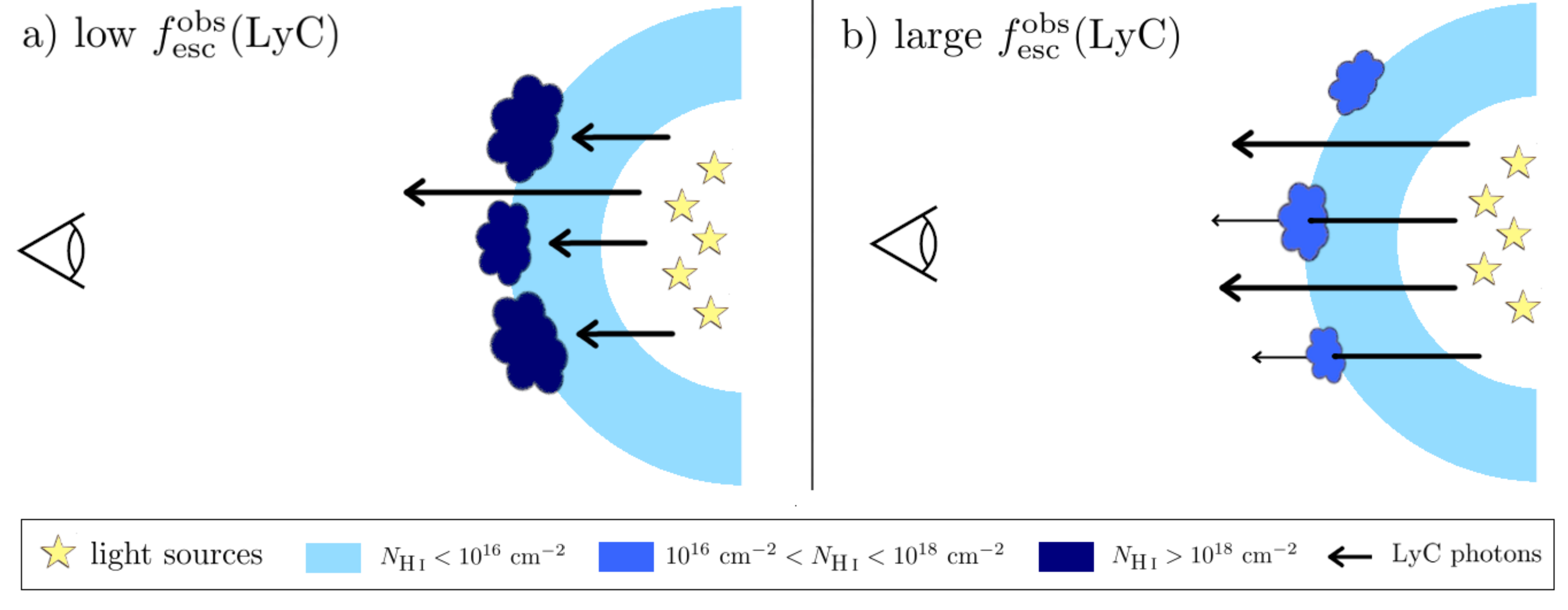}
 \caption{ An illustration of the physical picture discussed in Sect.~\ref{sect:physpic}. Left:  LyC photons escape through channels covered by \nhchan\ < $10^{16}$ cm$^{-2}$, while \ion{H}{i} clouds with \nhcloud\ larger than $10^{18}$ cm$^{-2}$ efficiently absorb the ionizing flux that passes through them. The ISM is ionization bounded such that the escape of ionizing photons is directly proportional to the fraction of low column density paths (1-$C_f$(\ion{H}{i})), and to the dust attenuation within or in front of these channels (note that we choose to exclude dust from this illustration because its spatial distribution is still unknown). This physical model likely describes the escape of ionizing photons from the low \fesclycobs\ LCEs of our sample. Right:  LyC photons escape because there is a higher fraction of low column density channels and because the densest \ion{H}{i} clouds do not efficiently absorb LyC photons (\nhcloud\  $< 10^{18}$ cm$^{-2}$). All the sightlines towards the observer are density-bounded such that the \fesclyc\ cannot be inferred only using the measured $C_f$(\ion{H}{i}). Our work suggests that this physical picture explains the escape of LyC photons in the large \fesclycobs\ LCEs (38, 46 and 73\%) of our sample.}
 \label{fig:fig:physpic}
\end{figure*}

The observations of partially covered but saturated Lyman series absorption lines suggestan inhomogeneous ISM with both high and low column density \ion{H}{i} gas. In this work, we studied the connection between the \lya\ and the neutral gas properties and provided insights to understand how this bi-modal distribution of \ion{H}{i} gas \citep[e.g see Fig 6 of][]{kakiichi2019} impacts the observed \lya\ emission and LyC escape. Similar to recent studies \citep{rivera2015, steidel2018, mckinney2019} that highlighted the connection between \lya\ properties and the neutral gas covering fraction, we found that the observed emission and escape of \lya\ photons of the 22 galaxies in our sample is closely connected to the fraction of sightlines between the galaxy and the observer covered by channels with low \nhchan\ (Sects.~\ref{sect:ew} and \ref{sect:$C_f$pred}). Additionally, Sect.~\ref{sect:$C_f$pred} showed the close connection between the escape of ionizing photons and 1-$C_f$(\ion{H}{i}), confirming that LCEs all have a non-unity \ion{H}{i} covering fraction \citep{gazagnes2018}. These results emphasize that there exists privileged escape roads for the \lya\ and LyC radiation. More interestingly, we found in Sect.~\ref{sect:$C_f$peak} that the \lya\ peak separation strongly scales with the \ion{H}{i} covering fraction. Because we expect $v^{\rm sep}_{\rm \lya}$ to scale with \nh\ \citep{verhamme2015, dijkstra2016}, this relation may suggest that a low $C_f$(\ion{H}{i}) not only indicates the presence of low column density channels, but also lower \nhchan. This is similarly highlighted in Sect.~\ref{sect:fmin}, where we showed that the \lya\ flux at the minimum of the \lya\ profile scales with $C_f$(\ion{H}{i}). Hence, more \lya\ photons escape unaffected by the \ion{H}{i} gas when a larger fraction of  low column density paths cut through the ISM. 

In \citet{gazagnes2018}, we constrained \nh\ in 6 low \fesclycobs\ leakers and found that the dense neutral regions have column densities large enough ($10^{18}$ cm$^{-2}$) to absorb all of the ionizing radiation. \citet{mckinney2019} reported similar insights from a sample of highly ionized Green Peas, and further emphasized that the \nh\ derived from \no\ is always a lower limit to other indirect approaches where \nh\ is derived from $N_{\ion{Si}{ii}}$, or from the fit of the observed \lya\ absorption profile. Hence, these results show that, in these LCEs,  ionizing photons escape through optically thin channels/holes in an ISM which is overall ionization bounded. Additionally, Sect.~\ref{sect:cfpred} highlighted that an indirect \fesclyc\ estimation method using $C_f$(\ion{H}{i}) and the dust extinction \citep{chisholm2018} accurately recovers the observed escape fractions in the low \fesclyc\ LCEs.

Nevertheless, the latter approach underestimates \fesclyc\ in LCEs with the largest \fesclycobs. This discrepancy could highlight that the escape mechanisms in these leakers can not be only explained by the presence of a low $C_f$(\ion{H}{i}). In Sect.~\ref{sect:ismvel}, we showed that the peak velocity separation is also tightly related to the velocity width of the maximal absorption of the saturated \ion{H}{i} absorption lines.  The connection between low $v^{\rm sep}_{\rm \lya}$ and low \nh\ \citep{verhamme2015, dijkstra2016} could suggest that $v^{\rm width}_{\rm \ion{H}{i}}$ indirectly probes the presence of lower \nhcloud\ within the densest neutral clouds in the ISM, such that a fraction of LyC photons escape  through these regions.  Interestingly, two of the three largest leakers (with \fesclycobs\ = 38 and 46 \%) have the lowest $v^{\rm width}_{\rm \ion{H}{i}}$ in our sample, and \ion{H}{i} is not detected in the largest LCE. Hence, for the largest leakers, a LyC prediction method using the covering fraction likely underestimates the total escape fraction of ionizing photons because all the lines of sight to the observer are density-bounded. 

These outcomes are supported by the recent radiation-hydrodynamics simulations of galaxies with a turbulent ISM by \citet{kakiichi2019}. The authors showed that the early evolution of the turbulent gas kinematics generates channels with low \nhchan, forming a ionization-bounded ISM where the leakage of LyC photons is directly driven by the fraction of paths covered by photo-ionized channels. Additionally, the evolution of the turbulent gas forms new channels and modifies the densities within these channels, increasing the fraction of low column density paths in the ISM. As a consequence, the escape mechanisms of LyC photons progressively evolve from a ionization bounded model (\fesclyc\ directly proportional to $C_f$(\ion{H}{i})) towards a density bounded scenario (low \nh\ in all the directions, weaker connection between \fesclyc\ and $C_f$(\ion{H}{i})). The authors show that markers of this trade-off can be found on the \lya\ profile, e.g using the red peak asymmetry ($A_f$). Their simulations support that the largest leakers from \citet{izotov2018a, izotov2018b} have $A_f$ consistent with a density bounded scenario, while leakers with \fesclycobs\ < 13\% have $A_f$ consistent with an ionization bounded model.

Overall these results provide new valuable insights to understand the dominant escape mechanisms of LyC photons. In the low LyC leakers, the ISM porosity enables the leakage of LyC and \lya\ photons, such that the ionising photons predominantly escape through low column density paths in the ISM, while the densest neutral clouds have \nhcloud\ too large and efficiently absorb the LyC photons passing through them (Panel (a) of Fig.~\ref{fig:fig:physpic}). However, the dynamical evolution of the ISM could shift the escape mechanisms towards a density bounded model, where all sightlines are covered by \nh\ < 10$^{17.9}$ cm$^{-2}$ (\fesclyc\ $\approx$ 1\%), enhancing the total escape fraction of LyC photons (Panel (b) of Fig.~\ref{fig:fig:physpic}). This outcome emphasizes the need to constrain the \nh\ in both the low and high density regions  (\nhchan\ and \nhcloud) to obtain accurate LyC predictions. Figure~\ref{fig:fig:physpic} illustrates the two different ISM configurations for small and large \fesclycobs\ LCEs discussed in this Section. Further low-redshift LyC observations should be able to assess this model.

\subsection{Indirectly constraining \fesclyc\ at high-z}
\label{sect:discpred}
The upcoming era of telescopes such as JWST and Extremely Large Telescopes (ELTs; GMT, TMT, ELT) will probe galaxies within the Epoch of Reionization and provide valuable clues to understand how the Universe was re-ionized. However, preparing for future observations requires finding pre-selection markers to efficiently survey large samples of LyC candidates and indirectly constrain their \fesclyc. To do this, low-redshift observations of the Lyman break in highly star-forming galaxies provides a unique laboratory of LCEs \citep{borthakur2014, izotov2016b,izotov2016a, izotov2018a, izotov2018b, wang2019} to explore the connection between the \lya, LyC and neutral gas properties. In this work, we highlighted that the neutral gas porosity is the plausible physical origin of the leakage of LyC and \lya\ photons, but that density-bounded mechanisms could dominate in the largest leakers. Because the ISM porosity affects the shape of the \lya\ emission, \lya\ properties can provide valuable insights about the leakage of LyC photons. The scaling relations between $v^{\rm sep}_{\rm \lya}$ , \nh\ and $C_f$(\ion{H}{i}) (Sect.~\ref{sect:$C_f$peak}) confirms that low peak velocity separation is likely a reliable probe of LyC leakage \citep{verhamme2017, izotov2018b, izotov2019}. However, deriving an accurate estimate of \fesclyc\ from it might be complicated because the \lya\ spectral shape is both sensitive to the covering fraction of low column density channels, and to the width of the maximal absorption of the saturated Lyman series (Sect.~\ref{sect:ismvel}). The positive correlation found between the red peak velocity and the \ion{H}{i} covering fraction might also support the use of  $v^{\rm red}_{\rm \lya}$ to probe the escape of LyC photons, but further theoretical simulations are required to understand its origin. 

Additionally, we showed that EW(\lya) or \fesclya\ scale with $C_f$(\ion{H}{i}) (Sect.~\ref{sect:ew} and \ref{sect:$C_f$pred}) and therefore could also indicate LyC leakage. Nevertheless, the EW(\lya) is also tightly connected to the velocity width of the maximal absorption of the Lyman series (Figs.~\ref{fig:vwidth} and \ref{fig:$C_f$vsew}), and is degenerate with different galaxy properties such as the star formation history or metallicity. Hence, galaxies with large EW(\lya) may have moderate \fesclya\ and \fesclyc\ \citep{jaskot2019}. On the other hand, \citet{izotov2019} derived an empirical relation between \fesclyc\ and \fesclya. Because both escape through channels with low \nh, it gives confidence that such connection could be useful to indirectly derive \fesclyc\ despite the complex task of accurately measuring \fesclya\ due to scattering mechanisms and biases introduced from aperture corrections. 

Finally, we showed in Sect.~\ref{sect:fmin} that the strength of the flux at the minimum of the \lya\ profile is, similarly to the peak velocity separation, a robust tracer of the leakage of ionizing photons. Hence, our work supports the use of the \lya\ properties to robustly pre-select LCEs candidate at low-redshift, and further constrain their \fesclyc\ using direct observations of the Lyman Break when possible.

However, probing the LyC escape with the \lya\ properties at high redshift is more challenging. Wavelengths bluer than \lya\ are highly contaminated by \lya\ forest, such that \lya\ diagnostics are less reliable \citep{schenker2014}. Quantities such as $v^{\rm sep}_{\rm \lya}$ and 
$\frac{\rm F_{trough}}{\rm F_{\rm cont}}$ may not be accurately measured and different approaches must be found. At these high redshifts, the study of the LIS metal lines likely provide better constrains on the presence of favorable ISM properties for the LyC escape. LIS lines are often used as proxies to extract the neutral gas properties \citep{shapley2003, jones2013, alexandroff2015, trainor2015, reddy2016stack, mckinney2019, jaskot2019}. Indeed, they are located at redder wavelengths and hence can be more easily observed at higher redshift. On the other hand, LIS metal lines also suffer some disadvantages, as they are potentially affected by scattering, fluorescent emission in-filling, and they often require high resolution observations to be fully resolved. Despite these caveats, they are currently one of the best LyC probes we have at high redshift. \citet{mckinney2019} and \citet{jaskot2019} recently showed that LIS lines provide similar insights as \lya\ properties.

In Sect.~\ref{sect:ismvel}, we highlighted the tight relation between the \lya\ peak velocity separation and the velocity width of the saturated Lyman series. \citet{chisholm2016} showed that different ionic transitions have similar velocity widths, thus, the \ion{Si}{ii} velocity width could provide an indirect estimation of $v^{\rm sep}_{\rm \lya}$ and indirectly probe the escape of ionizing photons. Similarly, in \citet{gazagnes2018} and \citet{chisholm2018}, we showed that $C_f$(\ion{Si}{ii}) can be used to empirically recover $C_f$(\ion{H}{i}) and then combined to the dust attenuation at 912 \AA\ to indirectly infer \fesclyc. Section~\ref{sect:cfpred} showed that the latter approach was less accurate for galaxies with large observed escape fractions, likely because of the presence of more complex leakage mechanisms, such that the total \fesclyc\ along the line of sight cannot be inferred only using the neutral gas coverage \citep{kakiichi2019}. Additionally, the strong trends found between the \ion{H}{i} covering fraction and the escape of \lya\ and LyC photons suggest that $C_f$(\ion{H}{i}) is a good proxy of the true geometric covering fraction of the optically thick neutral gas clouds along the line of sight for the LCEs in our sample. As mentioned above, this might not be always the case because kinematic effects might bias the $C_f$ measurements towards lower values. Despite these caveats, it remains a reliable approach to derive a lower limit to the true LyC escape fraction in compact star-forming galaxies where these effects might have less of an impact. Hence, it should provide valuable constraints on the expected impact of these galaxies during the epoch of reionization.

Upcoming observations will allow us to confirm the robustness of the different pre-selection and indirect estimation approaches discussed in this section.


\section{Summary}
\label{sect:conc}

We have examined the neutral gas properties of a sample of 22 star-forming galaxies that  have \lya\ and Lyman series observations. We fitted the stellar continua, dust attenuations, \ion{H}{i} covering fractions, \ion{H}{i} velocity shifts and measured the \lya\ properties, the \ion{H}{i} velocity widths and depths of the Lyman series using a Monte-Carlo approach. We investigated the relations between the neutral gas properties in galaxies with the escape and emission of \lya\ photons and the observed LyC escape fractions. We summarize our results as follow:
\begin{itemize}
    \item \lya\ photons are less scattered in galaxies with with a larger fraction of low density sightlines towards the observer (Sect.~\ref{sect:$C_f$peak}). Additionally, the scaling relation found between the \lya\ peak velocities and the \ion{H}{i} covering fractions ($C_f$(\ion{H}{i})) suggests that galaxies with lower neutral gas coverage also have lower \ion{H}{i} column densities in the low-density channels (Fig.~\ref{fig:velsep$C_f$}). Thus, low-z LCEs have small \lya\ peak separations because photons find low column density escape roads out of the galaxy. 
    \item The velocity width of the \ion{H}{i} gas, measured at the maximal depth of the \ion{H}{i} absorption lines, has a significant impact on the \lya\ emission, escape and velocity shift (Sect.~\ref{sect:ismvel}). Galaxies with narrower \ion{H}{i} absorption lines have higher EW(\lya) and larger \fesclya\ (Fig~\ref{fig:$C_f$vsew}). Additionally, $v^{\rm sep}_{\rm Ly\alpha}$ linearly scales with $v^{\rm width}_{\rm \ion{H}{i}}$, and is lower in galaxies with lower \ion{H}{i} covering fraction (Fig.~\ref{fig:vwidth}). This highlights that the shape of the \lya\ profile relates both to the presence of the low column density channels (low $C_f$(\ion{H}{i})) and to the properties of the denser neutral clouds imprinted from the saturated \ion{H}{i} absorption lines ($v^{\rm width}_{\rm \ion{H}{i}}$). The presence of low  $v^{\rm width}_{\rm \ion{H}{i}}$ in the galaxies with the highest \fesclycobs\ may suggest that they have overall less \ion{H}{i} gas.
    \item Galaxies with lower covering fractions and/or dust attenuations have larger \lya\ and LyC escape fractions (Figs.~\ref{fig:fesc$C_f$} and \ref{fig:fescebv}). The presence of a porous ISM ($C_f$(\ion{H}{i}) < 1) is a necessary condition to observe a significant leakage of ionizing radiation (Fig.~\ref{fig:$C_f$pred}). At high-redshift, the $C_f$(\ion{Si}{ii}) can be used to estimate $C_f$(\ion{H}{i}) and provide valuable insights about the presence of LyC leakage. Additionally, using 6 recently observed LCEs, we found that the indirect approach to measure the line of sight \fesclyc\ proposed in \citet{chisholm2018} accurately recovers their observed \fesclyc\ within $\pm\ 3 \sigma$ for 9 leakers with \fesclycobs\ < 13\%, and yields a lower limit on the total observed escape fractions in the LCEs with the largest LyC escape fractions. We suggest that recovering \fesclycobs\ in these extreme leakers is more complex because their LyC leakage combines ionization and density bounded mechanisms (Fig.~\ref{fig:fig:physpic}). Hence, an indirect prediction methods based only on $C_f$(\ion{H}{i})  sets a lower limit to the true LyC escape fractions.
    \item Galaxies with lower $C_f$(\ion{H}{i}) have a larger flux at the minimum of the \lya\ profile (Fig.~\ref{fig:fminfcont$C_f$}), and the \lya\ profile minima strongly scales with the escape fraction of LyC photons  (Fig.~\ref{fig:fminfcontfesc}). Therefore, the presence of low-density paths in the ISM is presumably the common origin of low $v^{\rm sep}_{\rm Ly\alpha}$ and high $\frac{\rm F_{trough}}{\rm F_{\rm cont}}$ in the LCEs in our sample. 
\end{itemize}

Our work emphasizes that the ISM porosity plays a major role in understanding the origin of the \lya\ properties and the escape of LyC photons in LCEs, but additional escape mechanisms might be needed to explain the LyC escape fractions observed in the leakers with the largest LyC escape fractions (Fig.~\ref{fig:fig:physpic}). Additionally, it shows that the \lya\ spectral shape, in particular through low $v^{\rm sep}_{\rm Ly\alpha}$ and high $\frac{\rm F_{trough}}{\rm F_{\rm cont}}$, probes ISM properties that favor the leakage of \lya\ and LyC photons. At high redshift, the use of LIS lines is likely more appropriate to detect and constrain the ionizing escape fraction in Reionization era galaxies. Indeed, \ion{Si}{ii} can be used to infer $C_f$(\ion{H}{i}) at high redshift \citep{gazagnes2018}, and hence combined with \ebv\ to derive a lower limit of \fesclyc\ using the procedure from \citet{chisholm2018}. Recent additional approaches could further constrain the LyC escape fractions in currently confirmed LCEs \citep[e.g using \ion{Mg}{ii}; Chisholm et al, in prep, ][]{henry2018}. Therefore LIS diagnostics seems promising for future prospects at higher redshift, to detect LyC leaking candidates at high redshift (z > 6) with the upcoming era of the telescopes probing the Epoch of Reionization.

\begin{acknowledgements}
The authors would like to thank the referee for the comments that helped improve the quality of this paper. The authors also thank Jed McKinney and Emil Rivera-Thorsen for the useful comments provided on a draft version of this paper. The position of SG was funded from a grant by the Center for Data Science and Systems Complexity (DSSC), University of Groningen. YI acknowledges support from the National Academy of Sciences of Ukraine by its priority project  No. 0120U100935 "Fundamental properties of the matter in the relativistic collisions of nuclei and in the early Universe".
Support for this work was provided by NASA through the NASA Hubble Fellowship grant \#51432 awarded by the Space Telescope Science Institute, which is operated by the Association of Universities for Research in Astronomy, Inc., for NASA, under contract NAS5-26555.
\end{acknowledgements}

\bibliographystyle{aa}
\bibliography{Bibliographie.bib}

\begin{appendix}

\section{Fits}
\label{sect:fits}
\begin{figure*}
  \centering
  \includegraphics[width=\hsize]{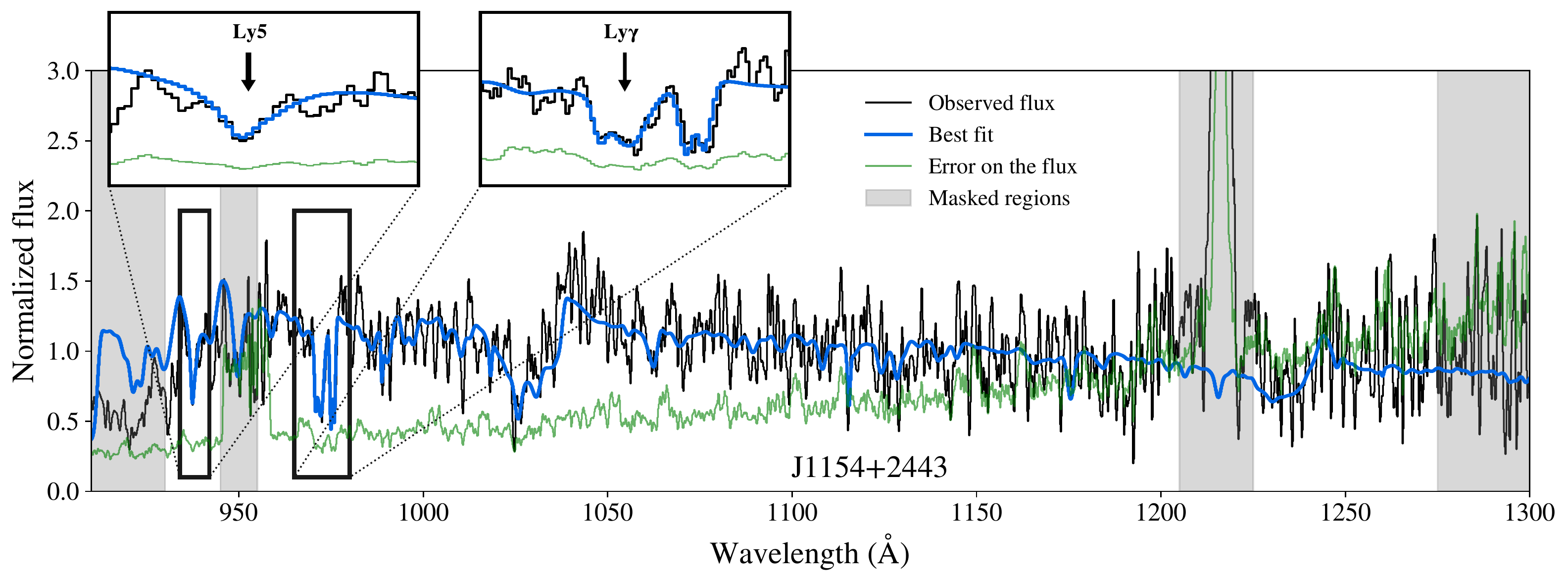}
     \caption{Fit (in blue solid line) of the COS G140L spectrum of the galaxy J1154+2443  \citep{izotov2018a}. The black line is the observed flux included in the routine either to fit the stellar continuum or the ISM absorption lines. Gray regions are masked out during the fit. The flux error array appears in green. The top panels present a zoom on the fitted Lyman series lines. }
         \label{fig:J1154}
  \end{figure*}

\begin{figure*}
  \centering
  \includegraphics[width=\hsize]{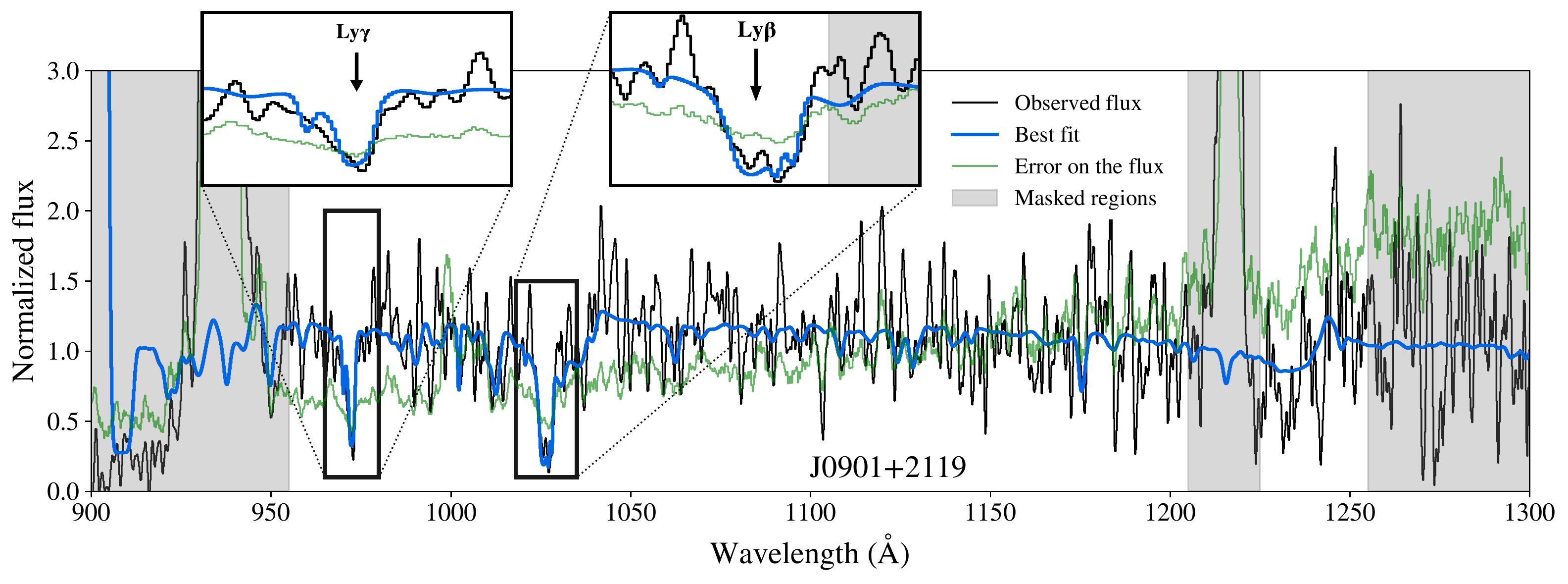}
     \caption{Same as Fig.~\ref{fig:J1154} but for the COS G140L spectrum of J0901+2119 \citep{izotov2018b}. }

         \label{fig:J0901}
  \end{figure*}

\begin{figure*}
  \centering
  \includegraphics[width=\hsize]{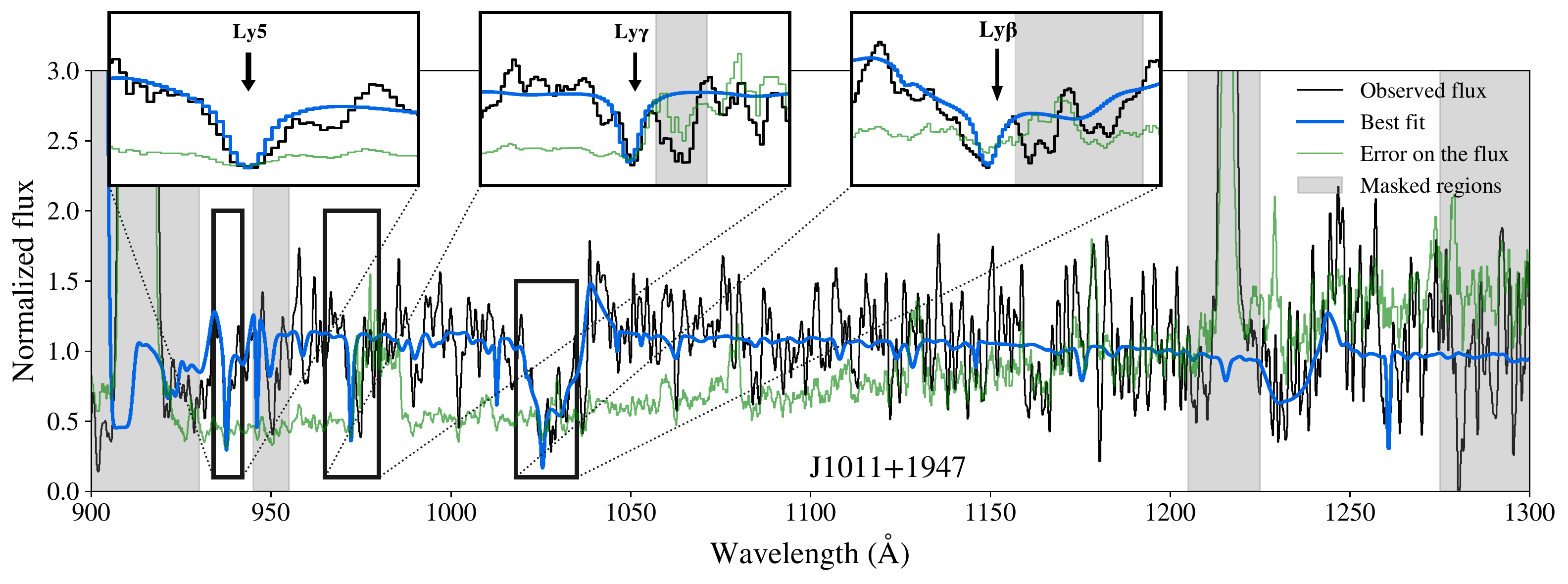}
   \caption{Same as Fig.~\ref{fig:J1154} but for the COS G140L spectrum of J1011+1947 \citep{izotov2018b}. }
         \label{fig:J1011}
  \end{figure*}

\begin{figure*}
  \centering
  \includegraphics[width=\hsize]{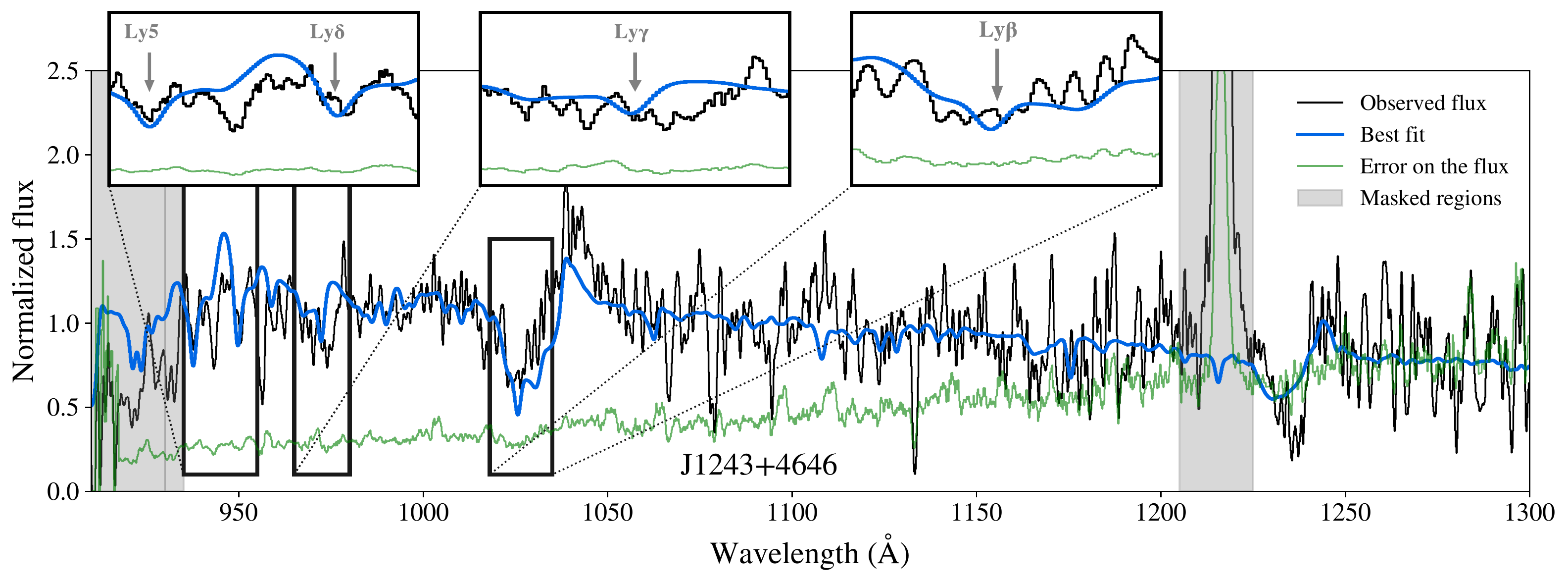}
   \caption{Same as Fig.~\ref{fig:J1154} but for the COS G140L spectrum of J1243+4646 \citep{izotov2018b}. The best fit only includes the attenuated stellar continuum because the \ion{H}{i} absorption lines were not found.}
         \label{fig:J1243}
  \end{figure*}

\begin{figure*}
  \centering
  \includegraphics[width=\hsize]{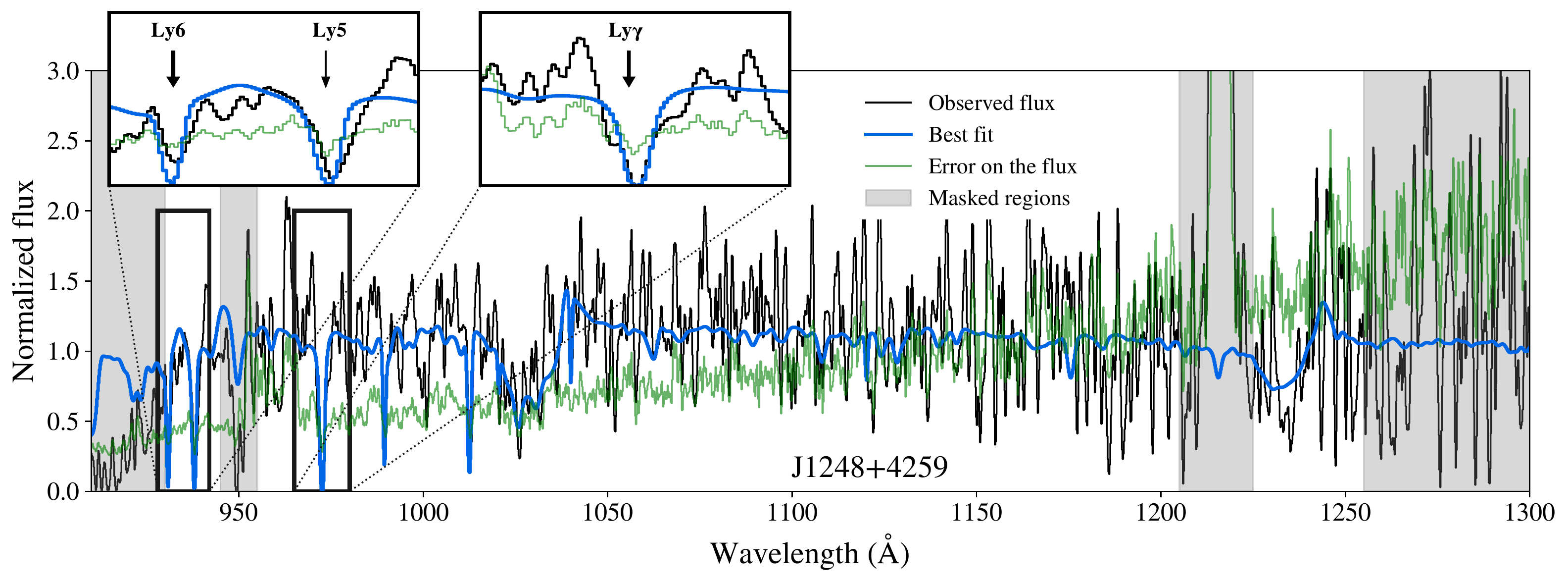}
  \caption{Same as Fig.~\ref{fig:J1154} but for the COS G140L spectrum of J1248+4259 \citep{izotov2018b}. }
         \label{fig:J1248}
  \end{figure*}

 In Figs~\ref{fig:J1154}--\ref{fig:J1248} we present the fits for the galaxies J1154+2443, J0901+2119, J1011+1947, J1243+4646 and J1248+4259 from \citet{izotov2018a, izotov2018b} that have been included in this sample. The fit of J1256+4509 is shown in Fig.~\ref{fig:fiitexample}, while the fits for the 16 other galaxies can be found in \citet{gazagnes2018}. Each figure shows the full wavelength coverage; the observed flux is in black, the green line shows the error on the observed flux and the resulting fit is in blue. The gray shaded areas show the main regions that were masked during the fit because of geo-coronal emission, low S/N, ISM or Milky Way absorption lines not included in the fit, or \lya\ emission. In the top panels, we zoom in on individual Lyman series lines and denote the fitted Lyman series lines. Note that for display purposes, the masks on the ISM or Milky Way absorption lines are only shown when they appear in the top panels.

 \section{Tables}
 
\subsection{Lyman series covering fraction measurements}

Table~\ref{table:$C_f$} lists the different measurements of the \ion{H}{i} covering fractions, from the fits and from the residual flux (see details in Sect.~\ref{sect:neutprop}). 

 \begin{table*}[!htbp]
    \caption{Measurement of \ion{H}{i} covering fraction derived from the fits and from the residual flux of the individual Lyman series. See Sect.~\ref{sect:neutprop}. }
     \label{table:$C_f$}
    \centering   
    \begin{tabular}{c c c c c | c  c || c}
    \hline \hline
    Galaxy name&   Ly$\beta$ & Ly$\gamma$ & Ly$\delta$ & Ly5  & Depth & Fits & Final \\ 
    (1) & (2) & (3) & (4) & (5) & (6) & (7) & (8) \\\hline 
    J1243+4646  &  - & 0.200 $\pm$ 0.121 & 0.020 $\pm$ 0.120 & 0.140 $\pm$ 0.120 & 0.120 $\pm$ 0.069 & - & <0.189 \\   
    J1154+2443  & -  & 0.471 $\pm$ 0.128 & - & 0.474 $\pm$ 0.184 & 0.472 $\pm$ 0.105 & 0.404 $\pm$ 0.152 & 0.450 $\pm$ 0.086 \\ 
    J1256+4509  &  0.369 $\pm$ 0.290 & - & 0.530 $\pm$ 0.165 & 0.329 $\pm$ 0.143 & 0.410 $\pm$ 0.101 & 0.407 $\pm$ 0.125 & 0.409 $\pm$ 0.079\\ 
    J1152+3400  & 0.619 $\pm$ 0.088 & - & - & - & 0.619 $\pm$ 0.088 & 0.629 $\pm$ 0.069  & 0.625 $\pm$ 0.054 \\ 
    J1442-0209  &  0.589 $\pm$ 0.049 & 0.471 $\pm$ 0.068 & - &  - & 0.549 $\pm$ 0.040 & 0.621 $\pm$ 0.120 & 0.556 $\pm$ 0.038 \\ 
    J0925+1409 &  0.635 $\pm$ 0.094 & - & - & - & 0.635 $\pm$ 0.094 & 0.652 $\pm$ 0.218 & 0.638 $\pm$ 0.086 \\ 
    J1011+1947 &  0.693 $\pm$ 0.376 & 0.616 $\pm$ 0.244 & - & 0.820 $\pm$ 0.210 & 0.727 $\pm$ 0.147 & 0.644 $\pm$ 0.265 & 0.708 $\pm$ 0.113 \\ 
    J1503+3644  &  0.847 $\pm$ 0.157 & 0.723 $\pm$ 0.129 & 0.741 $\pm$ 0.112 & 0.744 $\pm$ 0.113 & 0.754 $\pm$ 0.062 &  0.611 $\pm$ 0.110 & 0.721 $\pm$ 0.055\\
    J1333+6246  & 0.773 $\pm$ 0.152 & 0.870 $\pm$ 0.103 & 0.804 $\pm$ 0.106 & - & 0.826 $\pm$ 0.066 & 0.731 $\pm$ 0.122 & 0.804 $\pm$ 0.058  \\
    J0901+2119  & 0.590 $\pm$ 0.461 & 0.742 $\pm$ 0.247 & - & - & 0.708 $\pm$ 0.218 & 0.539 $\pm$ 0.257 & 0.637 $\pm$ 0.166 \\
    J1248+4259 & - & 1.000 $\pm$ 0.170 & - &  0.903 $\pm$ 0.157 & 0.948 $\pm$ 0.115 & 0.980 $\pm$ 0.241 & 0.954 $\pm$ 0.104  \\
    J0921+4509  &  0.769 $\pm$ 0.116 & - & - & - & 0.769 $\pm$ 0.116 & 0.754 $\pm$ 0.111 & 0.761 $\pm$ 0.080 \\ 
    Tol1247-232 & 0.543 $\pm$ 0.135 & - & 0.644 $\pm$ 0.115  &  0.414 $\pm$ 0.175  & 0.564 $\pm$ 0.078 & 0.486 $\pm$ 0.061 & 0.518 $\pm$ 0.046 \\
    J0926+4427  &  0.817 $\pm$ 0.057 & 0.807 $\pm$ 0.087 &  - & -  & 0.814 $\pm$ 0.048 & 0.673 $\pm$ 0.040 & 0.731 $\pm$ 0.031\\ 
    J1429+0643  &  0.955 $\pm$ 0.061 & - & - & - & 0.955 $\pm$ 0.061 & 0.898 $\pm$ 0.071 & 0.931 $\pm$ 0.046 \\ 
    GP0303-0759  & - & - & - & - & - & 0.908 $\pm$ 0.207 & 0.908 $\pm$ 0.207\\ 
    GP1244+0216  &  0.985 $\pm$ 0.211 & 0.894 $\pm$ 0.204 & 1.000 $\pm$ 0.292 & -  & 0.950  $\pm$ 0.131 & 0.909 $\pm$ 0.357 & 0.946 $\pm$ 0.123 \\ 
    GP1054+5238  & 0.936 $\pm$ 0.318 & - & 0.891 $\pm$ 0.203 & 0.798 $\pm$ 0.420  & 0.889  $\pm$ 0.158 & 0.778 $\pm$ 0.131 & 0.823 $\pm$ 0.101 \\ 
    GP0911+1831  &  0.718 $\pm$ 0.361 & 0.781 $\pm$ 0.189 & 0.825 $\pm$ 0.198 & 0.635 $\pm$ 0.282 & 0.765  $\pm$ 0.116 & 0.731 $\pm$ 0.150 & 0.752 $\pm$ 0.092\\
    \sone  &  1.000 $\pm$ 0.010 & 1.000 $\pm$ 0.030 & 0.981 $\pm$ 0.057 & 0.940 $\pm$ 0.060 & 0.998  $\pm$ 0.009 & 0.932 $\pm$ 0.038 & 0.994 $\pm$ 0.009\\
    \stwo  &  0.858 $\pm$ 0.141 & 1.000 $\pm$ 0.045 & 1.000 $\pm$ 0.087 & 1.000 $\pm$ 0.201 & 0.990  $\pm$ 0.038 & 0.993 $\pm$ 0.078 & 0.990 $\pm$ 0.034\\
    Cosmic Eye  &  1.000 $\pm$ 0.043 & 1.000 $\pm$ 0.063 & 1.000 $\pm$ 0.033 & 0.899 $\pm$ 0.166 & 0.998  $\pm$ 0.024 & 0.920 $\pm$ 0.070 & 0.990 $\pm$ 0.023\\
 \hline     
    \end{tabular}
    \tablefoot{(1) Galaxy name; (2)-(5) Residual flux of the individual Lyman series (from Ly$\beta$ to Ly5); (6) \ion{H}{i} covering fraction from the residual flux of the individual Lyman series derived as the weighted average of the columns (2)-(5). Dashes indicate that these transitions were not observed/included owing to Milky Way absorption, geocoronal emission, or low S/N. (7) \ion{H}{i} covering fraction obtained from the fitting method. (8) \ion{H}{i} covering fraction of the galaxy, derived as the weighted average of the columns (6) and (7).}
    \end{table*}
 
 \subsection{Lyman series velocity width measurements}

 Table~\ref{table:vel} lists the measurements of the \ion{H}{i} velocity width for the Lyman series in each galaxy (see details in Sect.~\ref{sect:neutprop}).
 
 \begin{table*}[!htbp]
    \caption{Measurement of \ion{H}{i} velocity width of the individual Lyman series. See Sect.~\ref{sect:neutprop}. }
     \label{table:vel}
    \centering   
    \begin{tabular}{c c c c c | c }
    \hline \hline
    Galaxy name&   $v^{\rm width}_{\rm Ly\beta}$ & $v^{\rm width}_{\rm Ly\gamma}$ & $v^{\rm width}_{\rm Ly\delta}$ & $v^{\rm width}_{\rm Ly5}$  & $v^{\rm width}_{\ion{H}{i}}$ \\ 
    & [km s$^{-1}$] & [km s$^{-1}$] & [km s$^{-1}$] & [km s$^{-1}$] & [km s$^{-1}$] \\
    (1) & (2) & (3) & (4) & (5) & (6)  \\\hline 
    J1243+4646  &  - & - & - & - & - \\   
    J1154+2443  & -  & 170 $\pm$ 100 & - & 170 $\pm$ 170 & 170 $\pm$ 86  \\ 
    J1256+4509  & 270 $\pm$ 80 & - & 220 $\pm$ 90  & 280 $\pm$ 90 & 250 $\pm$ 50 \\ 
    J1152+3400  & 420 $\pm$ 60 & - & - & - & 420 $\pm$ 60 \\ 
    J1442-0209  & 490 $\pm$ 80 & 280 $\pm$ 70 & - &  - & 371 $\pm$ 53  \\ 
    J0925+1409  & 320 $\pm$ 60 & - & - & - & 320 $\pm$ 60 \\ 
    J1011+1947  & 420 $\pm$ 80 & 170 $\pm$ 90 & - & 170 $\pm$ 130 & 285 $\pm$ 54 \\ 
    J1503+3644  & 410 $\pm$ 90 & 380 $\pm$ 100 & 280 $\pm$ 100 & 340 $\pm$ 100 & 356 $\pm$ 49 \\
    J1333+6246  & 320 $\pm$ 80 & 280 $\pm$ 100 & 230 $\pm$ 90 & - & 280 $\pm$ 51\\
    J0901+2119  & 320 $\pm$ 110 & 280 $\pm$ 110 & - & - & 300 $\pm$ 78 \\
    J1248+4259  & - & 260 $\pm$ 90 & - & 220 $\pm$ 60 & 232 $\pm$ 50  \\
    J0921+4509  & 440 $\pm$ 20 & - & - & - & 440 $\pm$ 20 \\ 
    Tol1247-232 & 530 $\pm$ 140 & - & 360 $\pm$ 180  &  430 $\pm$ 150  & 453 $\pm$ 89\\
    J0926+4427  & 400 $\pm$ 60 & 360 $\pm$ 70 &  - & -  & 383 $\pm$ 46 \\ 
    J1429+0643  & 420 $\pm$ 50 & - & - & - & 420 $\pm$ 50  \\ 
    GP0303-0759  & 380 $\pm$ 50 & - & - & - & 380 $\pm$ 50\\ 
    GP1244+0216  &  380 $\pm$ 50 & 370 $\pm$ 205 & - & -  & 379 $\pm$ 49  \\ 
    GP1054+5238  & 580 $\pm$ 50 & - & 400 $\pm$ 50 & 460 $\pm$ 50  & 480  $\pm$ 29 \\ 
    GP0911+1831  & 410 $\pm$ 60 & 410 $\pm$ 50 & 370 $\pm$ 20 & 350 $\pm$ 40 & 374  $\pm$ 16 \\
    \sone  &  560 $\pm$ 40 & 600 $\pm$ 80 &  - & 510 $\pm$ 50 & 548  $\pm$ 29 \\
    \stwo  &  610 $\pm$ 60 & 520 $\pm$ 70 & - & 410 $\pm$ 40 & 480  $\pm$ 30 \\
    Cosmic Eye  &   700 $\pm$ 380 &  740 $\pm$ 210 & 320 $\pm$ 140 & 450 $\pm$ 220 & 467  $\pm$ 99 \\
 \hline     
    \end{tabular}
    \tablefoot{(1) Galaxy name; (2)-(5) Velocity width of the individual Lyman series (from Ly$\beta$ to Ly5); (6) \ion{H}{i} velocity width from the residual flux of the individual Lyman series derived as the weighted average of the columns (2)-(5). Dashes indicate that these transitions were not observed/included owing to Milky Way absorption, geocoronal emission, low S/N, or irregular line shape. }
    \end{table*}
 
\subsection{Significance levels of the trends reported in this work}
\label{app:sigma}
 
In this Section, we investigate the dependence of the significance levels of the trends reported in Sect.~\ref{sect:results} on a few specific observations, or on different assumptions of the intrinsic \lya\ H$\alpha$ flux ratio adopted to derive \fesclya. Table~\ref{table:sigma} lists the $\sigma$ values quoted in the paper (Col (2)) and the new significance levels when one observation is excluded, quoted as a jackknife test interval (Col (3)), with the lowest significance and largest significance level obtained when one observation is removed from the sample. Col (3) shows that the lower bound is always larger than 2.5$\sigma$, except for three trends: $v^{\rm red, rel}_{\rm \lya}$ -- $C_f$(\ion{H}{i}),  \fesclycobs\ -- $C_f$(\ion{H}{i}) and $\frac{\rm F_{trough}}{\rm F_{\rm cont}}$ -- $C_f$(\ion{H}{i}). In the latter, the lowest $\sigma$ value is only moderately significant (2$\sigma$) and is obtained when the largest \fesclyc\ galaxy, J1243+4646, is excluded from the sample. We note that two of these trends suffer some scatter (illustrated in the Figs.~\ref{fig:vel$C_f$} and \ref{fig:fminfcont$C_f$}) while the trend between \fesclycobs\ and  $C_f$(\ion{H}{i}) only has a limited number of irregularly-spaced points (Fig.~\ref{fig:fesc$C_f$}). Future observations should confirm or rule out the existence of these trends. 

In Col (4), we compute the new significance levels when  $C_f$(\ion{H}{i}) is fixed to 0 for galaxies where we only measure an upper limit of the  $C_f$(\ion{H}{i}). Overall, the significances vary by $\pm 0.5 \sigma$, except for the EW(\lya) and \fesclya\ correlations which reach 1.5 and 5.0 $\sigma$, respectively. This is because the galaxies with an upper limit on  $C_f$(\ion{H}{i})  \citep[1 in our sample, 3 in the sample included in][]{mckinney2019} have similar \fesclya, but varying EW(\lya). Hence, fixing their  $C_f$(\ion{H}{i}) to the same value strengthens the correlation with the \fesclya, but weakens the one with EW(\lya). In Col (5), we excluded the three z\textasciitilde3 \megasaura\ galaxies to investigate the impact they have on the significance of the trends. All the trends, except for EW(\lya) -- $C_f$(\ion{H}{i}), remain significant at least at the 2.5$\sigma$ significance level. 

Finally, we tested the impact of the Case-A/B assumptions on the trends that involve \fesclya\ in Cols (6), (7) and (8). In the Case A recombination case (optically thin ISM), the intrinsic ratio of the \lya\ and H$\alpha$ flux can be 1.5 times higher \citep[the effective recombination coefficient of the Balmer emission lines is $\approx$ 1.5 times higher than in the Case B assumption,]{osterbrook1989} than in Case-B recombination (optically thick ISM, low-density case). For galaxies emitting ionizing photons, Case-B assumption (adopted in this work) is not likely  valid, hence, an optimal approach would consist of scaling down all \fesclya\ values from LCEs by 1.5. Nevertheless, this is not trivial because some galaxies in our sample, or in the sample from \citet{mckinney2019} and \citet{jaskot2019}, do not have observations at $\lambda < 912 \AA$ (unknown LyC leakage). Consequently, we choose to use a cut in the peak separation velocity of the \lya\ profile to select galaxies with a Case-A ISM. The choice of $v^{\rm sep}_{\rm Ly\alpha}$ is motivated by the strong observational correlation found between \fesclycobs\ and $v^{\rm sep}_{\rm Ly\alpha}$  \citep{verhamme2017, izotov2018b}. Because low \fesclyc\ leakers might still be consistent with a Case-B recombination assumption, we select three cuts in  $v^{\rm sep}_{\rm Ly\alpha}$ of 400, 300 and 200 km s$^{-1}$, to successively select different sub-samples with larger \fesclyc, and scaled-down their \fesclya\ by a factor 1.5. We found that the \fesclya\ trends are always significant at $\geq 2.5 \sigma$. Therefore, we conclude that variations in the intrinsic \lya-H$\alpha$ flux ratio should not affect the conclusions drawn in this work.

 \begin{table*}[!htbp]
    \caption{The impact of different assumptions on the reported significance level of observed trends.}
     \label{table:sigma}
    \centering   
    \begin{tabular}{c c c c c c c c c}
    \hline \hline
    Trend&   \multicolumn{7}{c}{Significance level ($\sigma$)}\\ 
    & All & Jackknife interval  & $C_f^{\rm limit}$(\ion{H}{i}) to 0 & Low z & $v^{\rm sep}_{\rm Ly\alpha}\ <$ 400 &  $v^{\rm sep}_{\rm Ly\alpha}\ <$ 300 & $v^{\rm sep}_{\rm Ly\alpha}\ <$ 200 \\
    (1) & (2) & (3) & (4) & (5) & (6) & (7) & (8)  \\\hline 
    $v^{\rm blue, rel}_{\rm \lya}$ -- $C_f$(\ion{H}{i})  &  3.0   & [2.5, 3.5]    & 2.5   & - & - & - & -\\   
    
    $v^{\rm red, rel}_{\rm \lya}$ -- $C_f$(\ion{H}{i})  & 3.0   & [2.0, 3.5]    & 3.0   & 2.5  & - & - & -\\ 
    $v^{\rm sep}_{\rm Ly\alpha}$ -- $C_f$(\ion{H}{i})   & 3.0   & [2.5, 3.5]    & 3.5   & - & - & - & -\\ 
    EW(\lya) -- $C_f$(\ion{H}{i})   & 2.5   & [2.5, 4.0]    & 1.5   & 2.0  & - & - & -\\ 
    \fesclya\ -- $C_f$(\ion{H}{i})  & 4.0   & [3.5, 5.5]    & 5.0   & 3.5  & 4.0   &  3.5   & 3.5   \\ 
    \fesclycobs\ -- $C_f$(\ion{H}{i}) & 3.0   & [2.0, 3.0]    & 3.5   & - & - & - & -\\ 
    $\frac{\rm F_{trough}}{\rm F_{\rm cont}}$ -- $C_f$(\ion{H}{i})   & 3.0   & [2.0, 4.5]    & 2.5   & - & - & - & -\\ 
    EW(\lya) -- $v^{\rm width}_{\ion{H}{i}}$   & 4.0   & [3.5, 4.0]    & -  & 3.0  & - & - & -\\ 
    \fesclya\ -- $v^{\rm width}_{\ion{H}{i}}$   & 3.0   & [2.5, 3.5]    & - & 2.5  & 2.5   & 3.0   & 2.5  \\ 

    \fesclya\ -- 10$^{-0.4A(1216\AA)}$   & 3.0   & [2.5, 3.5]    & -  & 2.5  &  2.5  &  2.5  &  2.5 \\ 
    \fesclycobs\ -- 10$^{-0.4A(912\AA)}$  & 3.0   & [2.5, 3.0]    & -  & - & - & - & -\\ 
    \fesclycobs\ -- $\frac{\rm F_{trough}}{\rm F_{\rm cont}}$    & 3.5   & [2.5, 4.5]    & -  & -& - & - & -\\ 
 \hline     
    \end{tabular}
    \tablefoot{(1) Tested correlation between the two variables, reported in Sect.~\ref{sect:results}; (2) trend significance level when all values are included, and when $C_f$(\ion{H}{i}) is fixed to the upper limit in galaxies with only an upper bound. These values correspond to the significance levels quoted in the paper. (3) The interval of $\sigma$ derived by performing a jackknife test; (4) same as (2) but fixing $C_f$(\ion{H}{i}) to 0 in galaxies with only a upper bound to the \ion{H}{i} covering fraction; (5) same as (2), but only including galaxies with z < 0.5; (6)-(8) significance level of the \fesclya\ trend when \fesclya\ is scaled-down by a factor of 1.5 for galaxies with (6) $v^{\rm sep}_{\rm Ly\alpha} < 400$ km s$^{-1}$; (7) $v^{\rm sep}_{\rm Ly\alpha} < 300$ km s$^{-1}$ and (8) $v^{\rm sep}_{\rm Ly\alpha} < 200$ km s$^{-1}$. All significance level tests include the sample from \citet{mckinney2019} and \citet{jaskot2019} when possible. Dashes indicate that the significance level remained unchanged with respect to (2) because the conditions applied did not affect the sample included.} 
    \end{table*}

 \section{Plots of \lya\ and \ion{H}{i} absorption line profiles.}
     \label{app:lya}

\begin{figure*}
\includegraphics[width = \hsize/3]{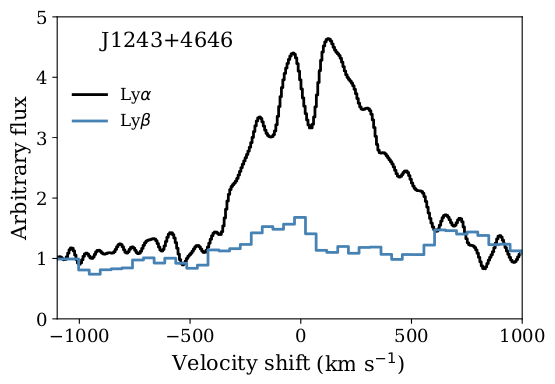}
\includegraphics[width = \hsize/3]{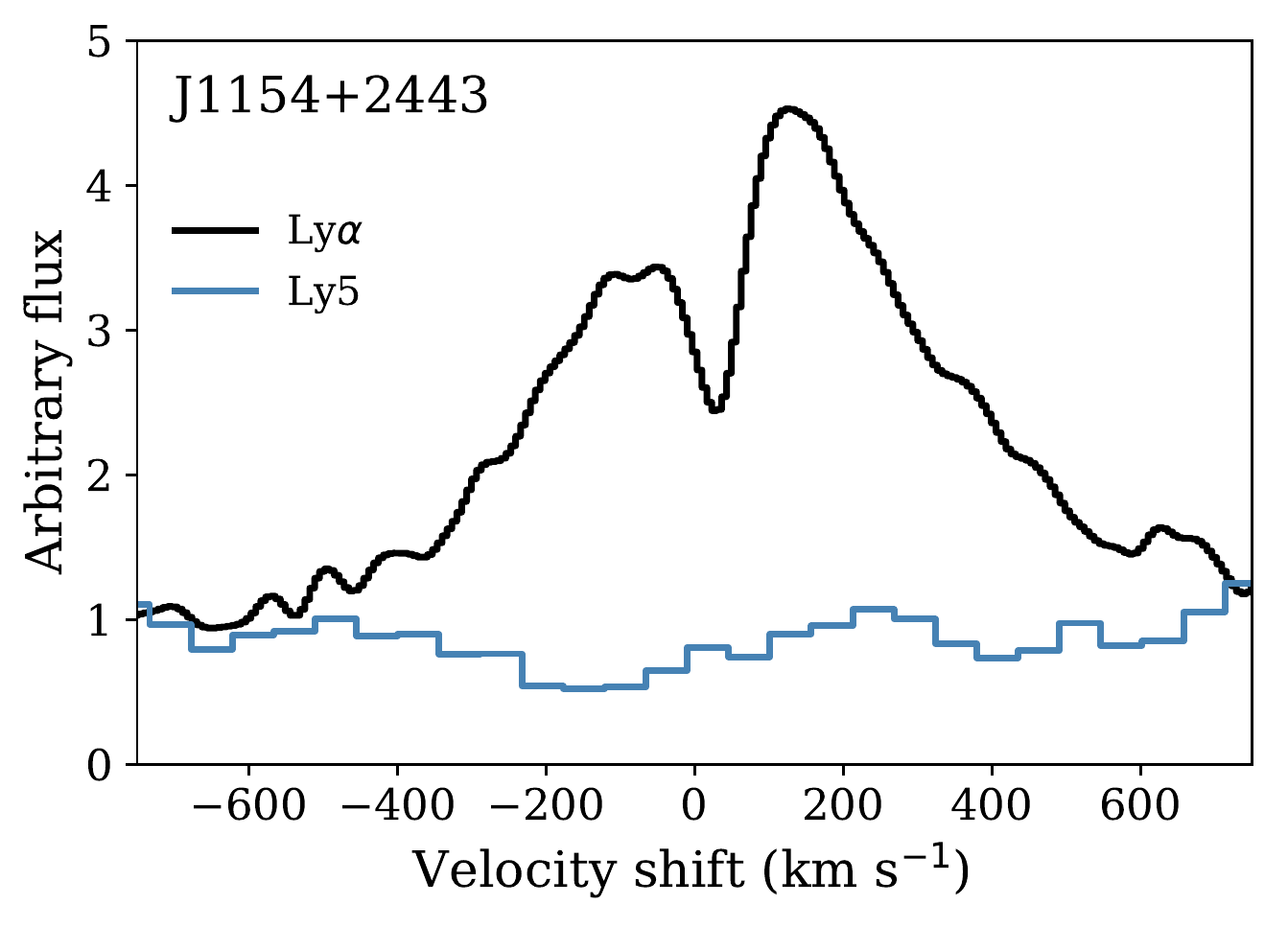}
\includegraphics[width = \hsize/3]{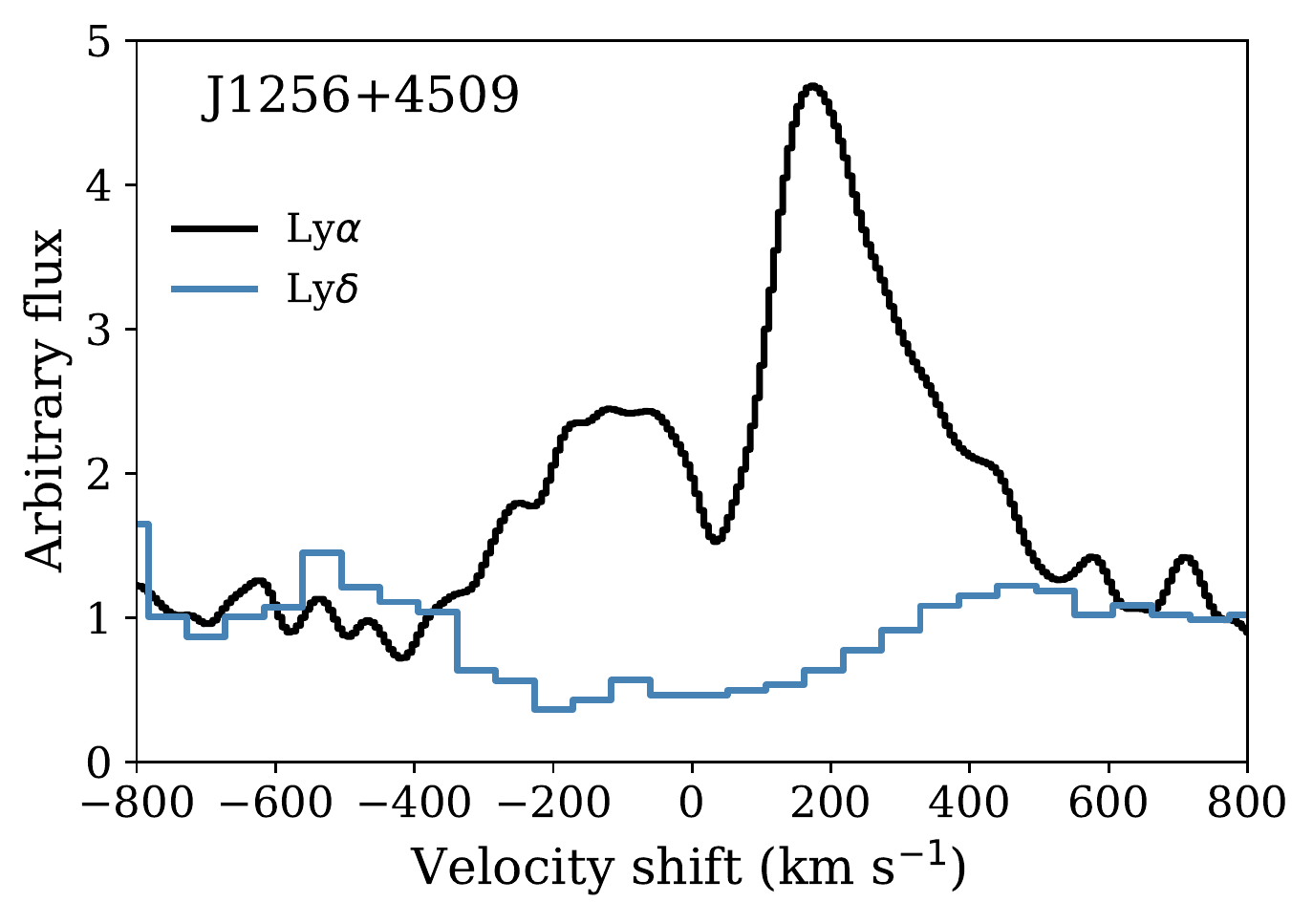}
\includegraphics[width = \hsize/3]{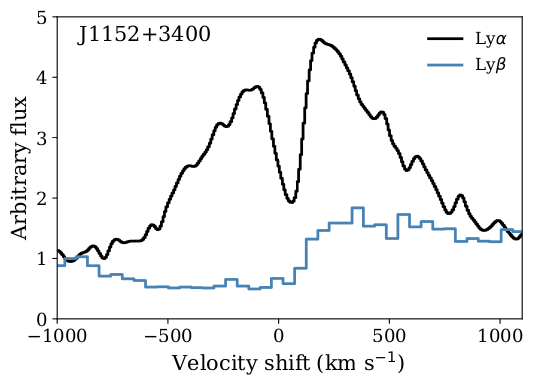}
\includegraphics[width = \hsize/3]{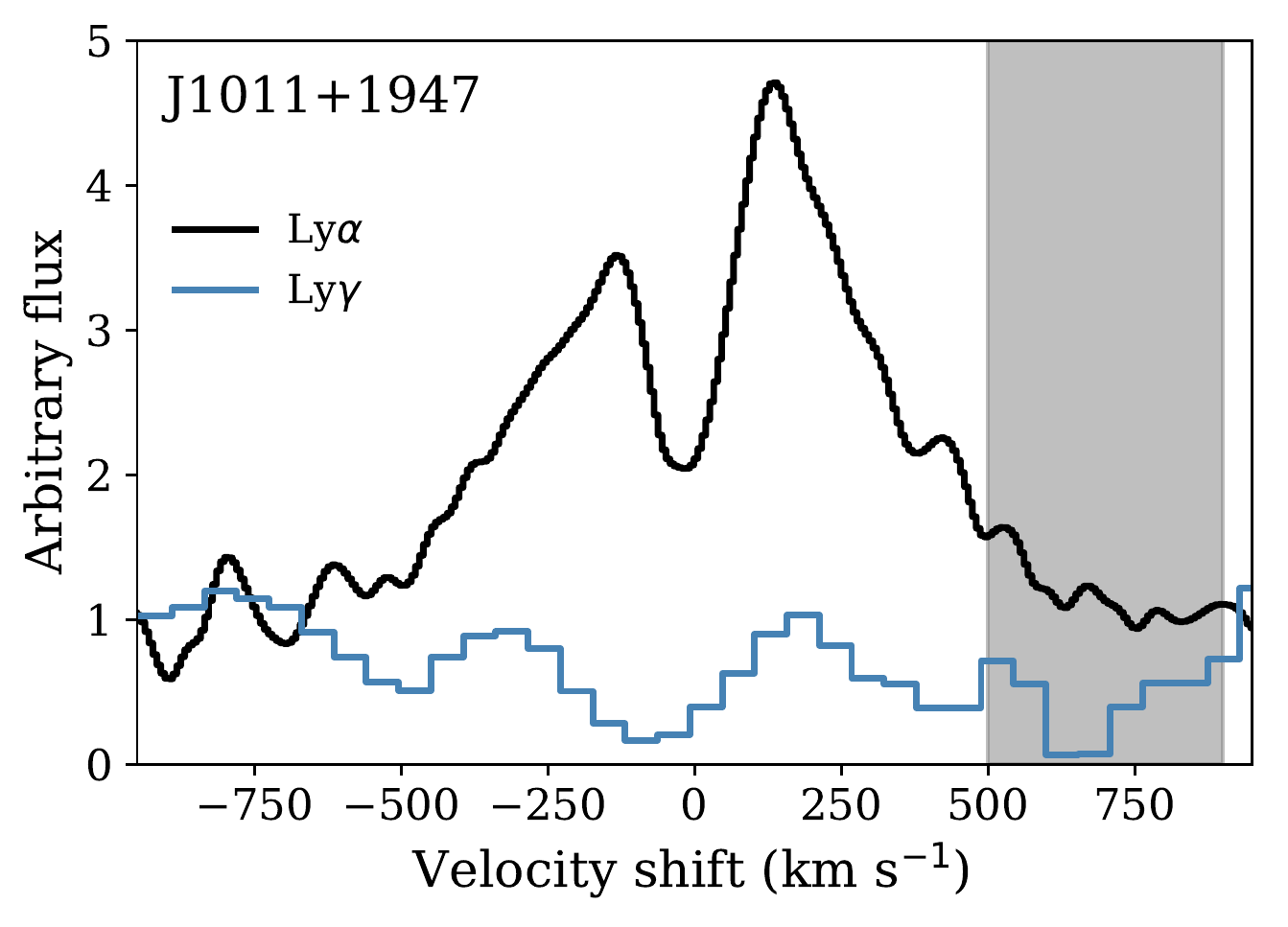}
\includegraphics[width = \hsize/3]{J1442-0209ly.png}
\includegraphics[width = \hsize/3]{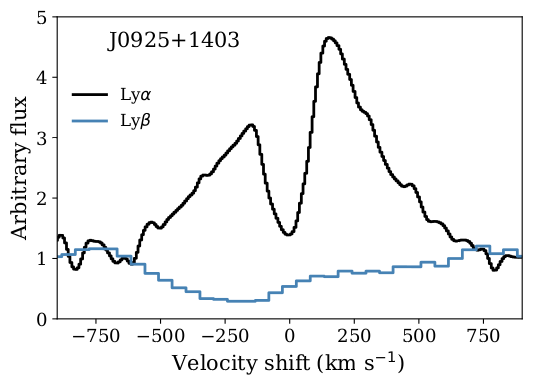}
\includegraphics[width = \hsize/3]{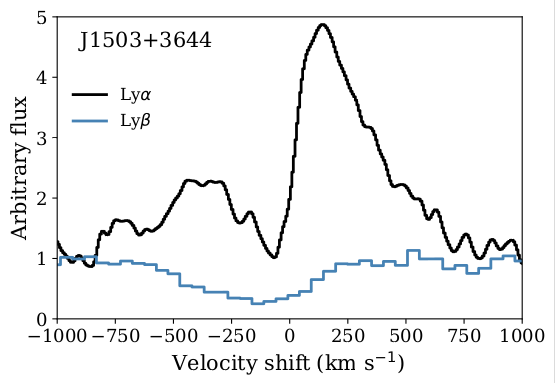}
\includegraphics[width = \hsize/3]{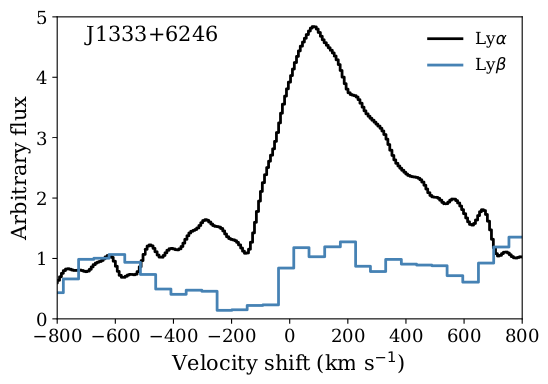}
\includegraphics[width = \hsize/3]{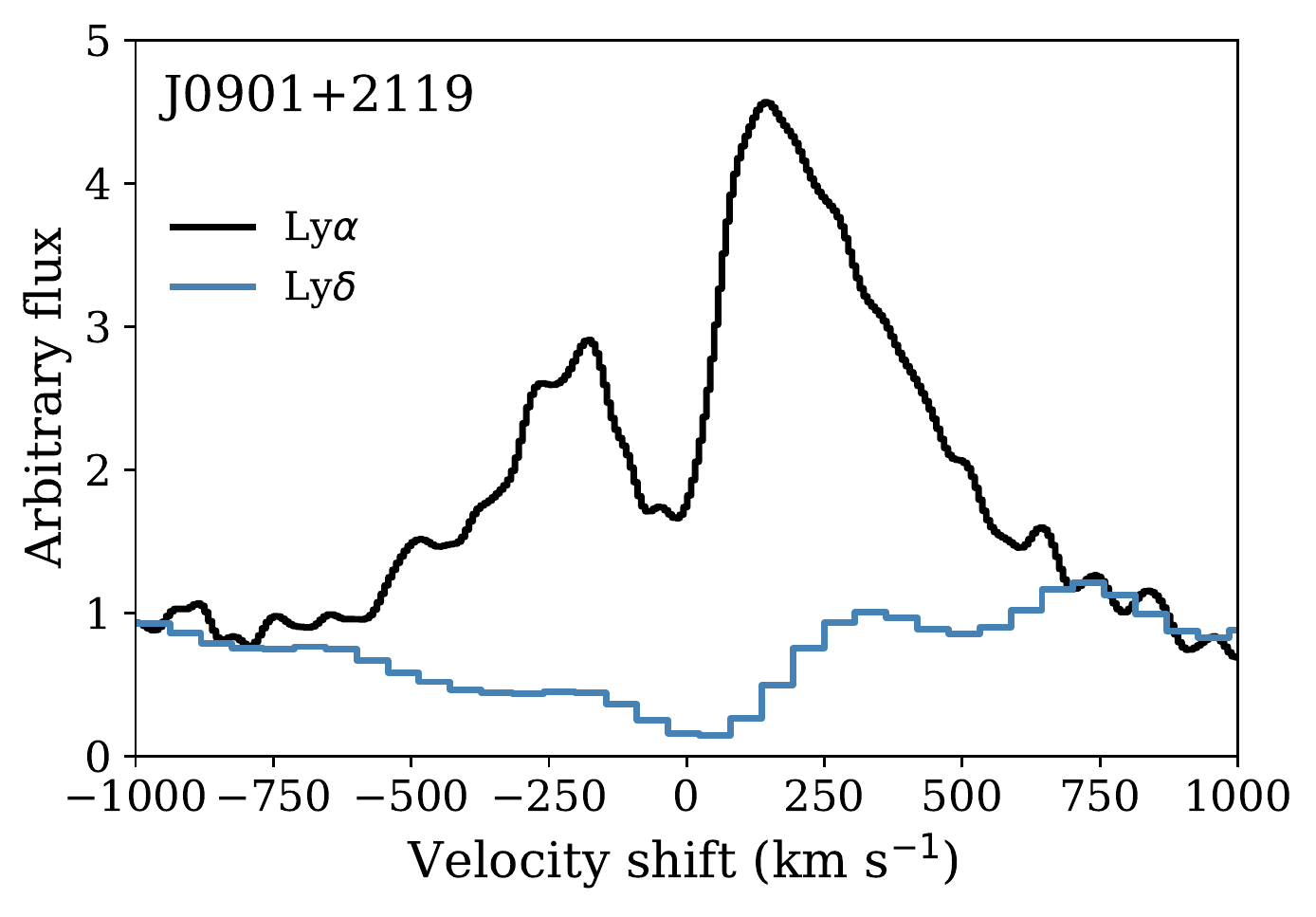}
\includegraphics[width = \hsize/3]{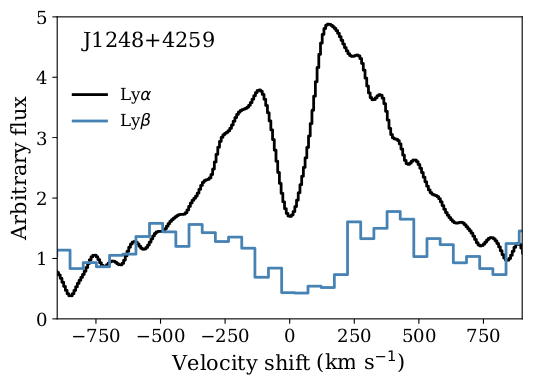}
\includegraphics[width = \hsize/3]{J0921+4509ly.png}
\includegraphics[width = \hsize/3]{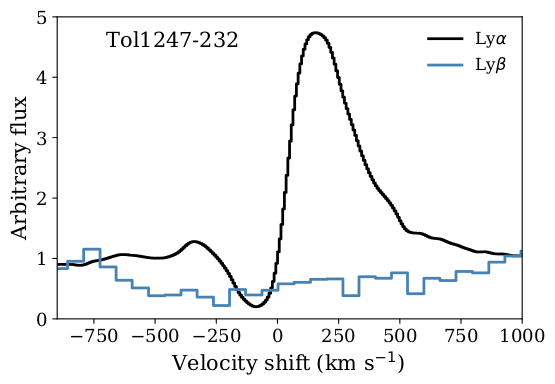}
\includegraphics[width = \hsize/3]{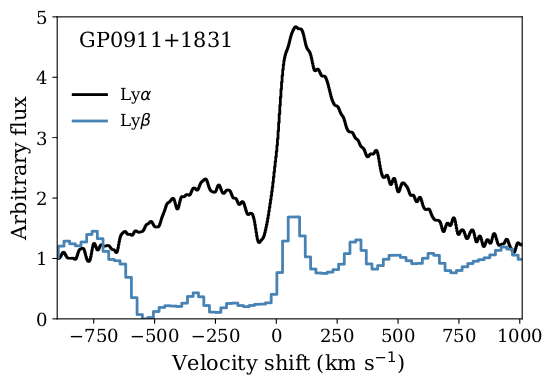}
\includegraphics[width = \hsize/3]{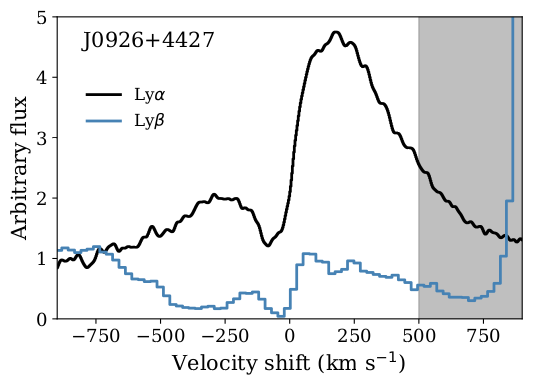}
        \caption{Plot of the \lya\ emission and the observed \ion{H}{i} Lyman series line with the larger S/N in velocity space for 15 of the 22 galaxies in our sample. The \lya\ flux has been smoothed to the resolution of the Lyman series, and scaled down by an arbitrary power law for display purpose. Gray shaded regions show contamination from geocoronal emission or Milky Way absorption lines next to the \ion{H}{i} absorption line.}
        \label{fig:lyaprof}
\end{figure*}

\begin{figure*}
\label{fig:lyaprof2}
\includegraphics[width = \hsize/3]{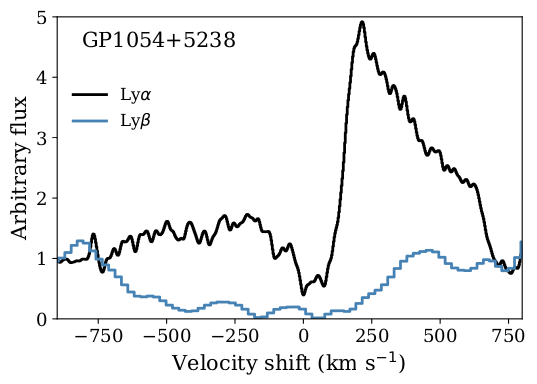}
\includegraphics[width = \hsize/3]{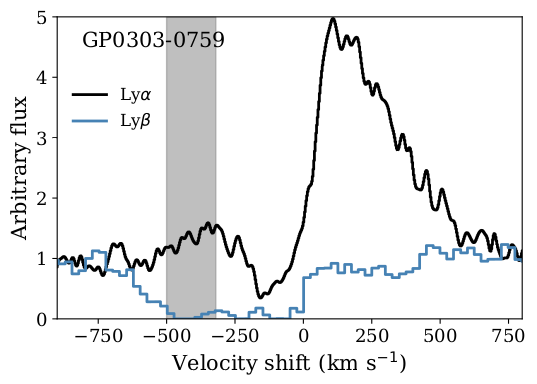}
\includegraphics[width = \hsize/3]{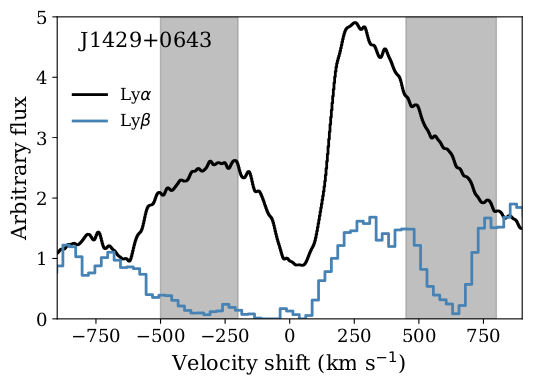}
\includegraphics[width = \hsize/3]{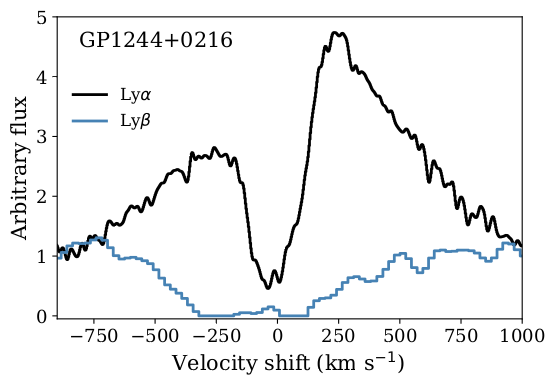}
\includegraphics[width = \hsize/3]{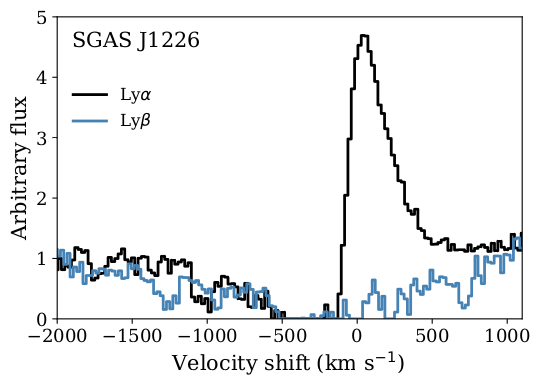}
\includegraphics[width = \hsize/3]{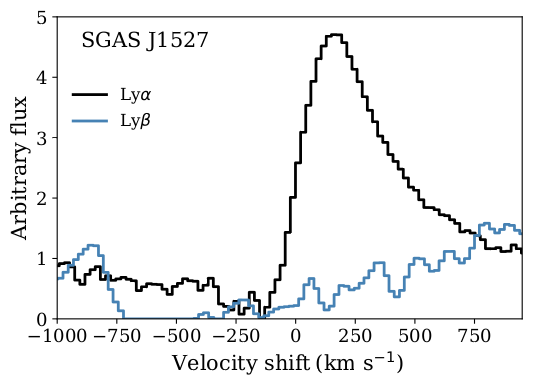}
\includegraphics[width = \hsize/3]{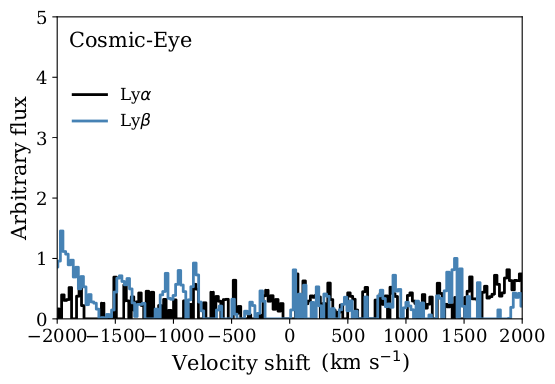}
\caption{Same as Fig.~\ref{fig:lyaprof} for the 7 remaining galaxies.}
\label{fig:lyaproff}
\end{figure*}

Figures~\ref{fig:lyaprof}~and~\ref{fig:lyaproff} show the  \lya\ emission and Ly$\beta$ (or Ly$\gamma$ when Ly$\beta$ is not observed) in velocity space for the 22 galaxies in our sample.

\end{appendix}

\end{document}